\documentclass[aps,pra,twocolumn,superscriptaddress,longbibliography]{revtex4-1}
\usepackage{float}
\usepackage{graphicx}
\usepackage{dsfont}
\usepackage[latin1]{inputenc}
\usepackage{version} 
\usepackage{color}
\usepackage{amssymb}
\usepackage{amsmath}
\usepackage{braket}
\usepackage{bm}
\usepackage{upgreek}
\usepackage{colortbl}
\usepackage{graphicx}
\usepackage{dcolumn}
\usepackage{bm}
\usepackage{bigints}
\usepackage{xcolor}  
\usepackage[pagebackref=false]{hyperref} 

\def\ket#1{|#1\rangle}
\definecolor{jrp}{rgb}{1,0,0}

\definecolor{mjg}{rgb}{.08,.05,.8}

\definecolor{yyl}{rgb}{.8,.05,.08}

\newcommand{\delete}[1]{{}} 
\newcommand{\nocontentsline}[3]{}
\newcommand{\tocless}[2]{\bgroup\let\addcontentsline=\nocontentsline#1{#2}\egroup}

\hypersetup{
    bookmarks=true,         
    unicode=false,          
    pdftoolbar=true,        
    pdfmenubar=true,        
    pdffitwindow=false,     
    pdfstartview={FitH},    
    pdftitle={OTOC_Main},    
    pdfauthor={X. Mi},     
    pdfkeywords={keyword1} {key2} {key3}, 
    pdfnewwindow=true,      
    colorlinks=true,       
    linkcolor=red,          
    citecolor=blue,        
    filecolor=magenta,      
    urlcolor=cyan,           
}

\makeatletter

\makeindex

\begin{document}
\title{Information Scrambling in Computationally Complex Quantum Circuits}

\newcommand{\xGoogle}{\affiliation{Google Research}}
\newcommand{\xUMass}{\affiliation{Department of Electrical and Computer Engineering, University of Massachusetts, Amherst, MA}}
\newcommand{\xUCSB}{\affiliation{Department of Physics, University of California, Santa Barbara, CA}}
\newcommand{\xUCR}{\affiliation{Department of Electrical and Computer Engineering, University of California, Riverside, CA}}
\newcommand{\xPritzker}{\affiliation{Pritzker School of Molecular Engineering, University of Chicago, Chicago, IL}}
\newcommand{\xCalTech}{\affiliation{California Institute of Technology, Pasadena, CA 91125}}
\newcommand{\xQUAIL}{\affiliation{QuAIL, NASA Ames Research Center, Moffett Field, California 94035, USA}}
\newcommand{\xKBR}{\affiliation{KBR, Inc., 601 Jefferson St., Houston, TX 77002, USA}}
\newcommand{\xUSRA}{\affiliation{USRA Research Institute for Advanced Computer Science, Mountain View, California 94043, USA}}

\author{Xiao Mi} \thanks{These authors contributed equally} \xGoogle 
\author{Pedram Roushan} \thanks{These authors contributed equally} \xGoogle 
\author{Chris Quintana} \thanks{These authors contributed equally} \xGoogle
\author{Salvatore Mandr{\` a}} \xQUAIL \xKBR
\author{Jeffrey Marshall} \xQUAIL \xUSRA
\author{Charles Neill}\xGoogle

\author{Frank Arute}\xGoogle
\author{Kunal Arya}\xGoogle
\author{Juan Atalaya}\xGoogle
\author{Ryan Babbush}\xGoogle
\author{Joseph C.~Bardin} \xGoogle \xUMass
\author{Rami Barends}\xGoogle
\author{Andreas Bengtsson}\xGoogle
\author{Sergio Boixo}\xGoogle
\author{Alexandre Bourassa} \xGoogle \xPritzker
\author{Michael Broughton}\xGoogle
\author{Bob B.~Buckley}\xGoogle
\author{David A.~Buell}\xGoogle
\author{Brian Burkett}\xGoogle
\author{Nicholas Bushnell}\xGoogle
\author{Zijun Chen}\xGoogle
\author{Benjamin Chiaro}\xGoogle
\author{Roberto Collins}\xGoogle
\author{William Courtney}\xGoogle
\author{Sean Demura}\xGoogle
\author{Alan R. Derk}\xGoogle
\author{Andrew Dunsworth}\xGoogle
\author{Daniel Eppens}\xGoogle 
\author{Catherine Erickson}\xGoogle
\author{Edward Farhi}\xGoogle
\author{Austin G.~Fowler}\xGoogle
\author{Brooks Foxen}\xGoogle
\author{Craig Gidney}\xGoogle
\author{Marissa Giustina}\xGoogle
\author{Jonathan A.~Gross}\xGoogle
\author{Matthew P.~Harrigan}\xGoogle
\author{Sean D.~Harrington}\xGoogle
\author{Jeremy Hilton}\xGoogle
\author{Alan Ho}\xGoogle
\author{Sabrina Hong}\xGoogle
\author{Trent Huang}\xGoogle
\author{William J. Huggins}\xGoogle
\author{L.~B.~Ioffe}\xGoogle
\author{Sergei V.~Isakov}\xGoogle
\author{Evan Jeffrey}\xGoogle
\author{Zhang Jiang}\xGoogle
\author{Cody Jones}\xGoogle
\author{Dvir Kafri}\xGoogle
\author{Julian Kelly}\xGoogle
\author{Seon Kim}\xGoogle
\author{Alexei Kitaev} \xGoogle \xCalTech
\author{Paul V.~Klimov}\xGoogle
\author{Alexander N.~Korotkov} \xGoogle \xUCR
\author{Fedor Kostritsa}\xGoogle
\author{David Landhuis}\xGoogle
\author{Pavel Laptev}\xGoogle
\author{Erik Lucero}\xGoogle
\author{Orion Martin}\xGoogle
\author{Jarrod R.~McClean}\xGoogle
\author{Trevor McCourt}\xGoogle
\author{Matt McEwen} \xGoogle \xUCSB
\author{Anthony Megrant}\xGoogle
\author{Kevin C.~Miao}\xGoogle
\author{Masoud Mohseni}\xGoogle
\author{Wojciech Mruczkiewicz}\xGoogle
\author{Josh Mutus}\xGoogle
\author{Ofer Naaman}\xGoogle
\author{Matthew Neeley}\xGoogle
\author{Michael Newman}\xGoogle
\author{Murphy Yuezhen Niu}\xGoogle
\author{Thomas E.~O'Brien}\xGoogle
\author{Alex Opremcak}\xGoogle
\author{Eric Ostby}\xGoogle
\author{Balint Pato}\xGoogle
\author{Andre Petukhov}\xGoogle
\author{Nicholas Redd}\xGoogle
\author{Nicholas C.~Rubin}\xGoogle
\author{Daniel Sank}\xGoogle
\author{Kevin J.~Satzinger}\xGoogle
\author{Vladimir Shvarts}\xGoogle
\author{Doug Strain}\xGoogle
\author{Marco Szalay}\xGoogle
\author{Matthew D.~Trevithick}\xGoogle
\author{Benjamin Villalonga}\xGoogle
\author{Theodore White}\xGoogle
\author{Z.~Jamie Yao}\xGoogle
\author{Ping Yeh}\xGoogle
\author{Adam Zalcman}\xGoogle
\author{Hartmut Neven}\xGoogle
\author{Igor Aleiner} \xGoogle

\author{Kostyantyn Kechedzhi} \email[Corresponding author: ]{kostyantyn@google.com} \xGoogle
\author{Vadim Smelyanskiy} \email[Corresponding author: ]{smelyan@google.com} \xGoogle
\author{Yu Chen} \email[Corresponding author: ]{bryanchen@google.com} \xGoogle

\begin{abstract}

Interaction in quantum systems can spread initially localized quantum information into the many degrees of freedom of the entire system. Understanding this process, known as quantum scrambling, is the key to resolving various conundrums in physics. Here, by measuring the time-dependent evolution and fluctuation of out-of-time-order correlators, we experimentally investigate the dynamics of quantum scrambling on a 53-qubit quantum processor. We engineer quantum circuits that distinguish the two mechanisms associated with quantum scrambling, operator spreading and operator entanglement, and experimentally observe their respective signatures. We show that while operator spreading is captured by an efficient classical model, operator entanglement requires exponentially scaled computational resources to simulate. These results open the path to studying complex and practically relevant physical observables with near-term quantum processors.

\end{abstract}

\maketitle
The inception of quantum computers was motivated by their ability to simulate dynamical processes that are challenging for classical computation \cite{Feynman_1982}. However, while the size of the Hilbert space scales exponentially with the number of qubits, quantum dynamics can be simulated in polynomial times when entanglement is insufficient \cite{Valiant_2002, Terhal_PRA_2002, Vidal_PRL_2004} or when they belong to special classes such as the Clifford group \cite{Gottesman_PRA_1996, Aaronson_PRA_2004, Bravyi_PRL_2016}. A physical process that fully leverages the computational power of quantum processors is quantum scrambling, which describes how interaction in a quantum system disperses local information into its many degrees of freedom \cite{Hayden_JHEP_2007, Sekino_JHEP_2008, Lashkari_JHEP_2013, AleinerIoffe_AP_2016, Zhuang_PRA_2019}. Quantum scrambling is the underlying mechanism for the thermalization of isolated quantum systems \cite{Deutsch_PRA_1991, Rigol_Nature_2008} and as such, accurately modeling its dynamics is the key to resolving a number of physical phenomena, such as the fast-scrambling conjecture for black holes \cite{Sekino_JHEP_2008, Lashkari_JHEP_2013}, non-Fermi liquid behaviors \cite{Sachdev2017, StrangeMetal2018} and many-body localization \cite{Aleiner_AP_2006}. Understanding scrambling also provides a basis for designing algorithms in quantum benchmarking or machine learning that would benefit from efficient exploration of the Hilbert space \cite{McClean_NC_2018, Knill_PRA_2008,CliffordTdesigns}.

A precise formulation of quantum scrambling is found in the Heisenberg picture, where quantum operators evolve and quantum states are stationary. Analogous to classical chaos, scrambling manifests itself as a ``butterfly effect'', wherein a local perturbation is rapidly amplified over time \cite{Roberts_JHEP_2015, AleinerIoffe_AP_2016}. More specifically, the perturbation is realized as an initially local operator (the ``butterfly operator'') $\hat{O}$, typically a Pauli operator acting on one of the qubits (the ``butterfly qubit''). When the quantum system undergoes a dynamical process $\hat{U}$, the butterfly operator $\hat{O}$ acquires a time-dependence and becomes $\hat{O} \left( t \right) = \hat{U}^\dagger  \hat{O} \hat{U}$, with $\hat{U}^\dagger$ being the inverse of $\hat{U}$. The resulting $\hat{O} \left( t \right)$ can be expanded as $\hat{O} \left( t \right) = \sum_{i = 1}^{n_\text{p}} w_i {\hat{B}_i}$, where $\hat{B}_i= \hat{\rho}_1^{(i)} \otimes \hat{\rho}_2^{(i)} \otimes ... $ are basis operators consisting of single-qubit operators $\hat{\rho}_j^{(i)}$ acting on different qubits and $w_i$ are their coefficients. 

Quantum scrambling is enabled by two different mechanisms: operator spreading and operator entanglement \cite{AleinerIoffe_AP_2016, Roberts_JHEP_2017, Nahum_PRX_2018, Keyserlingk_PRX_2018, ParticleConservingOTOCPollman,  Khemani_PRX_2018}. Operator spreading refers to the transformation of basis operators such that on average, each $\hat{B}_i$ involves a higher number of non-identity single-qubit operators. Operator entanglement, on the other hand, refers to the increase in $n_\text{p}$, i.e. the minimum number of terms required to expand $\hat{O} \left( t \right)$.  Independent characterizations of these two mechanisms are essential for a complete understanding of the nature of quantum scrambling. Quantifying the degree of operator entanglement also holds the key to assessing the classical simulation complexity of quantum observables \cite{plenio_PRL_2009}. However, operator spreading and operator entanglement are generally intertwined and indistinguishable in past experimental studies of quantum scrambling \cite{Li_PRX_2017, Garttner2017, Yao_Nature_2019, blok_arxiv_2020, Manoj_PRL_2020}.

\begin{figure}[t!]
	\centering
	\includegraphics[width=1\columnwidth]{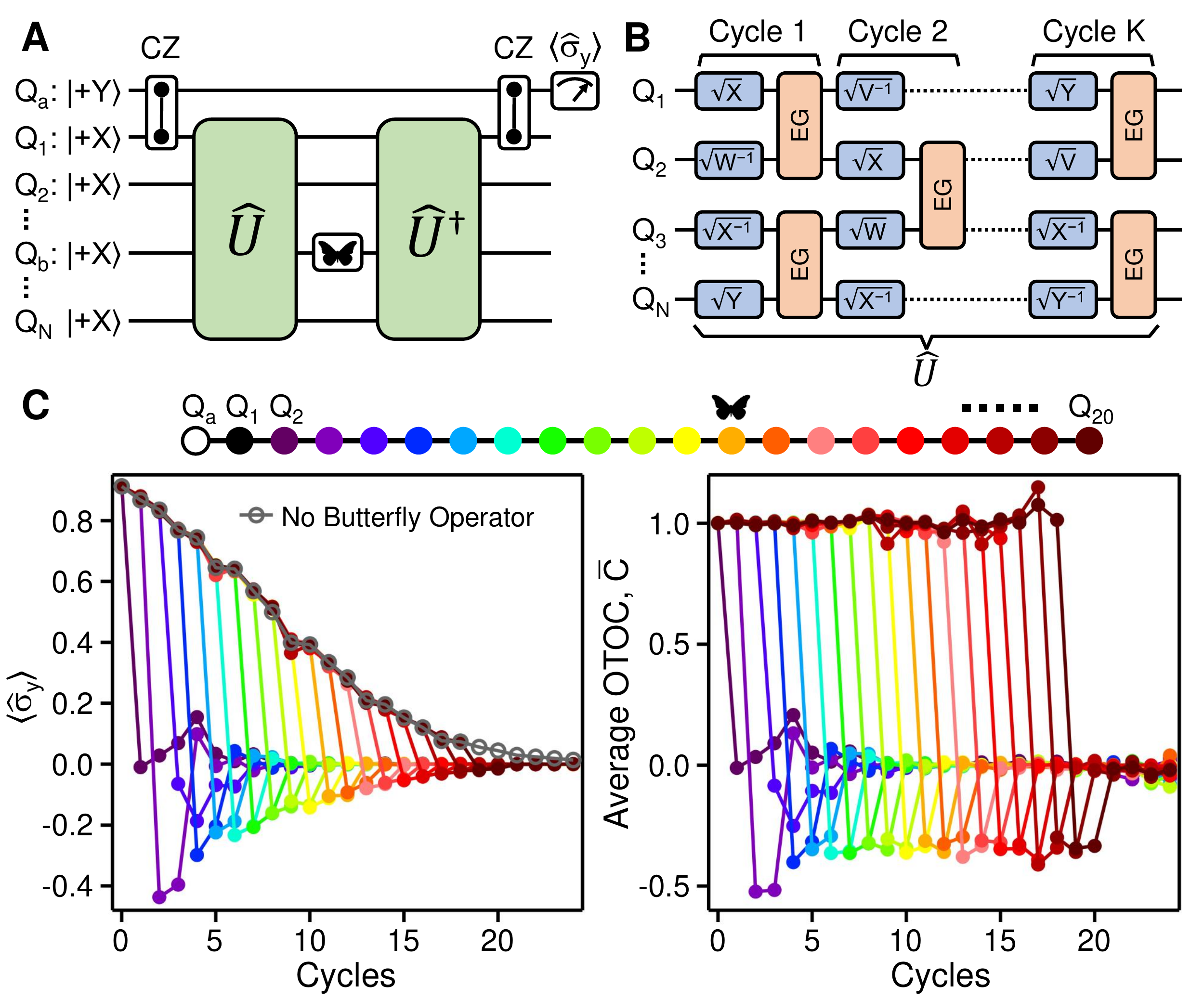}
	\caption{\textbf{OTOC measurement protocol.} (A) Measurement scheme for the OTOC of a quantum circuit $\hat{U}$. A quantum system (qubits $Q_1$ through $Q_N$) is first initialized in a superposition state $\prod_{j = 1}^{j = N}\ket{+X}_j$, where $\ket{+X}_j = \frac{1}{\sqrt{2}} \left( \ket{0}+ \ket{1}\right)_j$ and $\ket{0}$ ($\ket{1}$) is the ground (excited) state of individual qubits. $\hat{U}$ and its inverse $\hat{U}^\dagger$ are then applied, with a butterfly operator (realized as an $X$ gate on qubit $Q_\text{b}$) inserted in-between. An ancilla qubit $Q_\text{a}$ is initialized along the y-axis of the Bloch sphere and entangled with $Q_1$ by a pair of CZ gates. The y-axis projection of $Q_\text{a}$, $\braket{\hat{\sigma}_\text{y}}$, is measured at the end. (B) The structure of $\hat{U}$ consists of $K$ cycles: each cycle includes one layer of single-qubit gates (randomly chosen from $\sqrt{X^{\pm 1}}$, $\sqrt{Y^{\pm 1}}$, $\sqrt{W^{\pm 1}}$ and $\sqrt{V^{\pm 1}}$) and one layer of two-qubit entangling gates (EG). Here $W = \frac{X + Y}{\sqrt{2}}$ and $V = \frac{X - Y}{\sqrt{2}}$. (C) Left panel: The filled circles represent $\braket{\hat{\sigma}_\text{y}}$ measured with $Q_\text{b}$ successively chosen from $Q_2$ through $Q_N$. The open grey circles are a set of normalization values, corresponding to $\braket{\hat{\sigma}_\text{y}}$ without applying the butterfly operator. The data are plotted over different numbers of cycles in $\hat{U}$ and averaged over 60 random circuit instances. Right panel: Experimental average OTOCs $\overline{C}$ for different $Q_\text{b}$, obtained from dividing the corresponding $\braket{\hat{\sigma}_\text{y}}$ by the normalization values.}
	\label{fig:1}
\end{figure} 

In this Article, we perform a comprehensive characterization of quantum scrambling in a two-dimensional (2D) quantum system of 53 superconducting qubits. Signatures of operator spreading and operator entanglement are clearly distinguished in our experiment. These results are enabled by quantum circuit designs that independently tune the degree of each scrambling mechanism, as well as extensive error-mitigation techniques that allow us to faithfully recover coherent experimental signals in the presence of substantial noise. Lastly, we find that while operator spreading can be efficiently predicted by a classical stochastic process, simulating the experimental signature of operator entanglement is significantly more costly with a computational resource that scales exponentially with the size of the quantum circuit.

\begin{figure*}[t!]
	\centering
	\includegraphics[width=2\columnwidth]{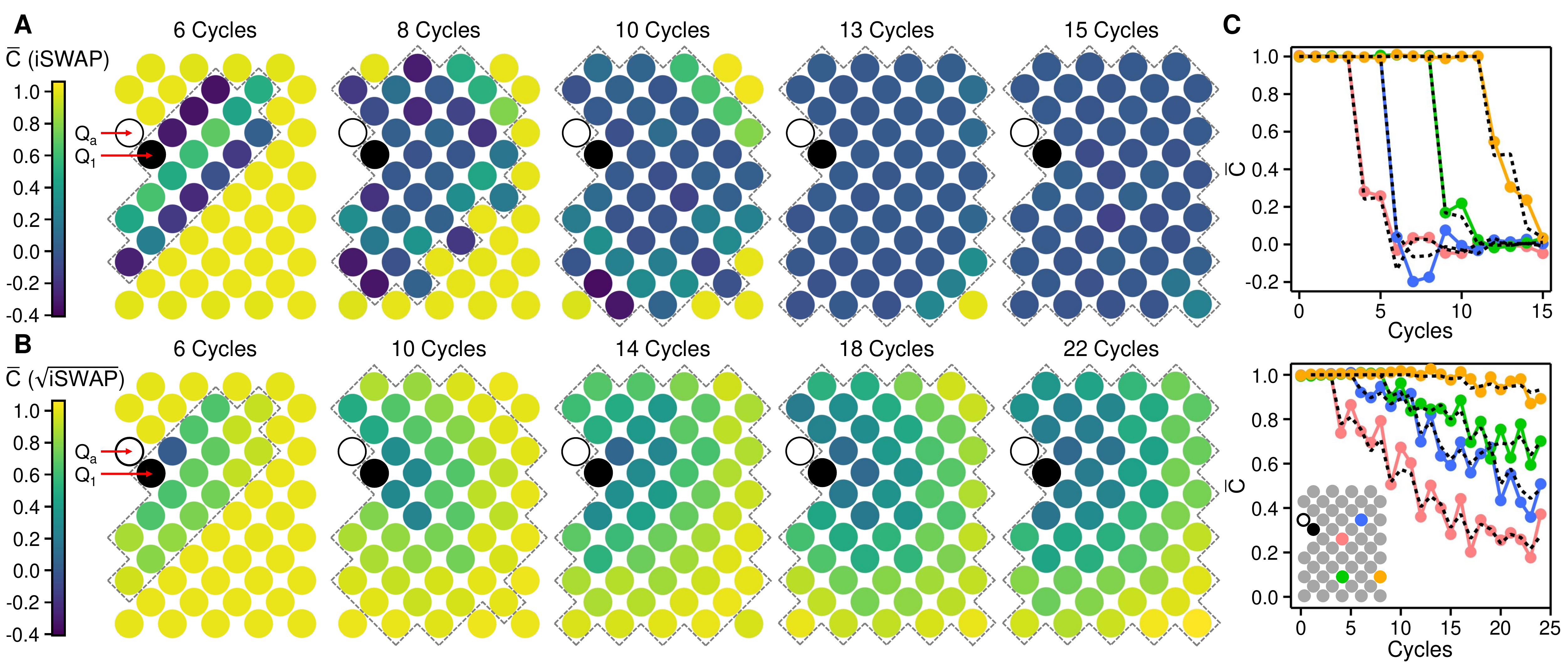}
	\caption{\textbf{OTOC propagation and speed of operator spreading.} (A) Spatial profiles of average OTOCs, $\overline{C}$, measured on the full 53-qubit processor. The ancilla qubit $Q_\text{a}$ and the the measurement qubit $Q_1$ are indicated by the red arrows. The colors of other filled circles represent $\overline{C}$ with different choices of the butterfly qubit $Q_\text{b}$. The two-qubit gates are iSWAP and applied between all nearest-neighbor qubits, with the same order as done in Ref.~\cite{Arute2019}. The dashed lines delineate the light-cone of $Q_1$. The data are averaged over 38 circuit instances. (B) Similar to (A) but with $\sqrt{\text{iSWAP}}$ as the two-qubit gates. Here $\overline{C}$ is averaged over 24 circuit instances. (C) Cycle-dependent $\overline{C}$ for 4 different choices of $Q_\text{b}$. The top (bottom) panel shows data with iSWAP ($\sqrt{\text{iSWAP}}$) being the two-qubit gates. The colors of the data points indicate the locations of $Q_\text{b}$ (inset to bottom panel). Dashed lines show theoretical predictions based on a classical population dynamics model.}
	\label{fig:2}
\end{figure*} 

Our experiment approach is based on evaluating the correlator between $\hat{O} \left( t \right)$ and a ``measurement operator'', $\hat{M}$, which is another Pauli operator on a different qubit (the ``measurement qubit''):
\begin{equation}
C(t) = \langle \hat{O}^\dagger(t)\,\hat{M}^\dagger\,\hat{O}(t)\,\hat{M}\rangle.
\end{equation}
Here $\langle ... \rangle$ denotes the expectation value over a particular quantum state. $C(t)$ is commonly known as the out-of-time-order correlator (OTOC) and related to the commutator $[\hat{O}(t),  \hat{M}]$ by $C(t) = 1 - \frac{1}{2} \langle | [ \hat{O}(t),  \hat{M} ]| ^ 2 \rangle$ \cite{Roberts_JHEP_2015, AleinerIoffe_AP_2016, Hosur_JHEP_2016, Swingle_PRA_2016, Yoshida_PRX_2019, Vermersch_PRX_2019}. Quantum scrambling is characterized by measuring $C$ over a collection of quantum circuits with microscopic differences, e.g. phases of individual gates. Operator spreading is then reflected in the average OTOC value, $\overline{C}$, which decays from 1 when $\hat{O}(t)$ and $\hat{M}$ overlap and no longer commute \cite{Yao_Nature_2019, blok_arxiv_2020}. In the fully scrambled limit where the commutation between $\hat{O}(t)$ and $\hat{M}$ is completely randomized, $\overline{C}$ becomes $\sim$0. If operator entanglement is also present (i.e. $n_\text{p} \gg 1$), $C$ approaches 0 for all circuits and their fluctuation $\delta_\text{C}$ vanishes as well. This is because each $C$ includes contributions from many basis operators $\hat{B}_i$ with different phases. It is therefore sufficient to identify operator spreading through the decay of $\overline{C}$, whereas any insight into operator entanglement necessitates an estimate of $\delta_\text{C}$.

The measurement protocol for OTOC is described in Fig.~\ref{fig:1}A and consists of a quantum circuit $\hat{U}$ and its inverse $\hat{U}^\dagger$, with a butterfly operator $\hat{O}$ (Pauli operator $\hat{\sigma}_\text{x}^{(\text{b})}$ on $Q_\text{b}$) inserted in-between. An ancilla qubit $Q_\text{a}$, connected to the measurement qubit $Q_1$ via a controlled-phase (CZ) gate, measures $C$ between $\hat{O}$ and $\hat{M}$ (Pauli operator $\hat{\sigma}_\text{z}^{(1)}$ on $Q_1$) through its $\braket{\hat{\sigma}_\text{y}}$ \cite{Swingle_PRA_2016, footnote}. For this work, we employ quantum circuits composed of random single-qubit gates and fixed two-qubit gates (Fig.~\ref{fig:1}B) due to the wide range of quantum scrambling that may be achieved with limited circuit depths \cite{Boixo_NatPhys_2018, Arute2019}. The OTOC measurement protocol is first implemented on a one-dimensional (1D) chain of 21 qubits (Fig.~\ref{fig:1}C). We use the qubit at one end of the chain as $Q_\text{a}$ and successively choose qubits $Q_2$ through $Q_{20}$ as $Q_\text{b}$. A two-qubit gate (iSWAP) is applied to each pair $(Q_j, Q_{j+1})$, where $j = 0, 2, 4...$ in odd circuit cycles and $j = 1, 3, 5...$ in even circuit cycles.

\begin{figure*}[t!]
	\centering
	\includegraphics[width=2\columnwidth]{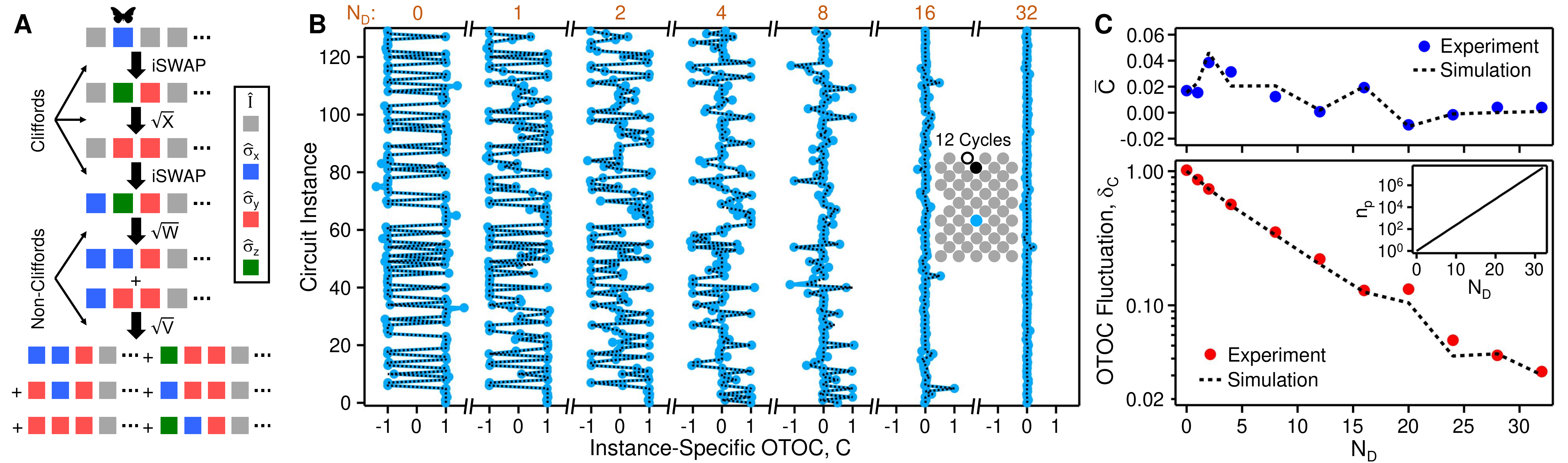}
	\caption{\textbf{OTOC fluctuation and signature of operator entanglement.} (A) Transformation of a butterfly operator into products of Pauli operators (Pauli strings) by quantum gates. In this example, an initial butterfly operator $\hat{I}^{(1)} \hat{\sigma}_\text{x}^{(2)} \hat{I}^{(3)} \hat{I}^{(4)} ...$ is mapped into $\hat{I}^{(1)} \hat{\sigma}_\text{z}^{(2)} \hat{\sigma}_\text{y}^{(3)} \hat{I}^{(4)} ...$ by an iSWAP gate, then into $\hat{I}^{(1)} \hat{\sigma}_\text{y}^{(2)} \hat{\sigma}_\text{y}^{(3)} \hat{I}^{(4)} ...$ by an $\sqrt{X}$ gate and so on. In the last two steps, the butterfly operator evolves into a superposition of multiple Pauli strings (coefficients not shown). (B) OTOCs of individual random circuit instances, $C$, measured with the number of non-Clifford gates in $\hat{U}$, $N_\text{D}$, fixed at different values. Dashed lines are numerical simulation results. The inset shows locations of $Q_\text{a}$ (open black circle), $Q_\text{1}$ (filled black circle), and $Q_\text{b}$ (filled blue circle) as well as the number of circuit cycles with which the data are taken. (C) The mean $\overline{C}$ (upper panel) and RMS values $\delta_\text{C}$ (lower panel) of $C$ for different $N_\text{D}$. Dashed lines are computed from the numerically simulated values in (B). Inset: Average numbers of Pauli strings in the time-evolved butterfly operator $\hat{O} \left( t \right)$, $n_\text{p}$, for different $N_\text{D}$. The scaling behavior is $n_\text{p} \approx 2^{0.79 N_\text{D}}$ \cite{sup}.}
	\label{fig:3}
\end{figure*} 

In the left panel of Fig.~\ref{fig:1}C, experimental values of $\braket{\hat{\sigma}_\text{y}}$ are shown for different numbers of cycles in $\hat{U}$. Here we average $\braket{\hat{\sigma}_\text{y}}$ over 60 circuit instances to first focus on operator spreading. It is seen that $\braket{\hat{\sigma}_\text{y}}$ decays as early as cycle 1, irrespective of the location of $Q_\text{b}$. This observation contradicts the 1D geometry which requires $h + 1$ circuit cycles before the time-evolved butterfly operator $\hat{O} \left( t \right)$ overlaps with $\hat{M}$, with $h$ being the number of qubits between $Q_\text{b}$ and $Q_1$. The signals are therefore complicated by errors in the quantum circuits, such as mismatch between $\hat{U}$ and $\hat{U}^\dagger$ or qubit decoherence \cite{Garttner2017, Yao_Nature_2019}. These deleterious effects are mitigated by additionally measuring $\braket{\hat{\sigma}_\text{y}}$ without applying the butterfly operator \cite{Swingle_PRA_2018_Renorm}. These data, referred to as normalization values, are also shown in the left panel of Fig.~\ref{fig:1}C and approximately equal to the total fidelities of $\hat{U}$ and $\hat{U}^\dagger$ \cite{Garttner2017}. We then divide $\braket{\hat{\sigma}_\text{y}}$ for each $Q_\text{b}$ by the normalization values to recover the effects of scrambling \cite{sup}.

The normalized data, equal to the average OTOCs $\overline{C}$ after error-mitigation, are shown in the right panel of Fig.~\ref{fig:1}C and exhibit features consistent with operator spreading: For each location of $Q_\text{b}$, $\overline{C}$ retains values near 1 before sufficient circuit cycles have occurred to allow an overlap between $\hat{O} \left( t \right)$ and $\hat{M}$. Beyond these cycles, $\overline{C}$ converges to 0, indicating that $\hat{O} \left( t \right)$ and $\hat{M}$ have overlapped and no longer commute. In addition, we observe that the time-evolution of $\overline{C}$ for each $Q_\text{b}$ resembles a ballistically propagating wave. The front of each wave coincides with the edge of the ``light-cone'' associated with $Q_1$, i.e. the set of qubits that have been entangled with $Q_1$. This profile is attributed to the iSWAP gates used in these circuits, which spread single-qubit operators at the same rate as their light-cones expand \cite{Claeys_PRReseach_2020}. For generic quantum circuits, the spreading velocity (a.k.a. the butterfly velocity) is typically slower. Using the full 2D system, we next demonstrate how the evolution of $\overline{C}$ may be used to diagnose the butterfly velocity of operator spreading.

In Fig.~\ref{fig:2}A, the spatial distribution of $\overline{C}$ is shown for five different numbers of cycles in $\hat{U}$, with iSWAP still being the two-qubit gate. It is seen that the number of qubits with $\overline{C} < 1$ rapidly increases with the number of cycles, consistent with the spatial spread of the time-evolved butterfly operator. Moreover, for each circuit cycle, the values of $\overline{C}$ abruptly change across the edge of the light-cone associated with $Q_1$ (dashed lines in Fig.~\ref{fig:2}A). In contrast, the spatiotemporal evolution of $\overline{C}$ shown in Fig.~\ref{fig:2}B is significantly different. Here the iSWAP gates are replaced with $\sqrt{\text{iSWAP}}$ gates and the decay of $\overline{C}$ is slower. Qubits far from $Q_1$ still retain average OTOC values closer to 1 even after 22 cycles. The sharp, step-like spatial transition seen with iSWAP is also absent for $\sqrt{\text{iSWAP}}$. Instead, $\overline{C}$ changes in a gradual fashion as $Q_\text{b}$ moves further away from $Q_1$.  

The different OTOC behaviors can alternatively be seen in the full temporal evolution of four specific qubits (Fig.~\ref{fig:2}C). For iSWAP, the shape of the OTOC wavefront remains sharp and relatively insensitive to the location of $Q_\text{b}$, similar to the 1D example in Fig.~\ref{fig:1}C. On the other hand, the wavefront propagates more slowly for $\sqrt{\text{iSWAP}}$ and also broadens as the distance between $Q_\text{b}$ and $Q_1$ increases. As a result, more circuit cycles are required before $\overline{C}$ reaches 0 for $\sqrt{\text{iSWAP}}$. The wavefront behavior seen with $\sqrt{\text{iSWAP}}$ is similar to generic quantum circuits analyzed in past works \cite{Nahum_PRX_2018, Keyserlingk_PRX_2018}.

The observed features of average OTOCs are quantitatively understood by mapping operator spreading to a classical Markov process involving population dynamics \cite{sup}. In this model, the 2D qubit lattice is populated by fictitious particles representing two copies of a single-qubit operator. The initial state of the entire system is a single particle at the site of $Q_\text{a}$. Whenever a two-qubit gate is applied to two neighboring lattice sites, their particle occupation changes between four possible states: $\lozenge \lozenge$ (both empty), $\lozenge \blacklozenge$ (left empty, right filled), $\blacklozenge \lozenge$ (right empty, left filled) and $\blacklozenge \blacklozenge$ (both filled). The transition probabilities are described by the stochastic matrix:
\begin{gather}
\Omega = \left(
\begin{array}{cccc}
  1 & 0 & 0 & 0 \\
  0 & 1-a-b & a & b \\
  0 & a & 1-a-b & b \\
  0 & \frac{b}{3} & \frac{b}{3} & 1-\frac{2}{3}b \\
\end{array}
\right), \label{eq:updateO}
\end{gather} 
where $a = \frac{1}{3}\sin^4\theta$, $ b = \frac{1}{3}\left(\frac{1}{2} \sin^2 2\theta + 2 \sin^2\theta \right)$ with $\theta$ being the swap angle of the two-qubit gate. The average probability of finding a particle at the site of $Q_1$ is then used to estimate $\overline{C}$. In this classical picture, the OTOC wavefront corresponds to the boundary separating the empty region from the region populated by particles.

The difference in OTOC propagation between iSWAP and $\sqrt{\text{iSWAP}}$ is then captured by the dependence of $\Omega$ on $\theta$: After each application of an iSWAP gate ($\theta = \pi/2$), the particle occupation always changes (with the exception of $\lozenge \lozenge$). In particular, any previously empty site will be filled and the region populated by particles always grows. This leads to the observed maximal butterfly velocity. In contrast, the application of any $\sqrt{\text{iSWAP}}$ gate ($\theta = \pi/4$) can leave the particle occupation unchanged with a probability $\frac{5}{12}$. $\overline{C}$ therefore decays more slowly in this case and its broadening is explained by the fact that the wavefront spreads at a different velocity for each trial of the Markov process. The predicted values of $\overline{C}$, plotted as dashed lines in Fig.~\ref{fig:2}C, agree well with the experimental data and indicate that the dynamics of operator spreading can be reliably predicted by classical models. We note that the effect of noise is included when $\overline{C}$ is calculated for $\sqrt{\text{iSWAP}}$ as it is found to introduce deformation to the observed signals \cite{sup}.

\begin{figure}[t!]
	\centering
	\includegraphics[width=1\columnwidth]{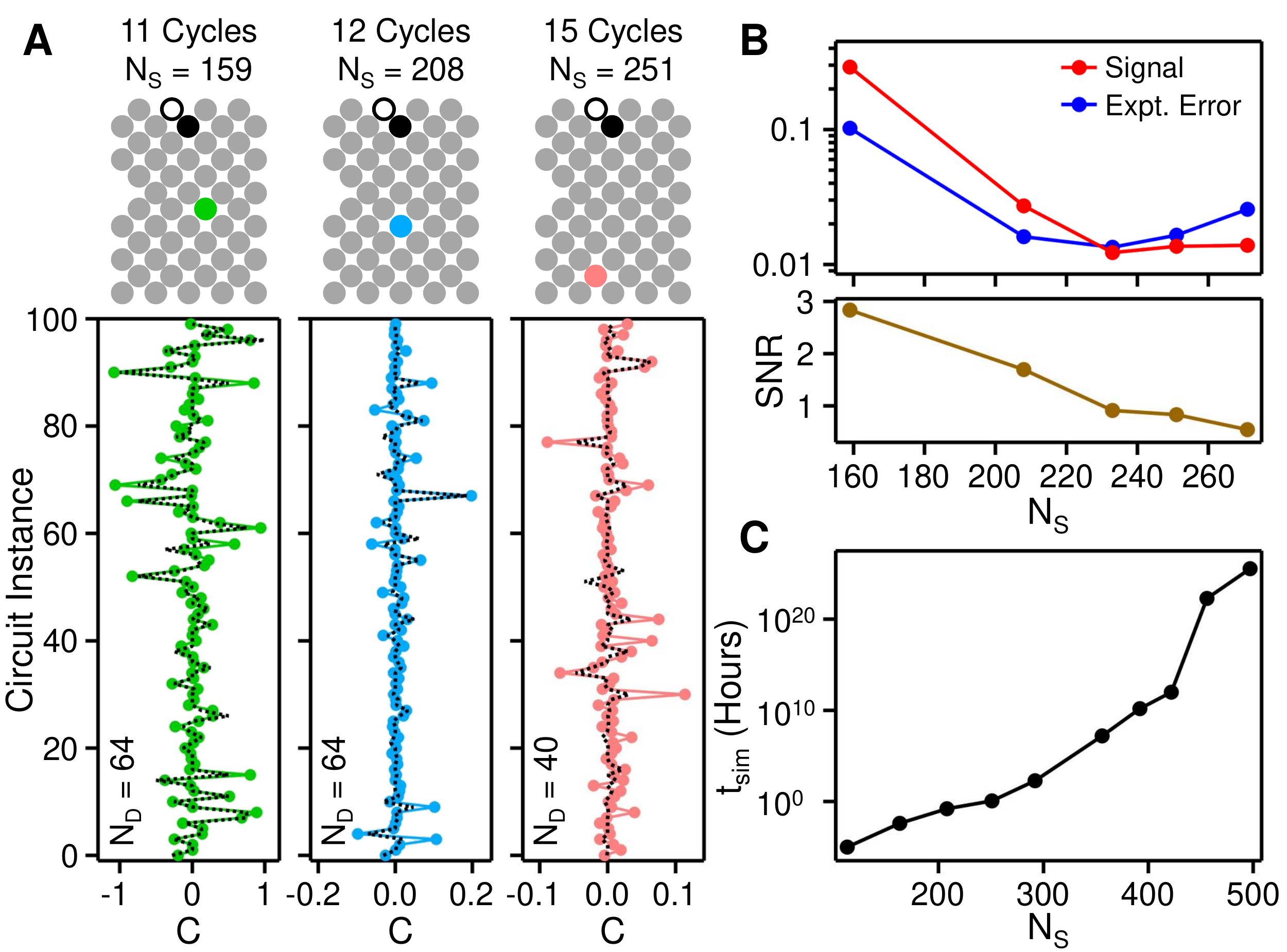}
	\caption{\textbf{Classical simulation complexity of quantum scrambling.} (A) Top panels show three different circuit configurations, each having the same $Q_\text{a}$ (black unfilled circle) and $Q_1$ (black filled circle) but different $Q_\text{b}$ (colored circle) and number of cycles in $\hat{U}$. The number of iSWAPs in $\hat{U}$ and $\hat{U}^\dagger$ that affect classical simulation costs, $N_\text{S}$, is indicated for each configuration. Bottom panels show the instance-dependent OTOCs measured for each configuration. The dashed lines are simulation results using tensor-contraction methods. $N_\text{D}$ is also indicated for each configuration. (B) Top: OTOC signal size and experimental error as functions of $N_\text{S}$. Bottom: Signal-to-noise ratio (SNR) as a function of $N_\text{S}$. SNR equals the ratio of OTOC signal size to experimental error. $N_\text{D} = 64$ for the first three values of $N_\text{S}$ and $N_\text{D} = 40$ for the last two values. The reason for decreasing $N_\text{D}$ is to increase the signal size such that it can be resolved with less statistical averaging \cite{sup}. (C) Estimated time $t_\text{sim}$ needed to simulate the OTOC of a single 53-qubit circuit with a variable number of iSWAPs, $N_\text{S}$, on a single CPU core (6 Gflop/s).}
	\label{fig:4}
\end{figure} 

Unlike operator spreading, an efficient classical description of operator spreading is not known to exist. In particular, population dynamics cannot be used to model the circuit-to-circuit fluctuation of OTOCs \cite{sup}. Resolving the growth of operator spreading is also difficult with a quantum processor since it is often accompanied by increased operator spreading \cite{AleinerIoffe_AP_2016, Roberts_JHEP_2017, Nahum_PRX_2018, Keyserlingk_PRX_2018}. We overcome this challenge by gradually adjusting the composition of $\hat{U}$ and $\hat{U}^\dagger$, realizing a group of circuits with predominantly Clifford gates (iSWAP, $\sqrt{X^{\pm 1}}$ or $\sqrt{Y^{\pm 1}}$) and a small number of non-Clifford gates ($\sqrt{W^{\pm 1}}$ and $\sqrt{V^{\pm 1}}$). As illustrated by the evolution of a butterfly operator in Fig.~\ref{fig:3}A, Clifford gates generate operator spreading by extending the butterfly operator to other qubits while preserving the total number of basis operators (which are products of Pauli operators, or Pauli strings, in these cases). In contrast, non-Clifford gates generate operator entanglement by transforming one single Pauli string into a superposition of multiple Pauli strings, maintaining the spatial extent of operator spreading in the process.

These distinctive properties of Clifford and non-Clifford gates therefore provide us a way to independently tune one scrambling mechanism without affecting the other. We now focus on operator entanglement and measure the circuit-to-circuit fluctuation of OTOCs, as shown in Fig.~\ref{fig:3}B. Here the number of circuit cycles is fixed at 12 and the number of non-Clifford gates in $\hat{U}$, $N_\text{D}$, is successively changed from 0 to 32. For each $N_\text{D}$, the individual OTOCs $C$ of 130 random circuit instances are measured using a modified normalized procedure \cite{sup}. At $N_\text{D} = 0$ where the circuits consist of only Clifford gates, we see that $C$ takes discrete values of 1 or $-1$. This is expected as the time-evolved butterfly operator $\hat{O} (t)$ is a single Pauli string and therefore either commutes or anti-commutes with the measurement operator $\hat{M}$. As more non-Clifford gates are introduced into the circuits, $C$ starts to assume intermediate values between $\pm 1$ and converges toward 0.

The mean $\overline{C}$ and fluctuation (i.e. root-mean-square value) $\delta_\text{C}$ of $C$ are then computed from experimental data and plotted against $N_\text{D}$ in Fig.~\ref{fig:3}C. We observe different behaviors for $\overline{C}$ and $\delta_\text{C}$: $\overline{C}$ remains largely constant and close to 0, confirming that operator spreading remains unaffected by the increasing number of non-Clifford gates. On the other hand, $\delta_\text{C}$ decays from an initial value of 1 and is almost suppressed by two orders of magnitude as $N_\text{D}$ increases from 0 to 32. Over the same range of $N_\text{D}$, we have numerically calculated the average numbers of Pauli strings in $\hat{O} (t)$, $n_\text{p}$, which are seen to increase exponentially (inset to Fig.~\ref{fig:3}C). These results demonstrate that the decay of OTOC fluctuation allows the growth of operator entanglement to be experimentally diagnosed.

To determine the accuracy of our measurements, the OTOCs of experimental circuits are simulated using a Clifford-expansion method and overlaid on the data in Fig.~\ref{fig:3}B and Fig.~\ref{fig:3}C \cite{sup}. We find good agreement between experiment and simulation even when $\delta_\text{C}$ is as small as 0.03, indicating the quantum processor's capability to resolve high degrees of operator entanglement. This may appear surprising given that other signatures of quantum entanglement, such as entropy, are highly susceptible to unwanted interaction with the environment \cite{Horodecki_RMP_2009}. Instead, we find that environmental effects are nearly absent from these data. This robustness is a result of the effective normalization protocol and a range of other error-mitigation techniques used in our experiment \cite{sup}.

Having identified means of characterizing both operator spreading and operator entanglement, it remains to be asked how the computational complexity of quantum scrambling, as well as our experimental error, scale with circuit size. This is addressed by systematically increasing the number of iSWAPs in the quantum circuits, $N_\text{S}$ (here $N_\text{S}$ counts only iSWAP gates that lie within the light-cones of $Q_\text{a}$ and $Q_1$ \cite{sup}). At the same time, $N_\text{D}$ is kept at a large value such that the Clifford-expansion simulation method used in Fig.~\ref{fig:3} is challenging to perform and tensor-contraction is the most efficient classical simulation method \cite{Arute2019, huang2020classical, sup}. Figure~\ref{fig:4}A shows representative data for three circuit configurations with different values of $N_\text{S}$, along with the corresponding numerical simulation results. As $N_\text{S}$ and the computational complexity for tensor-contraction increase, we observe that the OTOC fluctuation decreases and the agreement between experiment and simulation also degrades.

To quantify these observations, we define an ideal OTOC signal as the fluctuation $\delta_\text{C}$ computed from the simulated values of $C$. We also define an experimental error as the RMS deviation between the simulated and the measured values of $C$. Both quantities are shown as functions of $N_\text{S}$ in the upper panel of Fig.~\ref{fig:4}B. It is seen that the OTOC signal generally decreases as $N_\text{S}$ increases (see Extended Data in SM for data at $N_\text{D} = 24$ \cite{sup}). This change in signal size is difficult to predict theoretically \cite{sup} and may be related to the fact that the Pauli strings in $\hat{O} (t)$ become less correlated with each other as the light-cone associated with the butterfly qubit grows \cite{Claeys_PRReseach_2020}. On the other hand, the experimental error first decreases to a minimum of $\sim$0.01 for $N_\text{S} = 232$ before starting to increase. The ratio of these quantities is the experimental signal-to-noise ratio for OTOC and plotted in the bottom panel of Fig.~\ref{fig:4}B. The SNR is seen to monotonically decay from a value of 3 at $N_\text{S} = 159$ to 0.55 at $N_\text{S} = 271$.

At last, we estimate the time needed to simulate the OTOC of one circuit on a single CPU core using tensor-contraction, $t_\text{sim}$ \cite{sup}. The result is plotted against $N_\text{S}$ in Fig.~\ref{fig:4}C. We observe an exponential increase in $t_\text{sim}$ as $N_\text{S}$ increases, confirming that simulating the scrambling of complex quantum circuits indeed demands exponentially scaled classical computational resources. In particular, although simulating the results in Fig.~\ref{fig:4}A currently requires $t_\text{sim} \approx 100$ hours, this cost becomes $t_\text{sim} > 10^\text{10}$ hours when $N_\text{S}$ reaches 400. Chartering a path to this experimental regime is the focus of our ongoing research. In the SM, we have provided numerical simulation results showing that a percentage decrease in the coherent or incoherent errors of iSWAP gates will lead to at least commensurate levels of reduction for the OTOC errors \cite{sup}. Of less certainty is the size of $\delta_\text{C}$ in this regime, which cannot be predicted by any known classical model \cite{sup}. Nevertheless, this difficulty also implies that resolving $\delta_\text{C}$ alone might reveal scrambling dynamics beyond the simulation capacity of classical computers.

In conclusion, we characterize quantum scrambling in a 53-qubit system and demonstrate that entanglement in the space of quantum operators is the key to computational complexity of quantum observables. This result highlights the importance of careful classical analysis in the ongoing quest to attain quantum computational advantage on various problems of interest. On the other hand, the challenge in predicting OTOC fluctuations even moderately beyond the experimental regime also indicates that quantum processors of today can already shed light on certain physical phenomena as well as classical computers. Another encouraging finding of our work is that the accuracy of quantum processors can be significantly improved through effective error-mitigation. For example, an SNR of $\sim$1 for OTOCs is achieved for $N_\text{S} = 251$ where the circuit fidelity is merely $3 \%$ \cite{sup}. In the immediate future, classical simulation of average OTOCs may be used to efficiently benchmark performance of quantum processors. The experimental framework established here can also be used to study other quantum dynamics of interest, such as the integrability of the XY model (see SM for preliminary data) and many-body localization \cite{Wang_PRA_2001, Aleiner_AP_2006}. As the fidelity of quantum processors continues to increase, modeling scrambling in quantum gravity and unconventional quantum phases may become a reality as well \cite{Sekino_JHEP_2008, Lashkari_JHEP_2013, Sekino_JHEP_2008, Lashkari_JHEP_2013, Sachdev2017, StrangeMetal2018}.

\textbf{Acknowledgements} P. R. and X. M. acknowledge fruitful discussions with P. Zoller, B. Vermersch, A. Elben, and M. Knapp. S. M. and J. M. acknowledge the support from the NASA Ames Research Center and the support from the NASA Advanced Division for providing access to the NASA HPC systems, Pleiades and Merope. S. M. and J. M. also acknowledge the support from the AFRL Information Directorate under grant number F4HBKC4162G001. J. M. is partially supported by NAMS Contract No. NNA16BD14C.

\newpage~\newpage
\onecolumngrid
\begin{center}
	\textbf{\large Supplemental Materials}
\end{center}
\twocolumngrid
\setcounter{equation}{0}
\setcounter{figure}{0}
\setcounter{table}{0}
\makeatletter
\renewcommand{\theequation}{S\arabic{equation}}
\renewcommand{\thefigure}{S\arabic{figure}}

\tableofcontents

\section{Extended Data} \label{ext_data}

In this section, we provide data not covered in the main text. These include the spatial distribution of average OTOCs on the 53-qubit system, average OTOC behaviors for additional types of quantum circuits and more detailed investigation of errors in experimental OTOC measurements.

\subsection{Full 53-Qubit Average OTOC Data}

\begin{figure*}[!t]
	\centering
	\includegraphics[width=2\columnwidth]{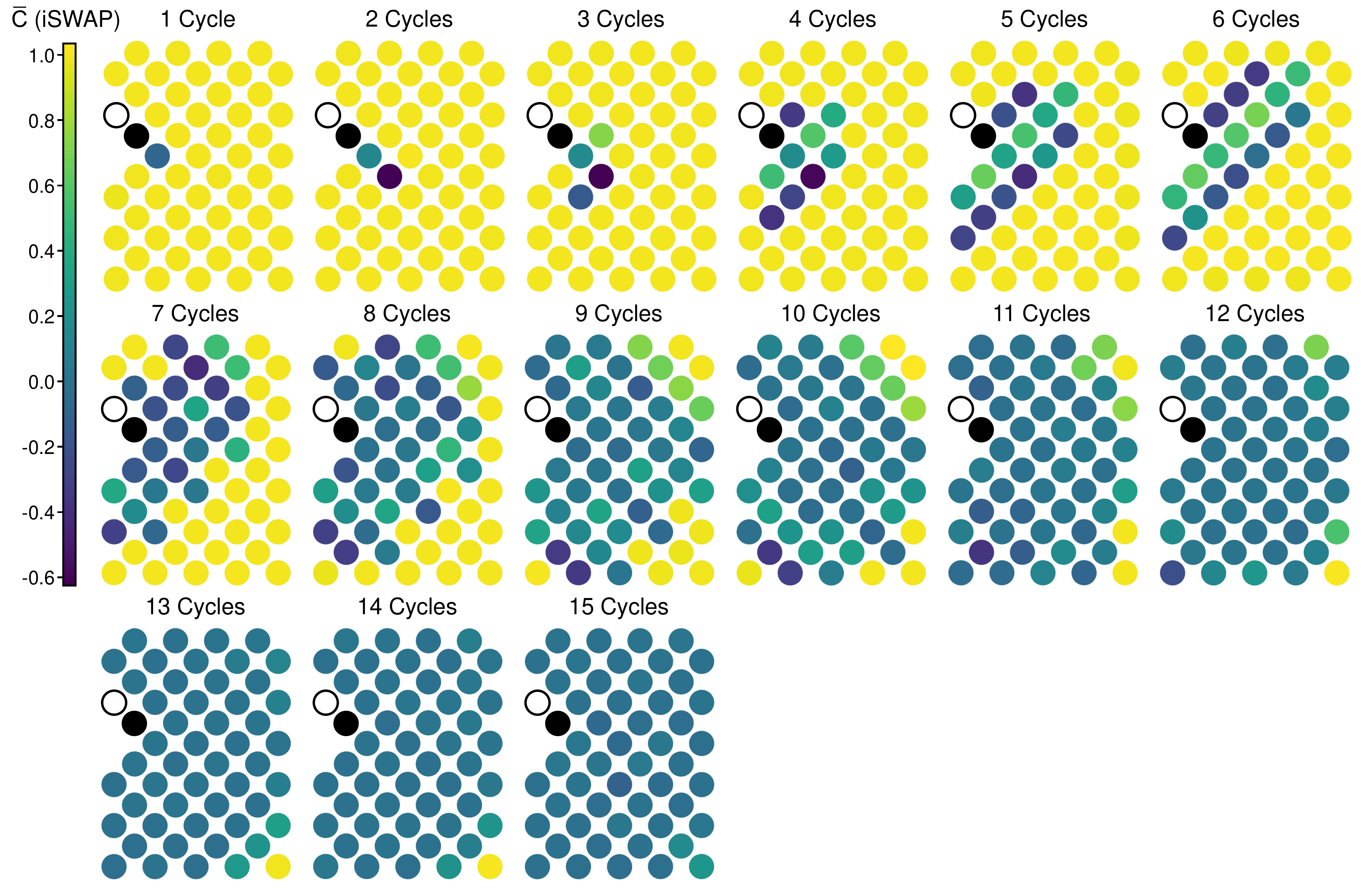}
	\caption{Full evolution of average OTOCs for iSWAP random circuits. The average OTOCs of the 53-qubit system shown for every cycle up to a total of 15. The black unfilled (filled) circle represents the location of the ancilla (measurement) qubit. The colors of the other filled circles represent the values of $\overline{C}$ for different locations of the butterfly qubit. The two-qubit gate used here is iSWAP and the data are averaged over 38 random circuit instances.}
	\label{fig:s1_1}
\end{figure*} 

\begin{figure*}[!t]
	\centering
	\includegraphics[width=2\columnwidth]{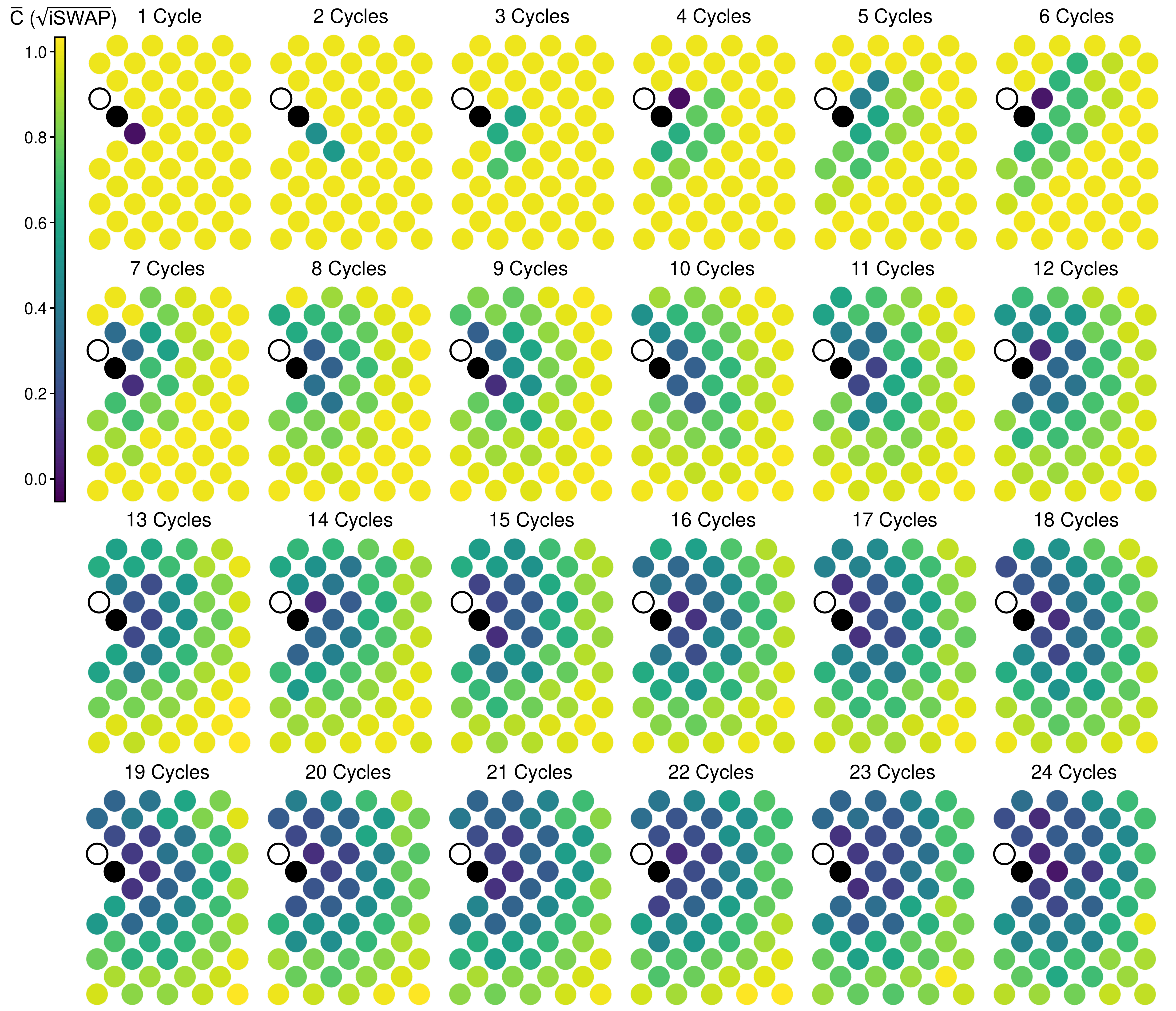}
	\caption{Full evolution of average OTOCs for $\sqrt{\text{iSWAP}}$ random circuits. The average OTOCs of the 53-qubit system shown for every cycle up to a total of 24. The black unfilled (filled) circle represents the location of the ancilla (measurement) qubit. The colors of the other filled circles represent the values of $\overline{C}$ for different locations of the butterfly qubit. The two-qubit gate used here is $\sqrt{\text{iSWAP}}$ and the data are averaged over 24 random circuit instances.}
	\label{fig:s1_2}
\end{figure*} 

Figure~\ref{fig:s1_1} shows experimentally measured average OTOCs, $\overline{C}$, for 51 different possible butterfly qubit locations. The data are also shown for every circuit cycle from 1 through 15. Here iSWAP is used as the two-qubit gate. The sharp nature of the wavefront propagation is readily visible, where a large reduction from $\overline{C} = 1$ is observed as soon as the lightcone of the measurement qubit (black filled circle) reaches a given qubit. In comparison, similar data are shown up to a total of 24 circuit cycles for random circuits containing $\sqrt{\text{iSWAP}}$ gates (Fig.~\ref{fig:s1_2}). The dynamics is seen to be slower than iSWAP, as mentioned in the main text. In particular, the wavefront is also broader, as seen by the more gradual spatial transition from $\overline{C} = 1$ to $\overline{C} \approx 0$.

\subsection{OTOCs for Non-Integrable and Integrable Quantum Circuits}

\begin{figure}[!t]
	\centering
	\includegraphics[width=1\columnwidth]{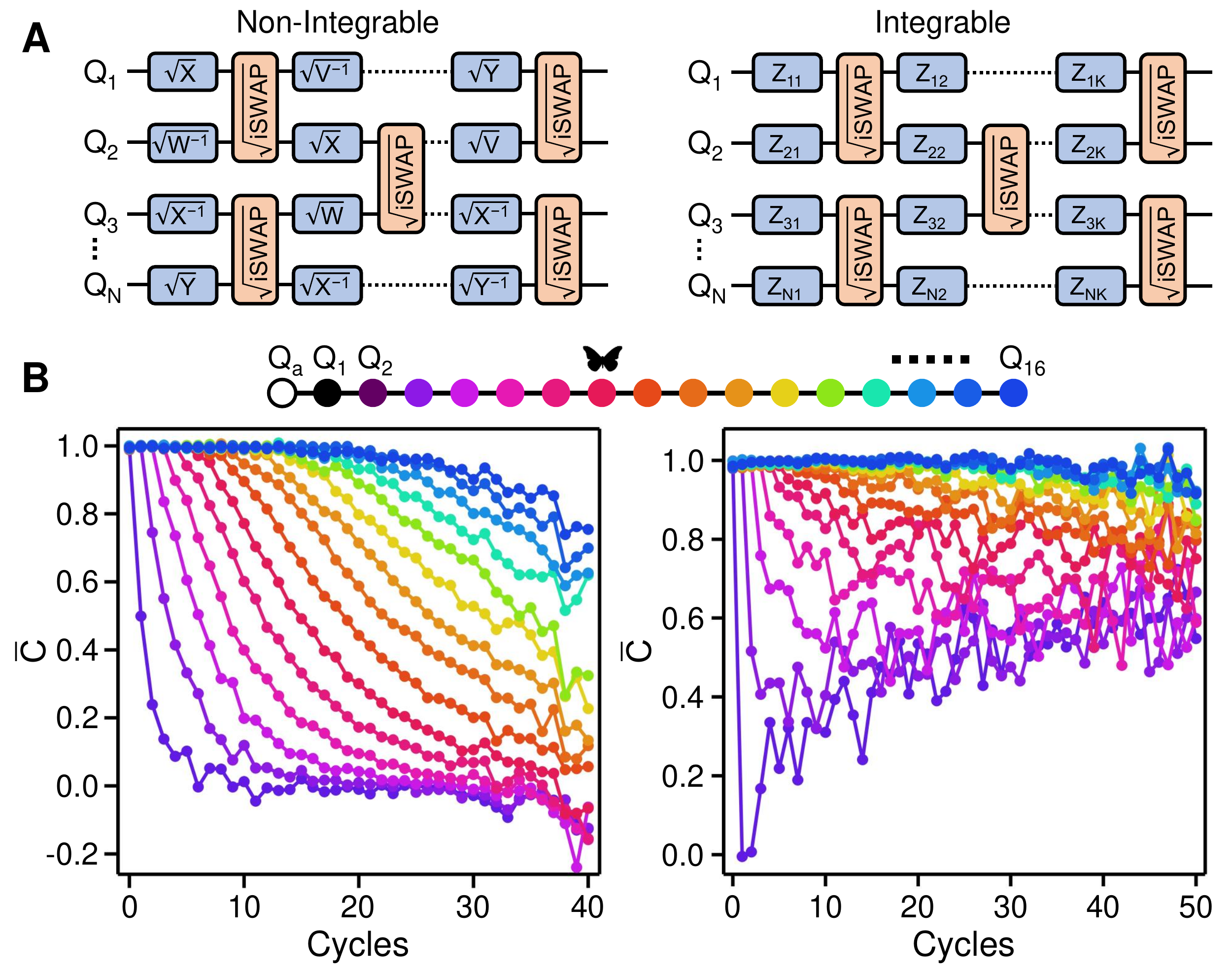}
	\caption{Average OTOCs for non-integrable and integrable quantum circuits. (A) Two different types of quantum circuits. Left panel shows a non-integrable quantum circuit composed of $\sqrt{\text{iSWAP}}$ and single-qubit gates randomly drawn from $\sqrt{X^\pm}$, $\sqrt{Y^\pm}$, $\sqrt{W^\pm}$, $\sqrt{V^\pm}$. Right panel shows an integrable quantum circuit composed of $\sqrt{\text{iSWAP}}$ and single-qubit gates that are rotations around the z-axis of the Bloch sphere with random angles. (B) Average OTOCs $\overline{C}$ for the two types of random circuits, implemented on a 1D chain of 16 qubits. The qubit configuration is shown on top, where the unfilled (filled) black circle represents the ancilla (measurement) qubit $Q_\text{a}$ ($Q_1$). Left panel shows $\overline{C}$ with the butterfly qubit corresponding to qubits $Q_2$ through $Q_{16}$, where the random circuits are of the type described in the left panel of (A). Right panel shows similar data for the random circuits of the type in the right panel of (A). The locations of the butterfly qubit are indicated by the colors of the data symbols and the legend on top. All data are averaged over 40 circuit instances.}
	\label{fig:s1_3}
\end{figure} 

\begin{figure}[!t]
	\centering
	\includegraphics[width=1\columnwidth]{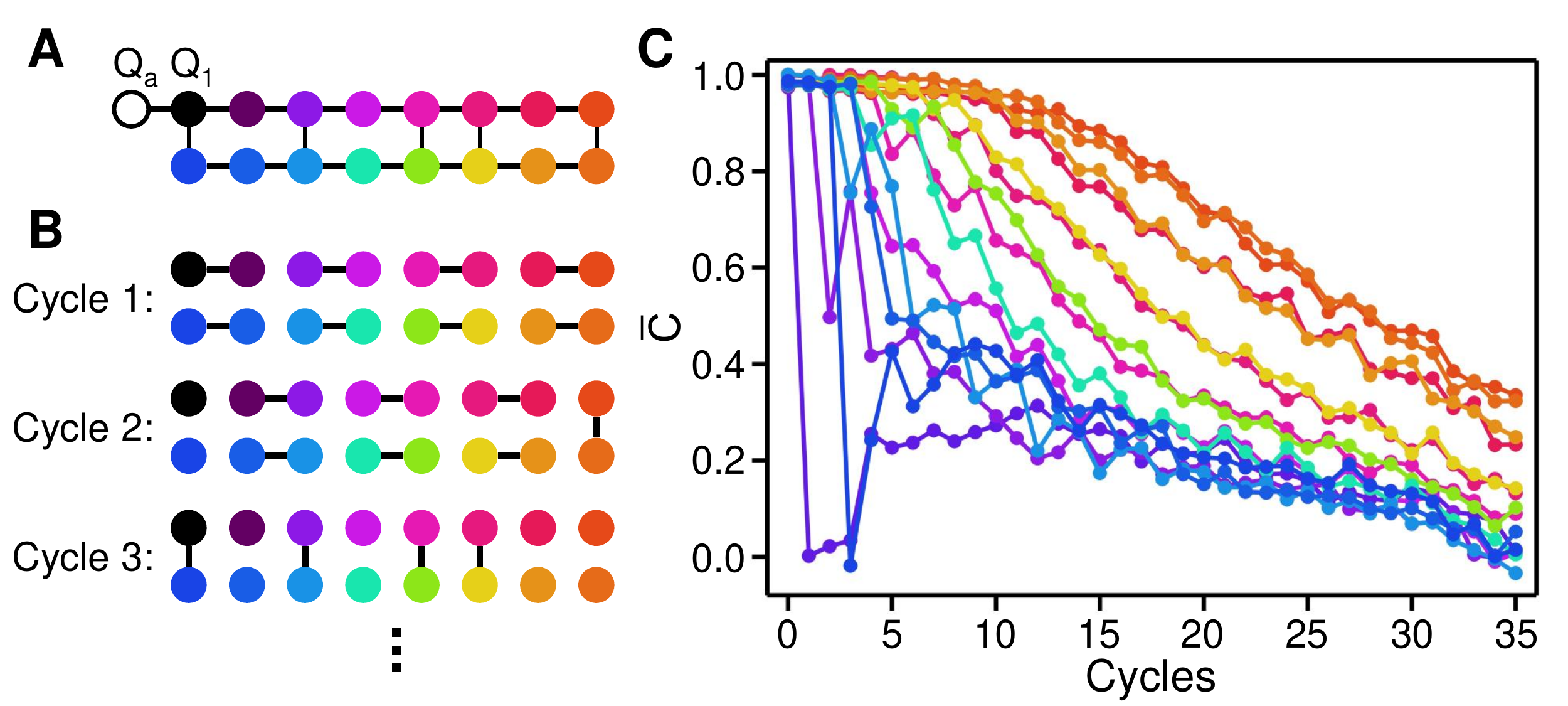}
	\caption{Transition into non-integrability in the XY Model. (A) A 2D arrangement of qubits in the form of two parallel 1D chains with 5 connections between each other. Black unfilled (filled) circle denotes the ancilla (measurement) qubit $Q_\text{a}$ ($Q_1$). (B) Order for applying the two-qubit gates $\sqrt{\text{iSWAP}}$. The qubit connections denote the $\sqrt{\text{iSWAP}}$ gates that are applied for a particular cycle. Additional cycles repeat the first three cycles, e.g. cycle 4 applies the same $\sqrt{\text{iSWAP}}$ gates as cycle 1 and so on. (C) Average OTOCs $\overline{C}$ with the butterfly qubit corresponding to qubits $Q_2$ through $Q_{16}$. The locations of the butterfly qubit are indicated by the colors of the data symbols and the colors of the qubits in (A) and (B). All data are averaged over 36 circuit instances.}
	\label{fig:s1_4}
\end{figure} 

In the main text of this work, we have primarily focused on quantum circuits that are non-integrable \cite{Calabrese_2016}, i.e. capable of evolving a quantum system into states with maximal degrees of scrambling. In general, many quantum circuits or dynamical processes are integrable and lead to small degree of quantum scrambling even at long time scales. Examples include Clifford circuits \cite{Gottesman_PRA_1996}, dynamics of free fermions \cite{Terhal_PRA_2002} and many-body localization \cite{Huse_ARCMP_2015}. For certain integrable, pseudo-random circuits such as Clifford circuits, OTOC fluctuation is needed to distinguish them from non-integrable circuits, as demonstrated in the main text. In many other cases, average OTOCs behave differently for integrable quantum circuits compared to non-integrable ones and are sufficient to differentiate between them, which we illustrate next.

Figure.~\ref{fig:s1_3} shows two different types of quantum circuits. The circuit in the left panel consists of $\sqrt{\text{iSWAP}}$ gates and single-qubit gates randomly chosen from $\sqrt{X^\pm}$, $\sqrt{Y^\pm}$, $\sqrt{W^\pm}$, $\sqrt{V^\pm}$. This is the example of a non-integrable circuit, which is expected to lead to maximal scrambling at long times. The circuit in the right panel has the same two-qubit gates, but the single-qubit gates are replaced with $Z$ gates that have angles randomly chosen from the interval $[-\pi, \pi]$. This is the example of an integrable circuit, where the dynamics does not lead to maximal scrambling if implemented in 1D.

The experimentally measured average OTOCs are shown in Fig.~\ref{fig:s1_3}B for both types of quantum circuits. Here, the two-qubit gates are applied in a brick-work pattern similar to what is used in Fig.~1C of the main text. For the non-integrable circuits, we observe a clear propagation of the OTOCs along with a diffusive broadening of the wavefronts, similar to what was observed in Fig.~2C of the main text. In particular, $\overline{C}$ monotonically decays toward 0 at large circuit cycles. On the other hand, $\overline{C}$ does not show the wave-like propagation for the integrable circuits. For qubits closer to the measurement qubit $Q_1$, $\overline{C}$ first decays but gradually increases for longer circuit cycles. Qubits further from $Q_1$ barely show any appreciate degree of OTOC decay up to 50 circuit cycles. Overall, $\overline{C}$ for different qubits converges toward values $> 0.5$ for the largest circuit cycles probed in this experiment. These results show how one might in some cases uses average OTOC behavior to distinguish non-integrable quantum dynamics from integrable ones.

The integrable circuits studied in Fig.~\ref{fig:s1_3} in fact are the digitized realizations of the so-called XY model \cite{Wang_PRA_2001} which is of wide interest in condensed-matter physics due to its ability to capture a variety of interesting physical phenomena such as quantum phase transitions. To demonstrate an immediate application of our work, we use OTOCs to study a particular feature of the XY-model, namely the transition from integrability to non-integrability due to geometry.

It is well-known that XY-model in 1D exhibits integrable dynamics, as demonstrated in Fig.~\ref{fig:s1_3}. Dynamics for XY-model in 2D remains a highly active area of research and is generally non-integrable. In Fig.~\ref{fig:s1_4}, we show that the transition into non-integrability for XY-model occurs as soon as the geometry changes from 1D to a ladder-like geometry (Fig.~\ref{fig:s1_4}A). Here we have arranged 16 qubits into two parallel chains of 8 qubits and connected them with 5 ``cross-links''. Next, we measure OTOCs of random circuits implemented with this new geometry where the single-qubit gates are again randomly chosen from $Z$ gates with angles in the interval $[-\pi, \pi]$ and the two-qubit gates are $\sqrt{\text{iSWAP}}$. The order for applying the $\sqrt{\text{iSWAP}}$ gates is shown in Fig.~\ref{fig:s1_4}B.

The average OTOCs for this ladder-geometry XY model are shown in Fig.~\ref{fig:s1_4}C. It is seen that the wavelike propagation and long-time limits of $\overline{C} = 0$ characteristic of non-integrable quantum circuits are both recovered in this modified geometry, indicating a transition from integrability to non-integrability for the XY-model. Detailed experimental studies of this transition with larger numbers of qubits is a subject of future work.

\subsection{Characterization of Experimental Errors}

\begin{figure}[!t]
	\centering
	\includegraphics[width=1\columnwidth]{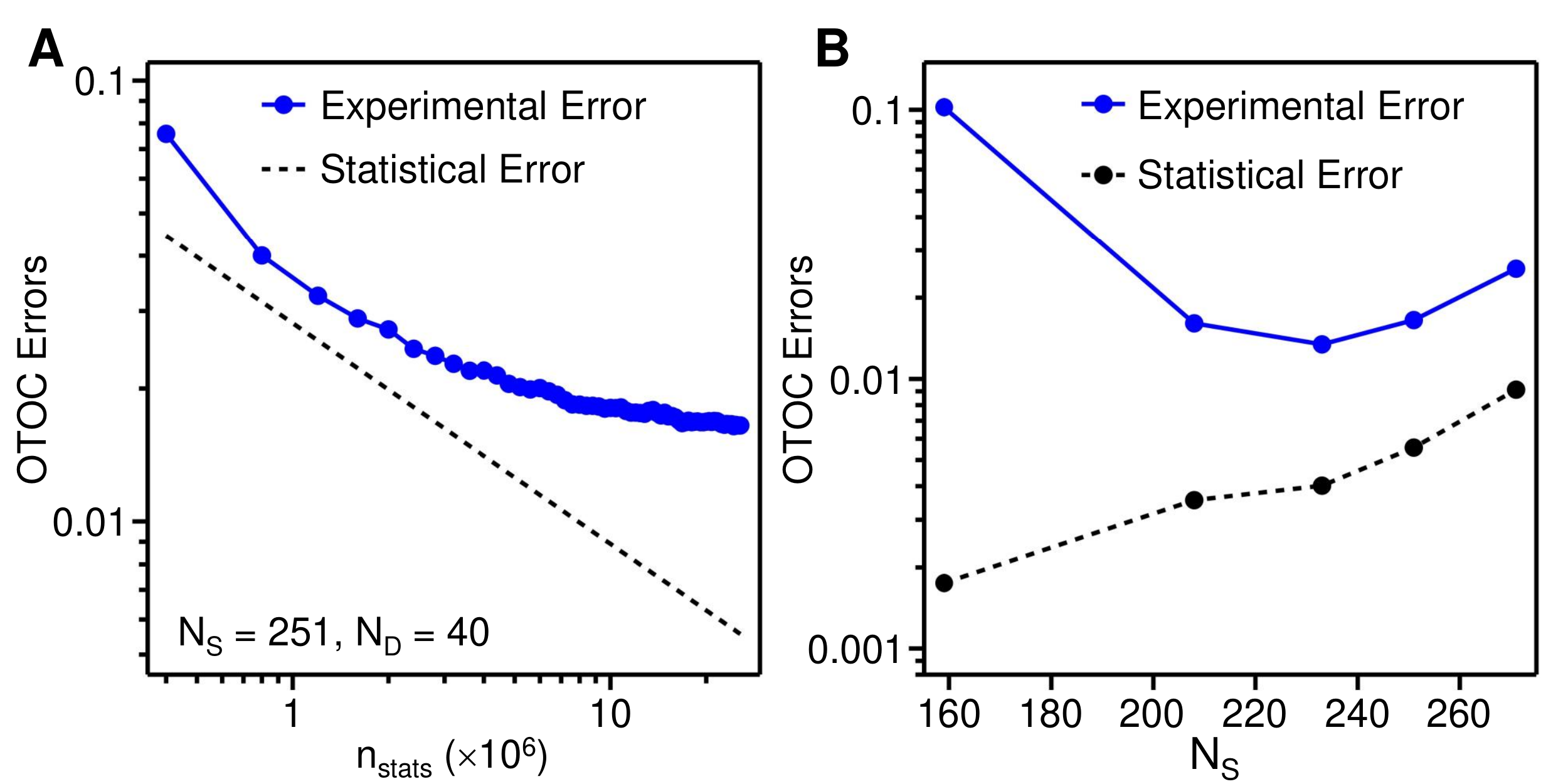}
	\caption{Contribution of finite sampling to experimental OTOC error. (A) Dependence of experimental OTOC error on the number of single-shots used to estimate $\braket{\hat{\sigma}_{\text y}}$, $n_\text{stats}$. The dashed line shows expected statistical errors for different values of $n_\text{stats}$. Here the number of iSWAPs is $N_\text{S} = 251$ and the number of non-Cliffords is $N_\text{D} = 40$. (B) Comparison of experimental OTOC errors and expected statistical errors for different values of $N_\text{S}$. The experimental errors are reproduced from Fig.~4C of the main text. The statistical errors are calculated based on $n_\text{stats}$ and the average normalization value for each $N_\text{S}$.}
	\label{fig:s1_5}
\end{figure} 

\begin{figure}[!t]
	\centering
	\includegraphics[width=1\columnwidth]{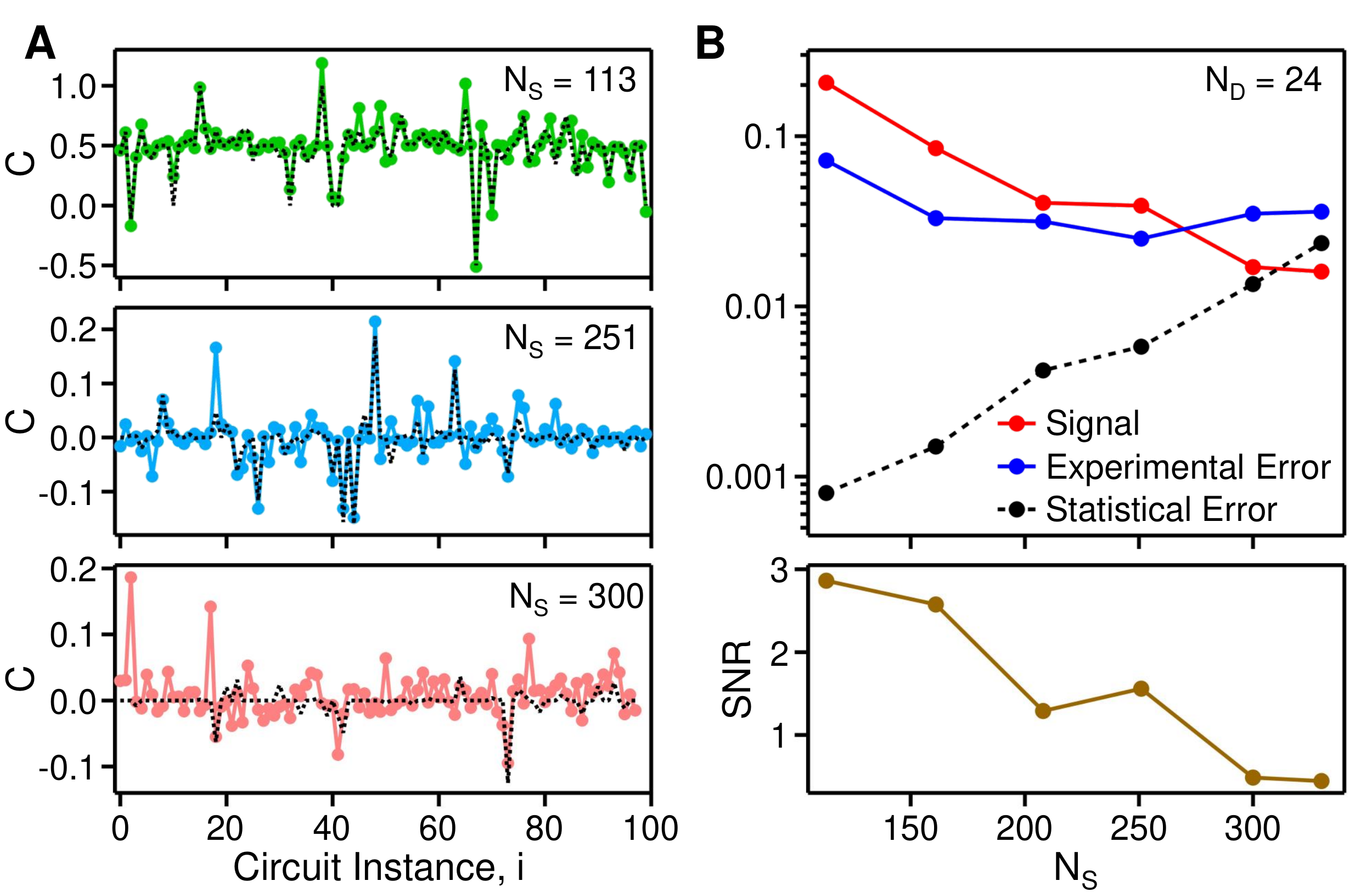}
	\caption{OTOC errors for $N_\text{D} = 24$. (A) OTOCs of 100 random circuit instances, $C$, for different values of $N_\text{S}$. $N_\text{D} = 24$ for all circuits. The dashed lines are exact numerical simulation results using the branching method. (B) Upper panel: OTOC signal size, experimental error and estimated statistical error as functions of $N_\text{S}$. $N_\text{D} = 24$ for all data included here. Lower panel: SNR (i.e. ratio of OTOC signal size to experimental error) as a function of $N_\text{S}$.}
	\label{fig:s1_6}
\end{figure} 

In this section, we present additional characterization data to further corroborate and understand the experimental results in Fig.~4 of the main text. In particular, we focus on answering two questions: 1. What fraction of the observed experimental errors can be attributed to statistical errors due to limited sampling? 2. How sensitive is the signal-to-noise ratio to the number of non-Cliffords in the circuits?

We first address question 1. The role of finite sampling in experimental OTOC measurement may be understood as follows: For $n_\text{stats}$ single-shot measurements, an average error of $\frac{1}{\sqrt{n_\text{stats}}}$ is expected to be present in the estimate for $\braket{\hat{\sigma}_{\text y}}$ of the ancilla qubit due to statistical uncertainty. In the presence of circuit errors and a normalization value $\braket{\hat{\sigma}_{\text y}}_\text{I} < 1$, this expected statistical error is amplified to a value of $\frac{1}{\braket{\hat{\sigma}_{\text y}}_\text{I} \sqrt{n_\text{stats}}}$, assuming the ideal OTOC value to be significantly smaller than 1 (i.e. $\braket{\hat{\sigma}_{\text y}}_\text{I} \gg |\braket{\hat{\sigma}_{\text y}}_\text{B}|$ where $\braket{\hat{\sigma}_{\text y}}_\text{B}$ is $\braket{\hat{\sigma}_{\text y}}$ measured with the butterfly operator applied). 

In Fig.~\ref{fig:s1_6}A, this expected statistical error is plotted as a function of $n_\text{stats}$ for $N_\text{S} = 251$ and $N_\text{D} = 40$ where $\braket{\hat{\sigma}_{\text y}}_\text{I} \approx 0.04$. On the same plot, we have included experimental OTOC errors (i.e. the RMS deviation between simulated and experimental OTOC values of 100 random circuits) obtained from reduced amounts of single-shot data. We observe that the experimental error initial decreases with increasing values of $n_\text{stats}$, suggesting that statistical uncertainty being a significant source of error for small numbers of single-shot measurements. At $n_\text{stats} > 10^7$, the experimental error has a significantly weaker dependence on $n_\text{stats}$ and is markedly higher than the expected statistical error, indicating other error sources are dominant in this regime. In Fig.~\ref{fig:s1_6}A, we have re-plotted the experimental errors in Fig.~4C of the main text along with the expected statistical errors calculated from $n_\text{stats}$ and $\braket{\hat{\sigma}_{\text y}}_\text{I}$ of each $N_\text{S}$. The increase in statistical error at higher $N_\text{S}$ is a result of decreasing normalization values. It is evident that the observed experimental errors are consistently larger than the expected statistical errors, indicating other error mechanisms are dominant. In Section \ref{sec:NoiseNumerics}, we provide numerical simulation results to further analyze the sources of experimental error.

Next, we focus on the scaling of experimental error vs number of iSWAPs for a different number of non-Cliffords in the quantum circuits. Fig.~\ref{fig:s1_6}A shows $C$ of 100 circuit instances for $N_\text{S} = 113$, 251 and 300, all of which have $N_\text{D} = 24$. On the same plots, exactly simulated OTOC values using the branching method (Section IV) are also plotted. Similar to Fig.~4B of the main text, we observe that the agreement between experimental and simulated OTOC values degrades as $N_\text{S}$ increases. In addition, the OTOC signal size (i.e. the fluctuation of the simulated OTOC values) also decreases as $N_\text{S}$ increases.

In Fig.~\ref{fig:s1_6}B, the OTOC signal size, experimental error and expected statistical error are plotted as functions of $N_\text{S}$ (all with $N_\text{D} = 24$). Here we see that the OTOC signal size indeed monotonically decreases as $N_\text{S}$ increases. The experimental error, on the other hand, shows less dramatic changes as a function of $N_\text{S}$. After taking the ratio of the OTOC signal size and the experiment error, we plot the resulting SNR as a function of $N_\text{S}$ in the lower panel of Fig.~\ref{fig:s1_6}B. The scaling of SNR vs $N_\text{S}$ is roughly consistent with Fig.~4 of the main text where higher values of $N_\text{D}$ (64 and 40) are used. In particularly, SNR falls below 1 after $N_\text{S}$ has increased beyond $\sim$250. The relative insensitivity of SNR to $N_\text{D}$ allows us to conveniently benchmark our system at lower values of $N_\text{D}$ where classical simulation is easy. This is especially important in the future when we conduct experiments with $N_\text{S} > 400$ where tensor-contraction may no longer be feasible (Section \ref{sec:largenumerics}).

\section{Experimental Techniques} \label{exp_tech}

In this section, we describe the calibration process and metrology of quantum gates used in the OTOC experiment. In addition, we also demonstrate a series of error-mitigation strategies used in the compilation of experimental circuits which significantly reduced errors from various sources such as those related to state preparation and readout (SPAM), cross-talk, or coherent control errors on two-qubit gates.

\subsection{Calibration of iSWAP and $\sqrt{\text{iSWAP}}$ Gates}

\begin{figure}[t]
	\centering
	\includegraphics[width=1\columnwidth]{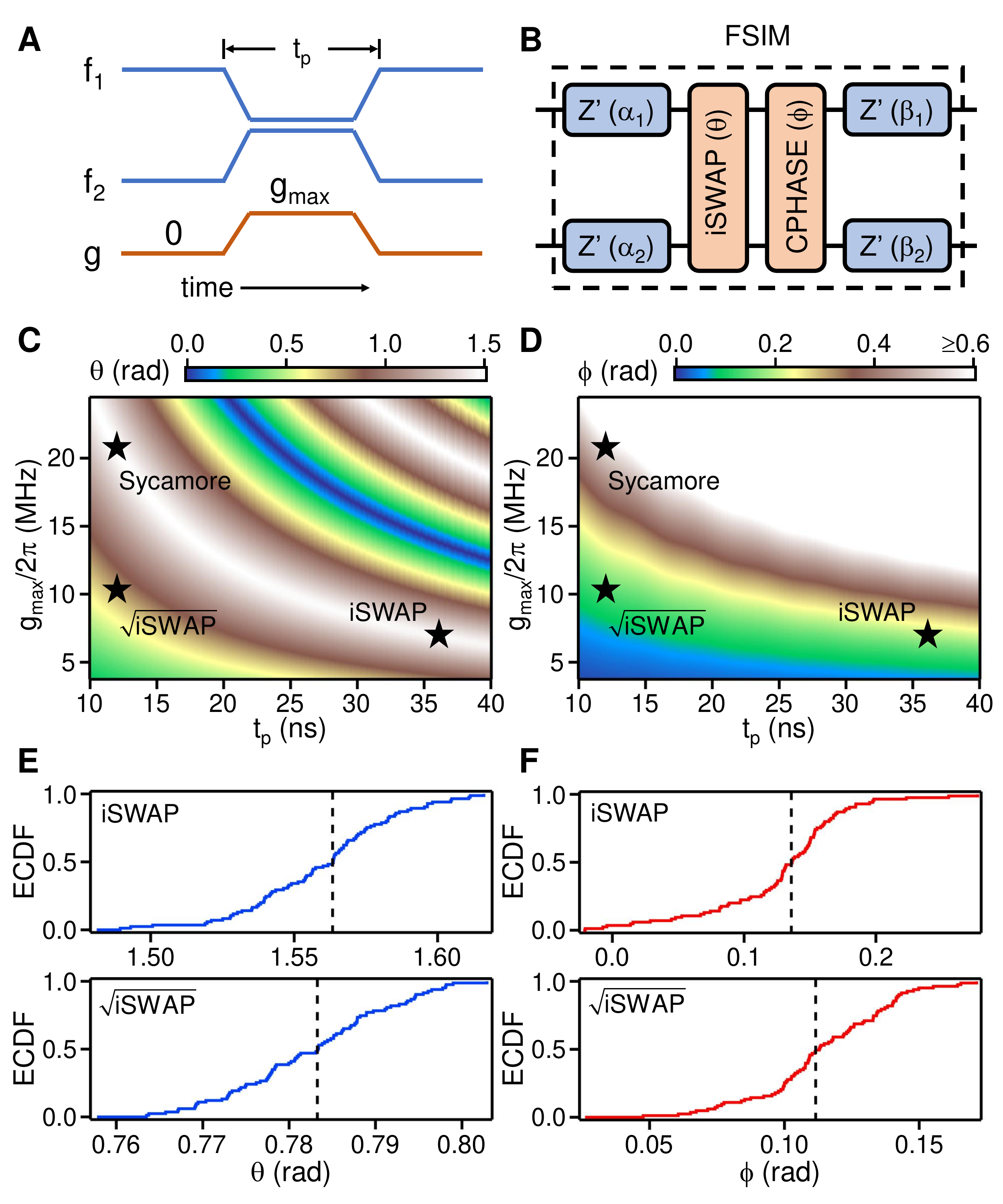}
	\caption{Calibrating ``textbook'' gates. (A) Schematic illustration of the waveforms used to realize pure iSWAP and $\sqrt{\text{iSWAP}}$ gates: The $\ket{0} \rightarrow \ket{1}$ transition frequencies of two transmon qubits, $f_1$ and $f_2$, are brought close to each other by adjusting the control fluxes to their superconducting quantum interference device (SQUID) loop. Concurrently, a pulse to the coupler's SQUID flux changes the qubit-qubit coupling $g$ from 0 to a maximum value of $g_\text{max}$. The total length of the pulses is $t_{\text p}$. (B) The generic FSIM gate realized by the waveforms in (A), composed of a partial-iSWAP gate with swap angle $\theta$, a CPHASE gate with a conditional-phase $\phi$ and four local Z-rotations with angles $\alpha_1$, $\alpha_2$, $\beta_1$, $\beta_2$ each. (C, D) Simulated $\theta$ (C) and $\phi$ (D) as functions of $g_\text{max}$ and $t_{\text p}$. The simulation assumes an average qubit frequency $\frac{1}{2} \left( f_1 + f_2 \right) = 6.0$ GHz and an anharmonicity of 200 MHz for each transmon qubit. The operating points for three different gates, Sycamore, iSWAP and $\sqrt{\text{iSWAP}}$, are indicated by the the star symbols. (E, F) Integrated histogram (empirical cumulative distribution function, ECDF) of $\theta$ (E) and $\phi$ (F) for both the iSWAP and $\sqrt{\text{iSWAP}}$ gates. The data include all 86 qubit pairs on the processor. The x-axis location of each dashed line indicates the median value of the corresponding angle.}
	\label{fig:s2_1}
\end{figure} 

The measurement of OTOCs requires faithful inversion of a given quantum circuit, $\hat{U}$. The ``Sycamore'' gate used in our previous work \cite{Arute2019}, equivalent to an iSWAP gate followed by a CPHASE gate with a conditional-phase of $\pi / 6$ radians (rad), is an ill-suited building block for $\hat{U}$ since its inversion cannot be easily created by combining Sycamore with single-qubit gates. On the other hand, iSWAP and $\sqrt{\text{iSWAP}}$ are commonly used two-qubit gates that can be readily inverted by adding local Z rotations:
\begin{equation}
G^{-1} = Z_1\left( \frac{\pi}{2} \right) Z_2\left(-\frac{\pi}{2}\right) G Z_1\left(-\frac{\pi}{2}\right) Z_2\left(\frac{\pi}{2}\right)
\end{equation}
Here $G$ is the two-qubit unitary corresponding to iSWAP or $\sqrt{\text{iSWAP}}$ and $Z_i (\varphi) = e^{-i\frac{\varphi}{2} \hat{\sigma}_z^{(i)}}$, where $\hat{\sigma}_z^{(i)}$ is the Pauli-$Z$ matrix acting on qubit $i$ ($i = 1$ or 2). Compared to Sycamore, realizing pure iSWAP and $\sqrt{\text{iSWAP}}$ gates requires the development of two additional capabilities on our quantum processor: 1. A significant reduction of the coherent error associated with the conditional-phase in Sycamore. 2. The abilility to implement arbitrary single-qubit rotations around the $Z$ axis.

\subsubsection{Reducing Conditional-Phase Errors}

We first describe the calibration technique for reducing conditional phase errors. As discussed in previous works \cite{Arute2019}, such a phase arises from the dispersive interaction between the $\ket{11}$ and $\ket{02}$ (or $\ket{20}$) states of two transmon qubits while their coupling strength $g$ is raised to a finite value to enable a resonant interaction between the $\ket{10}$ and $\ket{01}$ states (Fig.~\ref{fig:s2_1}A). Although eliminating this phase is possible by concatenating a Sycamore gate with a pure CPHASE gate having an opposite conditional-phase \cite{Foxen_PRL_2020}, this process would involve complicated waveforms and large qubit frequency excursions which are demanding to implement on a 53-qubit processor. Instead, we adopt an alternative approach based on the relatively simple control waveform used in Ref.~\cite{Arute2019} (Fig.~\ref{fig:s2_1}A). This waveform sequence produces a Fermionic Simulation (FSIM) gate comprising a partial-iSWAP gate with swap angle $\theta$, a CPHASE gate with conditional-phase $\phi$ and four local $Z$-rotations (Fig.~\ref{fig:s2_1}B). 

Next, we consider how the angles $\theta$ and $\phi$ depend on readily tunable waveform parameters such as the pulse duration $t_{\text p}$ and maximum qubit-qubit coupling $g_\text{max}$. This is done by numerically solving the time-evolution of two coupled transmons with typical device parameters. The dependences of $\theta$ and $\phi$ on $t_{\text p}$ and $g_\text{max}$ are shown in Fig.~\ref{fig:s2_1}C and Fig.~\ref{fig:s2_1}D, respectively. We observe that although both $\theta$ and $\phi$ increase linearly with $t_{\text p}$, their scaling with respect to $g_\text{max}$ is different: $\theta \propto g_\text{max}$ whereas $\phi \propto g_\text{max} ^ 2$. For a given value of $\theta$, it is therefore possible to reduce $\phi$ by increasing $t_{\text p}$ and decreasing $g_\text{max}$ while keeping $t_{\text p} g_\text{max}$ constant. However, since longer gate operation is more susceptible to decoherence effects such as relaxation and dephasing, it is also desirable to minimize values of $t_{\text p}$. As a result, we use $t_{\text p} = 12$ ns, $g_\text{max} / 2 \pi \approx 10$ MHz for calibrating the $\sqrt{\text{iSWAP}}$ gate and $t_{\text p} = 36$ ns, $g_\text{max} / 2 \pi \approx 7$ MHz for calibrating the iSWAP gate. Based on the simulation results, these choices would yield values of $\theta = \pi/4$ rad, $\phi = 0.14$ rad for the $\sqrt{\text{iSWAP}}$ gate and $\theta = \pi/2$ rad, $\phi = 0.19$ rad for the iSWAP gate.

The calibrated values of $\theta$ and $\phi$ associated with every qubit pair on the 53-qubit processor are shown in Fig.~\ref{fig:s2_1}E and Fig.~\ref{fig:s2_1}F, respectively. Each angle is measured via cross entropy benchmarking (XEB), similar to previous works \cite{Arute2019, Foxen_PRL_2020}. The median value of $\theta$ is 0.783 rad (standard deviation =  0.010 rad) for $\sqrt{\text{iSWAP}}$ and 1.563 rad (standard deviation =  0.027 rad) for iSWAP. These median values are very close to the target values of $\pi/4 = 0.785$ rad for $\sqrt{\text{iSWAP}}$ and $\pi/2 = 1.571$ rad for iSWAP. In the case of $\phi$, the median value is 0.112 rad for $\sqrt{\text{iSWAP}}$ (standard deviation = 0.026 rad) and 0.136 rad for iSWAP (standard deviation = 0.055 rad). The median values of $\phi$ are close to predictions from numerical simulation and 4 to 5 times lower compared to the Sycamore gate.

The coherent error introduced by remnant values of $\phi$ is further reduced by adjusting other phases of a FSIM unitary $U_\text{FSIM} (\theta, \phi, \Delta_+, \Delta_-, \Delta_{-, \text{off}})$, defined as:
\begin{equation}
\begin{pmatrix}
1 & 0 & 0 & 0 \\
0 & e^{i (\Delta_+ + \Delta_-)} \cos{\theta} & -ie^{i (\Delta_+ - \Delta_{-, \text{off}})} \sin{\theta} & 0 \\
0 & -ie^{i (\Delta_+ + \Delta_{-, \text{off}})} \sin{\theta} & e^{i (\Delta_+ - \Delta_-)} \cos{\theta} & 0 \\
0 & 0 & 0 & e^{i (2 \Delta_+ - \phi)}
\end{pmatrix}.
\end{equation}
Here $\Delta_+$, $\Delta_-$ and $\Delta_{-, \text{off}}$ are phases that can be freely adjusted by local $Z$-rotations. Imperfect gate calibration often results in an actual two-qubit unitary $U_{\text a}$ that differs slightly from the target unitary $U_{\text t}$, leading to a Pauli error \cite{Arute2019, nielsen_chuang_2010}:
\begin{equation}
r_{\text p} = 1 - \frac{1}{D^2} \left| \text{tr}\left( U_{\text a}^\dagger U_{\text t} \right) \right|^2
\end{equation}
Here $D = 4$ is the dimension of a two-qubit Hilbert space. 

Given that our target unitaries are $U_{\text t} = U_\text{FSIM} (\pi / 4, 0, 0, 0, 0)$ for $\sqrt{\text{iSWAP}}$ and $U_{\text t} = U_\text{FSIM} (\pi / 2, 0, 0, 0, 0)$ for iSWAP, one may naively expect that $\Delta_+$, $\Delta_-$ and $\Delta_{-, \text{off}}$ should all be set to 0 in $U_{\text a}$ in order to minimize $r_{\text p}$. While this is indeed the case if $\phi = 0$ in $U_{\text a}$, it is not true when $\phi$ assumes a finite value $\phi_{\text a}$ in $U_{\text a}$. In fact, simple algebraic calculation shows that in such a case, the minimum value of $r_p$ occurs at $(\Delta_+, \Delta_-, \Delta_{-, \text{off}}) = (\phi_{\text a} / 2, 0, 0)$, where it is a factor of 3 smaller compared to $(\Delta_+, \Delta_-, \Delta_{-, \text{off}}) = (0, 0, 0)$. By calibrating our system such that $\Delta_+ = \phi_{\text a} / 2$ for every qubit pair, we estimate that the median Pauli error introduced by the conditional-phase to be only 0.07 \% for $\sqrt{\text{iSWAP}}$ and 0.12 \% for iSWAP.

\subsubsection{Arbitrary $Z$-Rotations}

\begin{figure}[t]
	\centering
	\includegraphics[width=1\columnwidth]{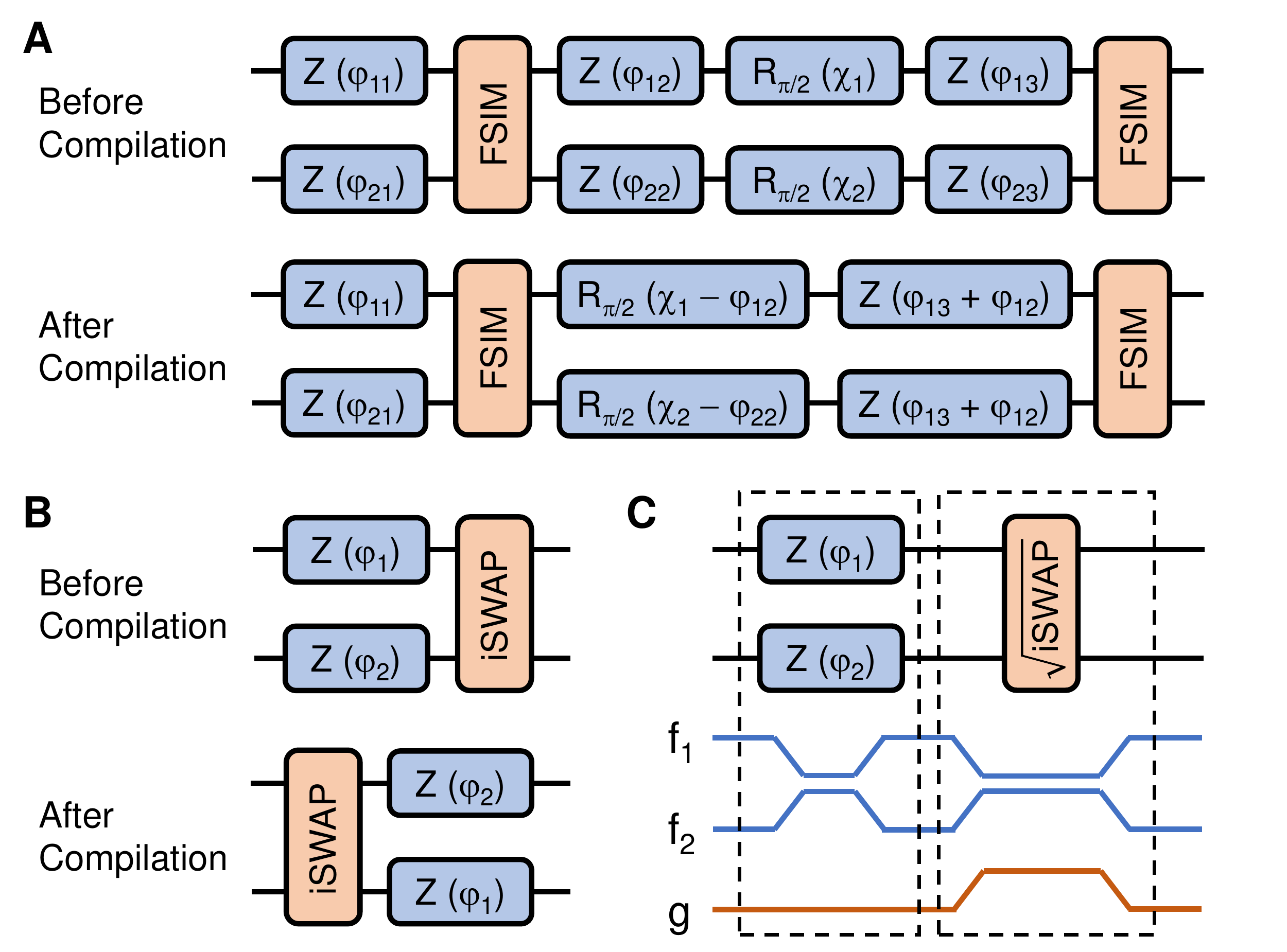}
	\caption{Implementation of $Z$-rotations. (A) Circuit compilation procedure for reducing number of required $Z$-rotations. $Z$-rotations occurring after each FSIM gate are combined with the $Z$-rotations before the next FSIM gate. Phases on the microwave-driven single-qubit gates are shifted to maintain the same overall quantum evolution. $\varphi_{ij}$ denotes the angles of various $Z$-rotations and $\chi_i$ denotes the phases of microwave single-qubit gates. (B) Circuit compilation procedure for pushing $Z$-rotations through an iSWAP gate. Combined with (A), $Z$-rotations in circuits containing only iSWAP and single-qubit gates become entirely virtual. (C) Schematic illustration of the waveforms used to implement $Z$-rotations before each $\sqrt{\text{iSWAP}}$ gate. Compared to the waveforms in Fig.~\ref{fig:s2_1}A, an additional control flux pulse is used to detune the qubits from their idle frequencies and thereby achieve a ``physical'' $Z$-gate.}
	\label{fig:s2_2}
\end{figure} 

We now describe the implementation of arbitrary single-qubit $Z$-rotations on the quantum processor. In addition to constructing the inverse gates $\sqrt{\text{iSWAP} ^ {-1}}$ and iSWAP$^{-1}$, $Z$-rotations are also used in removing native $Z$-rotations $Z'$ in the FSIM gate (Fig.~\ref{fig:s2_1}B) as well as adjusting values of $\Delta_+$ to minimize conditional-phase errors. The procedure for incorporating $Z$-rotations into quantum circuits is two-fold: First, we recompile the circuits and combine all $Z$-rotations occurring after each two-qubit FSIM gate with the $Z$-rotations before the next FSIM gate, as shown in Fig.~\ref{fig:s2_2}A. If any microwave-driven single-qubit gate such as $R_{\pi/2} (\chi) = e^{-i\frac{\pi}{4} \left( \cos \chi \hat{\sigma}_x + \sin \chi \hat{\sigma}_y \right)}$ occurs between the two FSIM gates, where $\hat{\sigma}_x$ and $\hat{\sigma}_y$ are Pauli-$X$ and Pauli-$Y$ matrices and $\chi$ is the phase of the microwave drive, we apply the equivalence $R_{\pi/2}(\chi) Z (\varphi) = Z (\varphi) R_{\pi/2}(\chi - \varphi)$ to shift the rotation axis of the single-qubit gate and ``push'' the $Z$-rotation through. This process effectively reduces the number of $Z$-rotations in the circuits by a factor of 2 and has been demonstrated to incur negligible degradation in the overall fidelity, since the only physical changes are modifications to the phases of the microwave drives of single-qubit gates \cite{McKay_PRA_2017}.

The second step for implementing $Z$-rotations is different for iSWAP and $\sqrt{\text{iSWAP}}$ gates. In the case of iSWAP, the equivalence $U_\text{FSIM} (\pi / 2, 0, 0, 0, 0) Z_1(\varphi_1) Z_2(\varphi_2) = Z_1(\varphi_2) Z_2(\varphi_1) U_\text{FSIM} (\pi / 2, 0, 0, 0, 0)$ allows us to push $Z$-rotations through each two-qubit gate by simply rearranging their phases (Fig.~\ref{fig:s2_2}B). As a result, the $Z$-rotations occurring in circuits with iSWAP gates are entirely virtual. In the case of $\sqrt{\text{iSWAP}}$, such equivalence does not exist and we implement $Z$-rotations before the two-qubit gates using additional control flux pulses, as illustrated in Fig.~\ref{fig:s2_2}C. Here, we detune the qubit frequencies from their idle positions by an variable amount $\Delta f$ and for a fixed duration $t_{\text z}$ = 20 ns before each $\sqrt{\text{iSWAP}}$ gate, leading to a $Z$-rotation $\approx Z \left( 2 \pi \Delta f t_{\text z} \right)$.

\subsection{Gate Error Benchmarking and Cross-Talk Mitigation}

\begin{figure}[t]
	\centering
	\includegraphics[width=1\columnwidth]{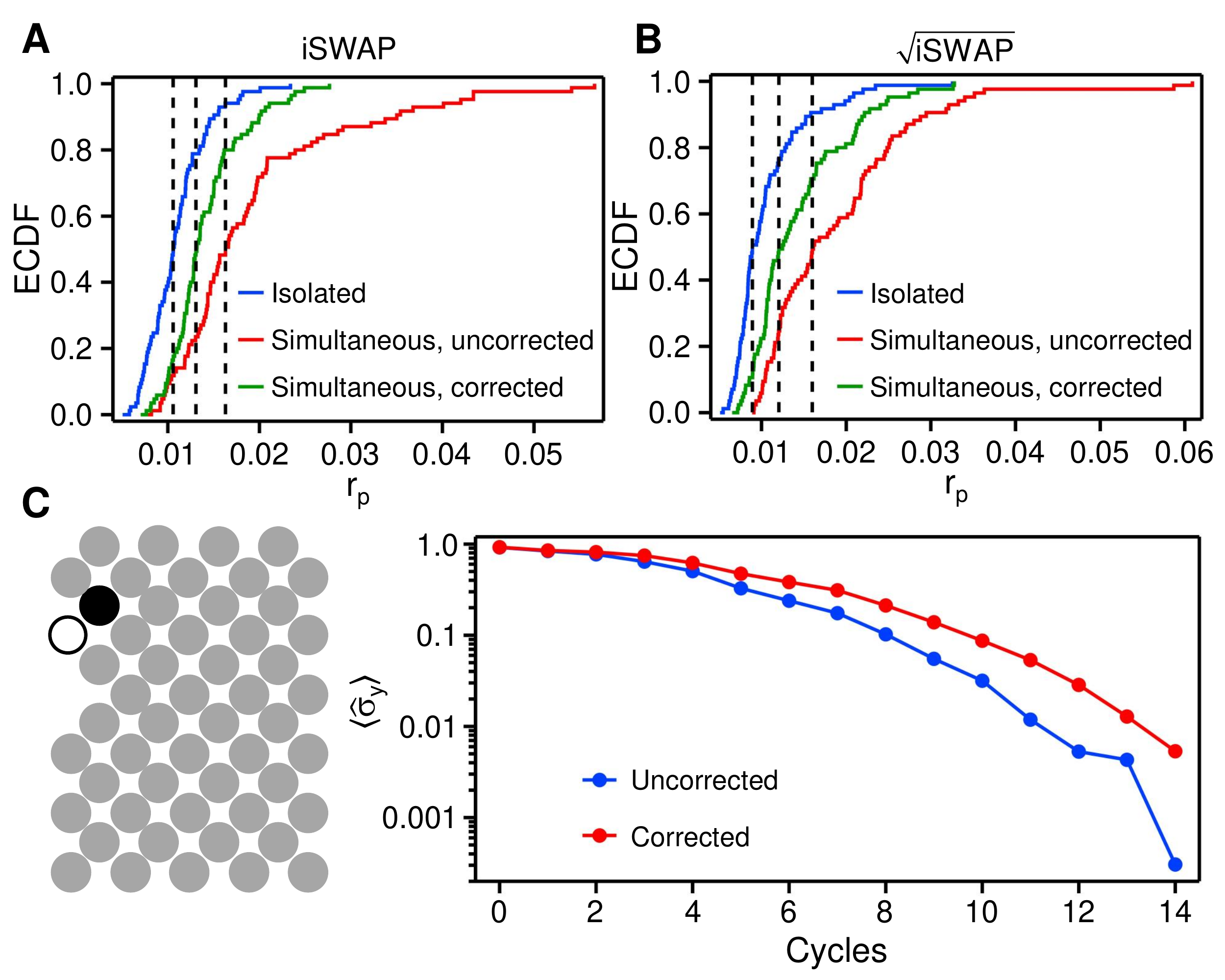}
	\caption{Fixed-Unitary XEB and cross-talk correction. (A, B) Integrated histograms (ECDF) of Pauli errors for both the iSWAP (A) and the $\sqrt{\text{iSWAP}}$ (B) gates. Three sets of values are shown for each case: the isolated (blue) curves represent the errors obtained by operating each qubit pair individually with all other qubits at ground state. The simultaneous, uncorrected (red) curves represent the errors obtained by operating all qubits at the same time, without any additional correction. The simultaneous, corrected (green) curves represent the simultaneous error rates after additional $Z$-rotations are included in the circuits to offset unitary shifts induced by cross-talk effects. Simultaneous error rates are obtained from four different configurations of parallel two-qubit operations, similar to Ref.~\cite{Arute2019}. Dashed lines indicate the median locations of various errors. All error rates are obtained from XEB with pure iSWAP or $\sqrt{\text{iSWAP}}$ gate used in simulation, and include contributions from two single-qubit gates and one two-qubit gate. (C) An OTOC experimental configuration for evaluating the effects of cross-talk correction. The empty circle represents the ancilla qubit and the black filled circle represents the measurement qubit. All other qubits are represented by purple spheres. (D) The OTOC normalization value $\braket{\hat{\sigma}_{\text y}}$ as a function of number of cycles in a quantum circuit $\hat{U}$ for the configuration shown in (C), measured both with and without applying the cross-talk corrections. $\hat{U}$ is composed of iSWAP and random single-qubit $\pi / 2$ rotations around axes on the $XY$ plane.}
	\label{fig:s2_3}
\end{figure} 

The Pauli errors of the calibrated iSWAP and $\sqrt{\text{iSWAP}}$ gates are measured through XEB \cite{Arute2019, Boixo_NatPhys_2018}, which uses a collection of random circuits comprising interleaved two-qubit and random single-qubit gates. For each random circuit, the probability distribution of all possible output bit-strings is both measured experimentally and computed numerically. The statistical correlation between the two distributions (``cross-entropy'') is then used to infer the total error of each quantum circuit. After measuring a sufficient number of circuit instances at different circuit depths, it has been shown that XEB reliably yields gate errors that are very consistent with values obtained from conventional characterization methods such as randomized benchmarking \cite{Arute2019, Knill_PRA_2008, Foxen_PRL_2020}. 

A key difference between the XEB process used in our current work compared to prior experiments \cite{Arute2019} is the unitaries used in the numerical computation of the benchmarking circuits. Previously, such unitaries are freely adjusted for each individual qubit pair, whereby values of various FSIM angles $\theta$, $\phi$, $\Delta_+$, $\Delta_-$ and $\Delta_{-, \text{off}}$ are optimized to obtain the lowest Pauli errors. In this work, we do not perform such an optimization step during gate error characterization and instead use a fixed unitary ($U_\text{FSIM} (\pi / 2, 0, 0, 0, 0)$ for iSWAP and $U_\text{FSIM} (\pi / 4, 0, 0, 0, 0)$ for $\sqrt{\text{iSWAP}}$) for all qubit pairs. The gate errors characterized by the ``fixed-unitary'' XEB process include contributions from both incoherent sources such as relaxation and dephasing and coherent sources such as remnant conditional-phases. The adoption of a more stringent benchmarking criterion for gate errors is motivated by the fact that both coherent and incoherent errors can lead to imperfect reversal of quantum circuits and adversely impact the accuracy of OTOC measurements, in contrast to previous experiment in which coherent errors are compensated by modifying circuits in simulation \cite{Arute2019}.

The Pauli error rate $r_\text{p}$ per cycle (aggregate error of two single-qubit gates and one two-qubit gate) associated with each qubit pair is first measured in isolation. The results are plotted as integrated histograms in Fig.~\ref{fig:s2_3}A and Fig.~\ref{fig:s2_3}B, where we observe a mean (median) error of 0.0109 (0.0106) for iSWAP and 0.0106 (0.0089) for $\sqrt{\text{iSWAP}}$. These higher errors rates compared to our previous work \cite{Arute2019} are a result of enhanced incoherent errors due to longer pulses used in iSWAP and additional detuning pulses in $\sqrt{\text{iSWAP}}$, as well as coherent errors arising from remnant conditional-phases.

We then repeat the same process but measure the error rates of different pairs simultaneously. We first observe, similar to previous work \cite{Arute2019}, a sizable increase in $r_{\text p}$, with the mean (median) being 0.0193 (0.0163) for iSWAP and 0.0190 (0.0161) for $\sqrt{\text{iSWAP}}$. To reduce these cross-talk effects, we first fit the XEB results to obtain the shifts in the two-qubit unitary associated with each individual qubit pair, which are often related to the single-qubit phases $\Delta_+$, $\Delta_-$ and $\Delta_{-, \text{off}}$. In a second step, instead of simply incorporating these shifts into classical simulation as was done previously \cite{Arute2019}, we add local $Z$-rotations into the quantum circuits to offset the unitary shifts. The parallel error rates are then re-measured with these $Z$-rotations. The mean (median) value of $r_{\text p}$ for simultaneous operation is reduced to 0.0140 (0.0131) for iSWAP and 0.0142 (0.0123) for $\sqrt{\text{iSWAP}}$.

The effects of the cross-talk correction can be readily seen in the ``normalization'' values used in the OTOC experiment. Figure~\ref{fig:s2_3}C shows the configuration for a 53-qubit OTOC experiment. The corresponding OTOC normalization $\braket{\hat{\sigma}_{\text y}}$ as a function of the number of cycles in a quantum circuit is shown in Fig.~\ref{fig:s2_3}D. Without applying the additional $Z$-rotations for cross-talk correction, $\braket{\hat{\sigma}_{\text y}}$ decays rapidly and falls below 0.1 after 8 cycles. After applying the additional $Z$-rotations, $\braket{\hat{\sigma}_{\text y}}$ decays at a visibly slower rate and falls below 0.1 after 10 cycles. The slower decay of $\braket{\hat{\sigma}_{\text y}}$ is indicative of more accurate inversion of the quantum circuit after cross-talk correction.

\subsection{Dynamical Decoupling}

\begin{figure}[t]
	\centering
	\includegraphics[width=1\columnwidth]{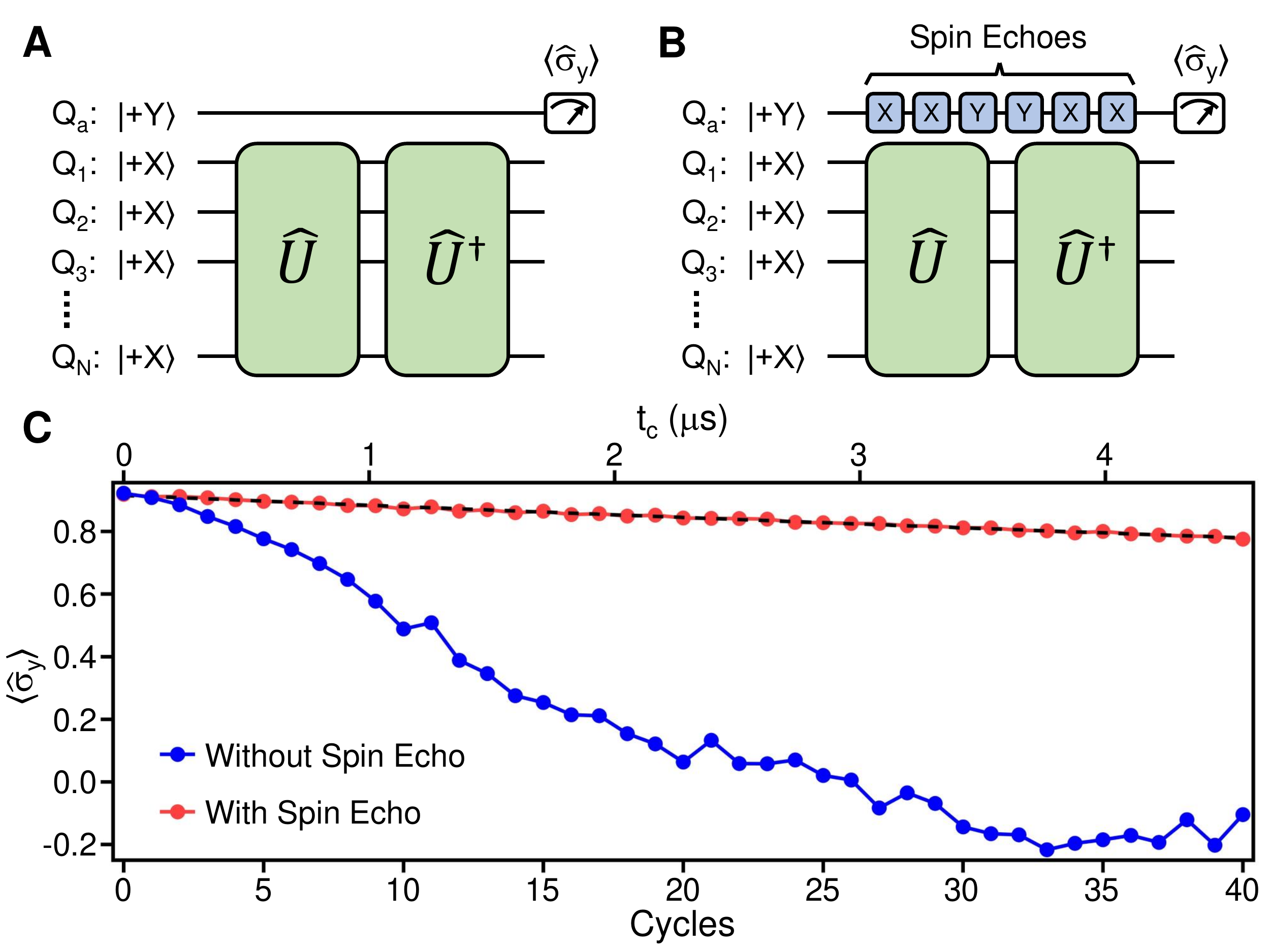}
	\caption{Dynamical coupling in OTOC experiments. (A) Test experimental circuit for benchmarking the intrinsic coherence of the ancilla qubit. Compared to an actual OTOC circuit, the CZ gates between the ancilla qubit and measurement qubit is removed. (B) Same circuit as (A) with the additional of consecutive spin echo pulses to the ancilla qubit during the quantum circuits $\hat{U}$ and $\hat{U}^\dagger$, in the form of random $X$ or $Y$ gates. (C) The projection of the ancilla qubit to the y-axis, $\braket{\hat{\sigma}_{\text y}}$, at the end of the circuit, measured with (red) and without (blue) spin echo. The bottom x-axis of the plot shows the number of cycles in $\hat{U}$ ($\hat{U}^\dagger$ has the same number of cycles), and the top x-axis shows the corresponding total circuit duration $t_\text{c}$.}
	\label{fig:s2_4}
\end{figure} 

We now describe a series of additional error-mitigation techniques used in the compilation of quantum circuits that further improve the accuracy of OTOC experiments. The first such technique is dynamical decoupling, motivated by long ``idling'' times at non-ground states for some of the qubits during the experiment. The most prominent of such qubits is the ancilla qubit, which remains idle throughout the time needed to implement the quantum circuit $\hat{U}$ and its inverse $\hat{U}^\dagger$. Intrinsic decoherence of the ancilla qubit can in principle limit the circuit depth at which OTOC can be resolved, especially if the characteristic time $T_2$ is comparable to the total duration of $\hat{U}$ and $\hat{U}^\dagger$.

To benchmark the intrinsic $T_2$ of the ancilla qubit during an OTOC experiment, we design the test circuits shown in Fig.~\ref{fig:s2_4}A and Fig.~\ref{fig:s2_4}B. In either case, we removed the two CZ gates otherwise present in an actual OTOC experiment such that the ancilla qubit does not interact with any other qubit apart from cross-talk effects. The difference between the two cases is that the ancilla qubit remains completely idle during $\hat{U}$ and $\hat{U}^\dagger$ in Fig.~\ref{fig:s2_4}A, whereas a train of spin echo pulses $X-X-Y-Y-X-X...$ is applied to the ancilla qubit during $\hat{U}$ and $\hat{U}^\dagger$ in Fig.~\ref{fig:s2_4}B.

The y-axis projection of the ancilla qubit, $\braket{\hat{\sigma}_{\text y}}$, at the end of $\hat{U}$ and $\hat{U}^\dagger$ is measured both with and without the added spin echoes. The results are shown in Fig.~\ref{fig:s2_4}C. We observe that without spin echo, $\braket{\hat{\sigma}_{\text y}}$ decays rather quickly despite no entanglement between the ancilla and other qubits in the system, falling to $\sim$0 after 25 circuit cycles (corresponding to an evolution time $t_\text{c} \approx 3.0$ $\mu$s). The Gaussian shape of the decay at earlier times suggests that low-frequency noise is likely the dominant source of decoherence \cite{Ithier_PRB_2005}. On the hand, with the addition of spin echo, $\braket{\hat{\sigma}_{\text y}}$ decays at a much slower rate maintaining a value of 0.78 even after 40 circuit cycles ($t_\text{c} \approx 4.6$ $\mu$s). By fitting $\braket{\hat{\sigma}_{\text y}}$ to a functional form $Ae^{-\frac{t_\text{c}}{T_2}}$ where $A$ and $T_2$ are fitting parameters, we obtain a coherence time $T_2 = 28.6$ $\mu$s for the ancilla qubit, which is close to the $2 T_1$ limit of our quantum processor \cite{Arute2019}. Given this value of $T_2$ is significantly longer than all OTOC experimental circuits used in this work (the longest circuit lasts $\sim$5 $\mu$s), we conclude that changes in the ancilla projection $\braket{\hat{\sigma}_{\text y}}$ in the OTOC experiments are indeed dominated by the many-body effects in $\hat{U}$ and $\hat{U}^\dagger$ rather than the intrinsic decoherence of the ancilla qubit itself. For all experimental results presented in the main text, spin echo is applied to the ancilla qubit.

\subsection{Elimination of Bias in $\braket{\hat{\sigma}_{\text y}}$}

\begin{figure}[t]
	\centering
	\includegraphics[width=1\columnwidth]{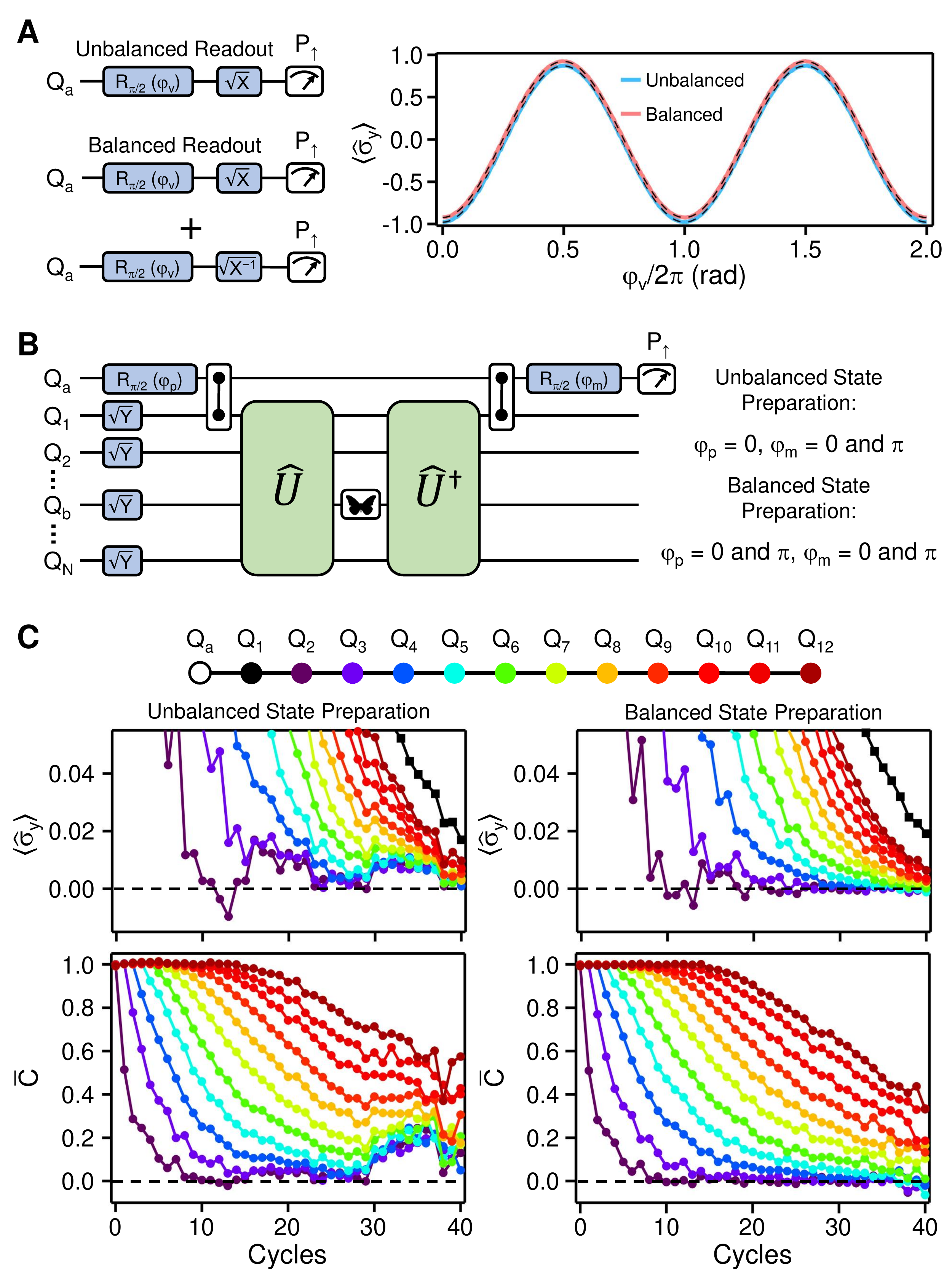}
	\caption{Unbiased measurements of $\braket{\hat{\sigma}_{\text y}}$. (A) Left panel shows the test circuits for characterizing $\braket{\hat{\sigma}_{\text y}}$ arising from asymmetry in readout fidelities. Two different schemes are implemented. The conversion between the excited-state population(s) $P_\uparrow$ and $\braket{\hat{\sigma}_{\text y}}$ is described in the main text. Right panel shows the experimental values of $\braket{\hat{\sigma}_{\text y}}$ as a function of the variable phase $\varphi_{\text v}$, obtained with both readout schemes. Dashed lines show fits to $\braket{\hat{\sigma}_{\text y}} = (1 - 2F_{\text r}) \cos \left( \varphi_{\text v} \right) + d_{\text r}$, where $F_{\text r}$ and $d_{\text r}$ are fitting parameters. (B) Schematic of the full OTOC circuit showing two different state preparation schemes: in the unbalanced scheme, only one phase is used for the first $R_{\pi / 2}$ gate on the ancilla qubit. In the balanced scheme, two different phases are used. Balanced readout is used in both schemes. (C) Results of a 12-qubit OTOC experiment. Upper panels show $\braket{\hat{\sigma}_{\text y}}$ at different number of circuit cycles. The black squares represent the normalization case and the colored spheres represent $\braket{\hat{\sigma}_{\text y}}$ obtained with the butterfly operator ($Z$) successively applied to qubits 2 (Q$_{2}$) through 12 (Q$_{12}$). The location of the butterfly operator for each curve is indicated by the color legend on top. The normalized OTOCs, $\overline{C}$, are shown in the lower panels. The data are the average values of 40 different random circuit instances.}
	\label{fig:s2_5}
\end{figure}

The accuracy of OTOC measurements is particularly susceptible to any bias in $\braket{\hat{\sigma}_{\text y}}$ of the ancilla qubit, i.e. a fixed offset $d$ to the ideal value. Such a bias can, for example, be introduced by the different readout fidelities for the $\ket{0}$ and $\ket{1}$ states of the ancilla qubit. To see the impact of the bias on OTOC, we consider an ideal OTOC value of $C_0$ and an ideal normalization value of $\braket{\hat{\sigma}_{\text y}}_{\text I}$. The ideal y-projection of the ancilla with the butterfly operator applied, $\braket{\hat{\sigma}_{\text y}}_{\text B}$, in such a case is $\braket{\hat{\sigma}_{\text y}}_{\text B} \approx C_0 \braket{\hat{\sigma}_{\text y}}_{\text I}$. However, in the presence of a bias, the measured projections become $\braket{\hat{\sigma}_{\text y}} = \braket{\hat{\sigma}_{\text y}}_{\text I} + d$ for the normalization value and $\braket{\hat{\sigma}_{\text y}} = \braket{\hat{\sigma}_{\text y}}_{\text B} + d$ with the butterfly operator applied. The experimental value for OTOC then becomes $C_1 = \frac{\braket{\hat{\sigma}_{\text y}}_{\text B} + d}{\braket{\hat{\sigma}_{\text y}}_{\text I} + d} \approx \frac{C_0 \braket{\hat{\sigma}_{\text y}}_{\text I} + d}{\braket{\hat{\sigma}_{\text y}}_{\text I} + d}$. Assuming typical values of $\braket{\hat{\sigma}_{\text y}}_{\text I} = 0.05$ and $C_0 = 0.1$, even a small asymmetry $d = 0.01$ would lead to a highly erroneous value of $C_1 \approx 0.25$. It is therefore crucial to identify and mitigate any bias in $\braket{\hat{\sigma}_{\text y}}$ of the ancilla qubit.

We begin by measuring $\braket{\hat{\sigma}_{\text y}}$-bias due to asymmetry in readout errors. The test circuit is shown in the left panel of Fig.~\ref{fig:s2_5}A under the label ``unbalanced readout''. We first project the qubit onto the equator of the Bloch sphere with a $\pi/2$ rotation, $R_{\pi/2} \left( \varphi_{\text v} \right)$, where $\varphi_{\text v}$ is a variable phase. A second $\pi/2$ rotation around a fixed axis, $\sqrt{X}$, is then applied and excited state population $P_\uparrow$ is finally measured. The population is then converted to $\braket{\hat{\sigma}_{\text y}}$ via $\braket{\hat{\sigma}_{\text y}} = 2 P_\uparrow - 1$. The right panel of Fig.~\ref{fig:s2_5}A shows $\braket{\hat{\sigma}_{\text y}}$ as a function of $\varphi_{\text v}$ and a fit to a functional form $\braket{\hat{\sigma}_{\text y}} = (1 - 2F_{\text r}) \cos \left( \varphi_{\text v} \right) + d_{\text r}$. Here $F_{\text r}$ is the average of the readout fidelities of the $\ket{0}$ and $\ket{1}$ states, and $d_{\text r}$ is their difference. For this unbalanced readout scheme, we obtain $F_{\text r} = 0.0962$ and $d_{\text r} = 0.055$. The observed difference in readout fidelities is consistent with past experiments where it was shown that energy relaxation of the qubit during dispersive readout generally leads to lower readout fidelity for the $\ket{1}$ state compared to the $\ket{0}$ state \cite{Jeffrey_PRL_2014}.

The bias $d_{\text r} = 0.055$ is significantly reduced by adopting a "balanced readout" scheme, also shown in the left panel of Fig.~\ref{fig:s2_5}A. Here, we measure the $P_\uparrow$ of the ancilla qubit with the second gate in the test circuit being either $\sqrt{X}$ or $\sqrt{-X}$. The results are then combined to obtain $\braket{\hat{\sigma}_{\text y}} = \left(P_{\uparrow,+} - P_{\uparrow,-} \right)$, where $P_{\uparrow,\pm}$ is $P_\uparrow$ measured with the second gate being $\sqrt{\pm X}$. The averaged $\braket{\hat{\sigma}_{\text y}}$ is shown in the right panel of Fig.~\ref{fig:s2_5}A as a function of $\varphi_{\text v}$. A similar fit as before yields the same average readout fidelity $F_{\text r} = 0.0962$ and a much reduced bias $d_{\text r} \approx 0.0002$.

Balanced readout alone, as we will demonstrate below, is insufficient for completely removing bias from $\braket{\hat{\sigma}_{\text y}}$. For all experiments reported in the main text, we apply a second symmetrization step shown in Fig.~\ref{fig:s2_5}B, which is the measurement scheme for OTOC with the gates related to SPAM explicitly shown. Here, we parametrize both the phase $\varphi_{\text p}$ of the first $\pi / 2$ gate and the phase $\varphi_{\text m}$ of the last $\pi / 2$ gate on the ancilla qubit. In an ``unbalanced state preparation'' scheme, we measure the average projections $\braket{\hat{\sigma}_{\text y}} = \left(P_{\uparrow,++} - P_{\uparrow,+-} \right)$, where $P_{\uparrow,++}$ and $P_{\uparrow,+-}$ are $P_\uparrow$ obtained with $(\varphi_{\text p}, \varphi_{\text m}) = (0, 0)$ and $(\varphi_{\text p}, \varphi_{\text m}) = (0, \pi)$, respectively. In a "balanced state preparation" scheme, we measure the average projections $\braket{\hat{\sigma}_{\text y}} = \frac{1}{2}\left( P_{\uparrow,++} - P_{\uparrow,+-} - P_{\uparrow,-+} + P_{\uparrow,--} \right)$, where $P_{\uparrow,-+}$ and $P_{\uparrow,--}$ are additional populations obtained with $(\varphi_{\text p}, \varphi_{\text m}) = (\pi, 0)$ and $(\varphi_{\text p}, \varphi_{\text m}) = (\pi, \pi)$, respectively.

The difference between the two state preparation schemes is illustrated in Fig.~\ref{fig:s2_5}C, where we present the results of a 12-qubit OTOC experiment. The quantum circuit $\hat{U}$ is non-integrable and composed of $\sqrt{\text{iSWAP}}$ and random single-qubit $\pi/2$ rotations around 8 different axes on the $XY$ plane. A total of 40 circuit instances are used and the data shown are average values over all instances. The top panels show the measured values of $\braket{\hat{\sigma}_{\text y}}$ for the normalization case and cases where a butterfly operator $Z$ is successively applied to qubits 2 to 12 in-between $\hat{U}$ and $\hat{U}^\dagger$. The y-axis scale is intentionally limited to $<0.05$. We observe in the case of unbalanced state preparation, the $\braket{\hat{\sigma}_{\text y}}$ values exhibit sudden rise from 0 to $>$0.006 at cycles 29 to 38. This behavior is inconsistent with the effects of scrambling and decoherence, either of which is expected to lead to monotonic decay of $\braket{\hat{\sigma}_{\text y}}$ toward 0 for a non-integrable process. In contrast, with balanced state preparation, $\braket{\hat{\sigma}_{\text y}}$ indeed monotonically decays toward 0 at large cycles for all curves. 

The bottom panels of Fig.~\ref{fig:s2_5}C show the normalized average OTOCs, $\overline{C}$, for each qubit. Here we observe that $\overline{C}$ obtained with the unbalanced state preparation scheme again manifests unphysical jumps from 0 at cycles 29 to 38, resulting from the finite bias $\braket{\hat{\sigma}_{\text y}}$ and the mechanism outlined at the beginning of this section. In contrast, $\overline{C}$ obtained with the balanced state preparation scheme decays monotonically toward 0, in agreement with the scrambling behavior of a non-integrable process. These data suggest that a symmetrization step duration the state preparation phase of the ancilla qubit is needed to completely remove the bias in $\braket{\hat{\sigma}_{\text y}}$, in addition to the symmetrization step before readout. The physical origin of the remnant bias seen in the left panels of Fig.~\ref{fig:s2_5}C is not completely understood at the time of writing, and could be related to control errors in the single-qubit gates on the ancilla qubit, incomplete depolarization of $T_1$ errors by the spin echoes, or other unknown mechanisms.

\subsection{Light-cone Filter}

\begin{figure}[!t]
	\centering
	\includegraphics[width=1\columnwidth]{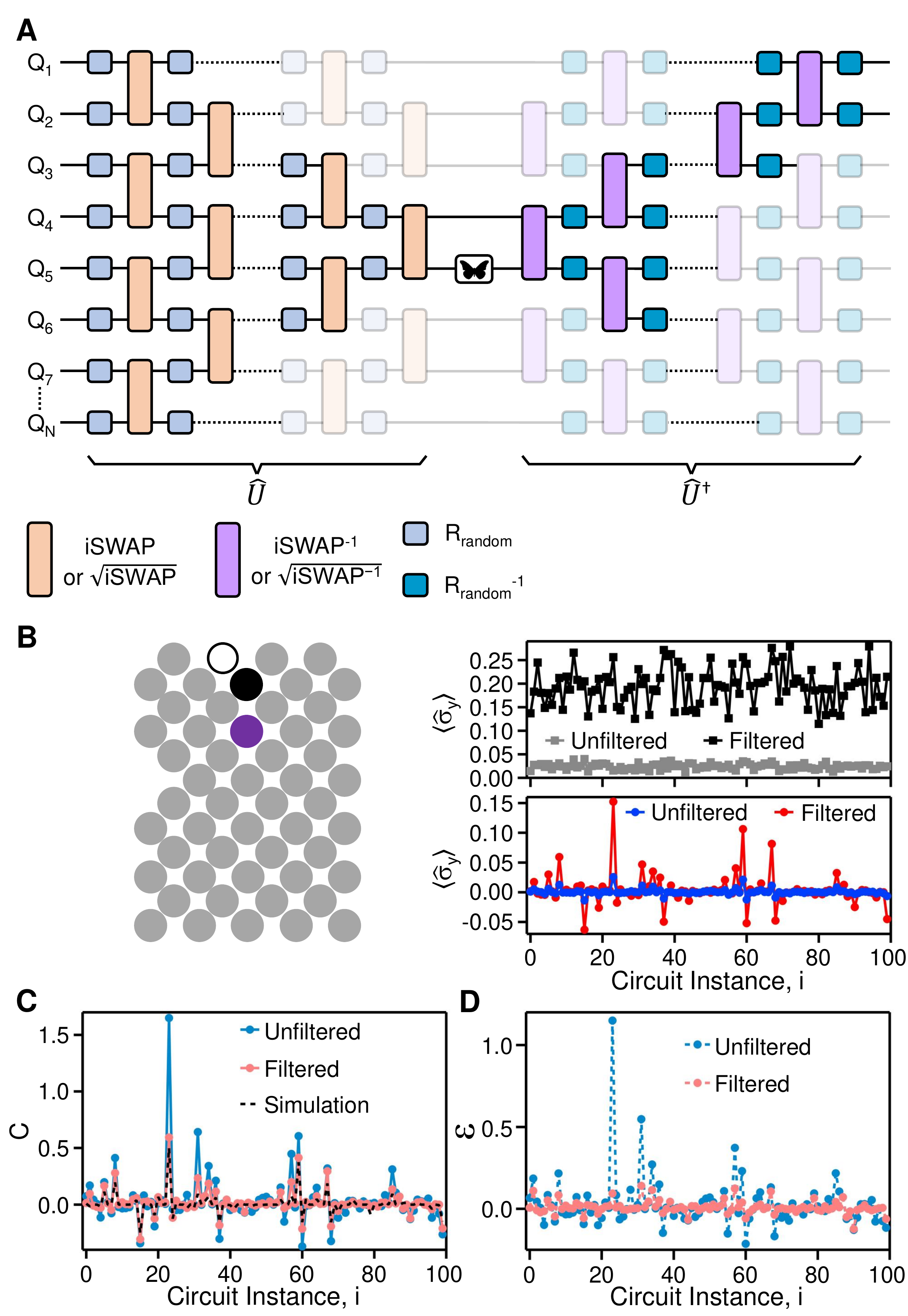}
	\caption{Re-compiling OTOC circuits with light-cone filter. (A) Schematic of an OTOC measurement circuit, including the component gates of the quantum circuits $\hat{U}$ and $\hat{U}^\dagger$. The ancilla qubit and its related gates, as well as the $\sqrt{Y}$ gates used in the state preparation of all qubits, are omitted for simplicity. Gates shown with semi-transparent colors can be removed from the OTOC measurement circuit without altering its output. (B) Left panel shows a configuration for evaluating experimental effects of the light-cone filter. The unfilled circle represents the ancilla qubit and the black filled circle represents the measurement qubit. The purple filled circle indicates the butterfly qubit. The total number of circuit cycles is 11. Right panel shows the measured values of $\braket{\hat{\sigma}_{\text y}}$ for 100 random circuit instances. Data obtained in the normalization case are shown on top and data obtained with the butterfly operator applied are shown at the bottom. The same number of repetitions (1 million) is used in all cases to estimate $\braket{\hat{\sigma}_{\text y}}$. (C) Normalized experimental OTOC values $C$ for different circuit instances, plotted alongside exact numerical simulation results. (D) Experimental errors $\epsilon$ for different circuit instances, corresponding to the differences between experimental and simulated values.}
	\label{fig:s2_6}
\end{figure}

Analogous to classical systems, quantum perturbations often travel at a limited speed (the ``butterfly velocity''). This typically results in a ``light-cone'' structure for many quantum circuits, which can be capitalized to reduce their classical simulation costs \cite{Hastings_PRB_2008}. Similarly, the light-cone structure of these quantum circuits may also be utilized to modify their implementations on a quantum processor and improve the fidelity of experimental results. In this section, we describe a light-cone-based circuit re-compilation technique that led to considerable improvements in the accuracy of experimental OTOC measurements.

Figure~\ref{fig:s2_6}A displays the generic structure of an OTOC measurement circuit, where the component gates of the quantum circuit $\hat{U}$ and its inverse $\hat{U}^\dagger$ are explicitly shown. The butterfly operator possesses a pair of triangular light-cones extending from the middle of the circuit into both $\hat{U}$ and $\hat{U}^\dagger$. Quantum gates outside these light-cones may be completely removed (``filtered'') without altering the output of the circuit. Furthermore, since the measurement at the end of the circuit is also localized at a single qubit (Q$_1$), one may additionally discard quantum gates outside the light-cone of Q$_1$ originating from the right-end of $\hat{U}^\dagger$, without altering circuit output. The gates removed by the light-cone filter are shown with semi-transparent colors in Fig.~\ref{fig:s2_6}A. In practice, some qubits have much longer idling times as a result of gate removal and become more susceptible to decoherence effects such as relaxation and dephasing. To mitigate such effects, we also apply spin echo to qubits with long idling times, similar to the approach to the ancilla qubit in Fig.~\ref{fig:s2_4}.

The effects of the light-cone filter on OTOC measurements are shown in Fig.~\ref{fig:s2_6}B and Fig.~\ref{fig:s2_6}C. The left panel of Fig.~\ref{fig:s2_6}B shows the configuration for a 53-qubit OTOC experiment. Here we choose a quantum circuit $\hat{U}$ with iSWAP and random single-qubit gates which are $\pi/2$ rotations around axes on the $XY$ plane. The axes of rotation are chosen such that exactly 48 non-Clifford rotations (randomly selected from $\sqrt{\pm W}$ and $\sqrt{\pm V}$) occur in $\hat{U}$ and $\hat{U}^\dagger$. All other single-qubit gates are Clifford rotations randomly selected from $\sqrt{\pm X}$ and $\sqrt{\pm Y}$). The number of circuit cycles is fixed at 11. The right panels of Fig.~\ref{fig:s2_6}B show experimental results for 100 individual instances of $\hat{U}$, whereby $\braket{\hat{\sigma}_{\text y}}$ for the normalization case is plotted at the top and $\braket{\hat{\sigma}_{\text y}}$ with the butterfly operator ($X$) applied is plotted at the bottom. We observe significant enhancements in the amplitudes of $\braket{\hat{\sigma}_{\text y}}$ after the application of the light-cone filter, with the normalization $\braket{\hat{\sigma}_{\text y}}$ values averaging to 0.024 without the filter and 0.194 with filter.

The experimental improvement facilitated by the light-cone filter is more clearly seen by comparing the normalized OTOC values $C$ with exact numerical simulation of the same circuits, as shown in Fig.~\ref{fig:s2_6}C. Without the light-cone filter, the experimental $C$ values are substantially different from simulation results. On the other hand, with the light-cone filter, the agreement between experimental and simulated values is much closer. We further quantify the effect of light-cone filter by plotting the differences between numerical and experimental values of $C$, $\epsilon$, in Fig.~\ref{fig:s2_6}D. Here we observe that the root-mean-square (RMS) value for $\epsilon$ is 0.156 without the light-cone filter, whereas it is reduced to 0.041 after filter is applied. This four-fold improvement in accuracy of OTOC measurements is a natural consequence of the reduction of number of gates in the overall quantum circuit (the number of iSWAP gates is reduced from 464 to 161 for the example in Fig.~\ref{fig:s2_6}). Although the additional qubit idling introduced by the light-cone filter carries errors as well, they are expected to be much less than the errors of the removed two-qubit gates, particularly when spin echo is also applied during the idling.

\subsection{Normalization via Reference Clifford Circuits}

\begin{figure}[t]
	\centering
	\includegraphics[width=1\columnwidth]{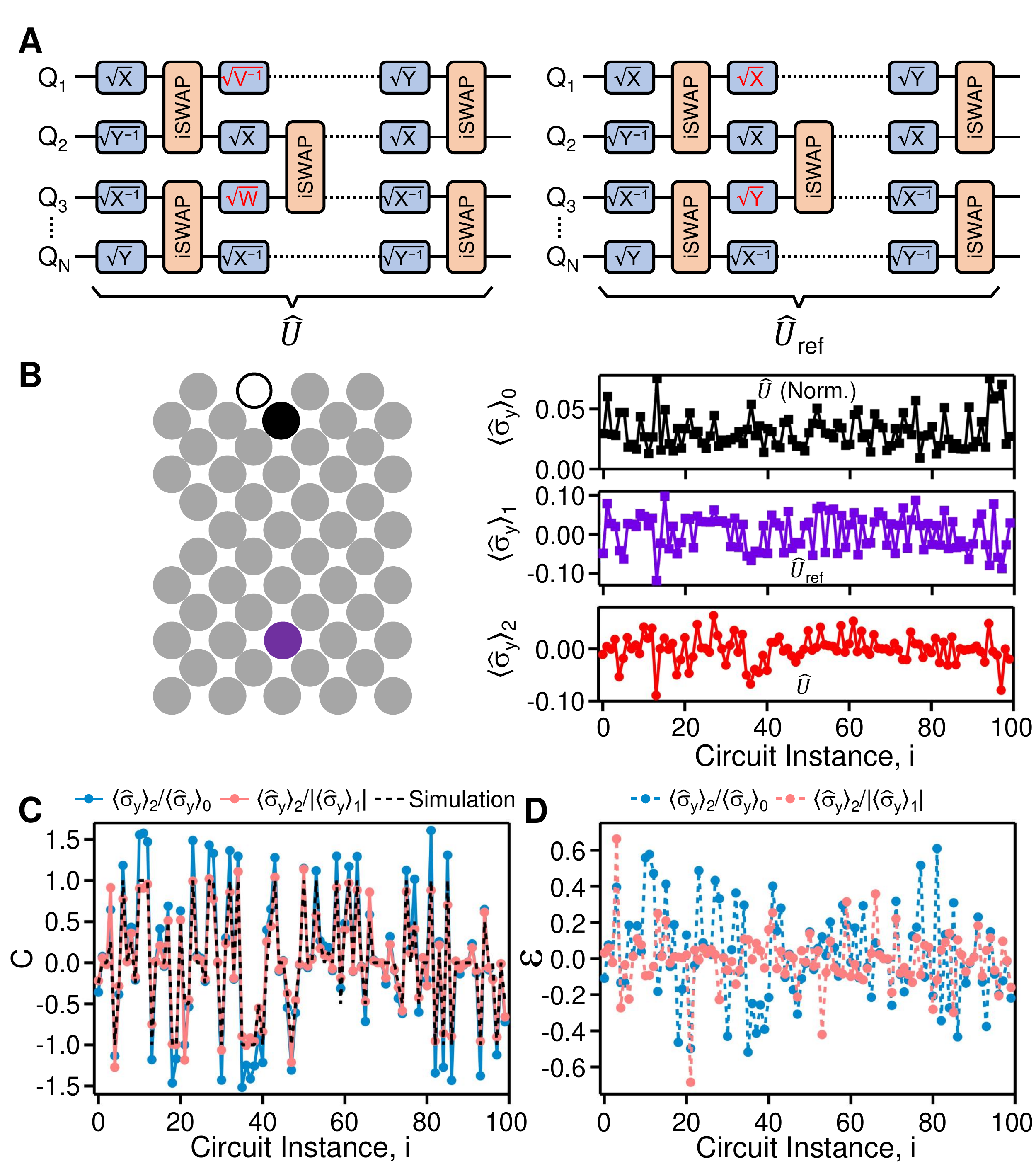}
	\caption{Normalization via reference Clifford circuits. (A) Schematic of two quantum circuits: $\hat{U}$ is the actual quantum circuit of interest, composed mostly of Clifford gates and a few non-Clifford gates ($\sqrt{\pm V}$ and $\sqrt{\pm W}$). $\hat{U}_\text{ref}$ is a reference circuit with the same Clifford gates as $\hat{U}$. The non-Clifford gates in $\hat{U}$ are replaced with random Clifford gates $\sqrt{\pm X}$ and $\sqrt{\pm Y}$ in $\hat{U}_\text{ref}$. (B) Left panel shows an experimental configuration for comparing two normalization procedures. The black unfilled (filled) circle represents the ancilla (measurement) qubit. The purple filled circle represents the butterfly qubit. The total number of circuit cycles is 14. Right panel shows the measured values of $\braket{\hat{\sigma}_{\text y}}$ for 100 random circuit instances. $\braket{\hat{\sigma}_{\text y}}_0$ denotes values obtained without applying butterfly operator to $\hat{U}$, $\braket{\hat{\sigma}_{\text y}}_1$ denotes values obtained with butterfly operator applied to $\hat{U}_\text{ref}$, and $\braket{\hat{\sigma}_{\text y}}_2$ denotes values obtained with butterfly operator applied to $\hat{U}$. The same number of repetitions (4 millions) is used in all cases to estimate $\braket{\hat{\sigma}_{\text y}}$. (C) Normalized experimental OTOC values $C$ for different circuit instances, plotted alongside exact numerical simulation results. (D) Experimental errors $\epsilon$ for different circuit instances, corresponding to the differences between experimental and simulated values.}
	\label{fig:s2_7}
\end{figure}

The last error-mitigation technique we use for the OTOC experiment is specific to quantum circuits $\hat{U}$ composed of predominantly Clifford gates with a small number of non-Clifford gates. For such circuits, it is found through numerical studies that a modified normalization procedure yields more accurate values of OTOC (Fig.~\ref{fig:s2_7}A): Consider a quantum circuit $\hat{U}$ composed of mostly Clifford gates (iSWAP, $\sqrt{\pm X}$ and $\sqrt{\pm Y}$) and a small number of non-Clifford gates ($\sqrt{\pm V}$ and $\sqrt{\pm W}$). We first measure the $\braket{\hat{\sigma}_{\text y}}$ of the ancilla with a butterfly operator applied between $\hat{U}$ and $\hat{U}^\dagger$ (we denote this value as $\braket{\hat{\sigma}_{\text y}}_2$), same as before. In a second step, instead of measuring $\braket{\hat{\sigma}_{\text y}}$ without applying the butterfly operator (denoted by $\braket{\hat{\sigma}_{\text y}}_0$), we measure $\braket{\hat{\sigma}_{\text y}}$ with the same butterfly operator but a different quantum circuit $\hat{U}_\text{ref}$ and its inverse $\hat{U}_\text{ref}^\dagger$ (denoted by $\braket{\hat{\sigma}_{\text y}}_1$). The reference circuit $\hat{U}_\text{ref}$ has the same Clifford gates as $\hat{U}$, whereas the non-Clifford gates in $\hat{U}$ are replaced with Clifford gates chosen randomly from $\sqrt{\pm X}$ and $\sqrt{\pm Y}$.

Example data showing $\braket{\hat{\sigma}_{\text y}}_1$, $\braket{\hat{\sigma}_{\text y}}_2$ and $\braket{\hat{\sigma}_{\text y}}_3$ are shown in Fig.~\ref{fig:s2_7}B, where $\hat{U}$ contains a total of 8 non-Clifford rotations and 14 cycles. Similar to the previous section, we present results from 100 circuit instances. Next, we process the data to obtain experimental values of OTOC, $C$, in two different ways: First, we apply $C = \braket{\hat{\sigma}_{\text y}}_2 / \braket{\hat{\sigma}_{\text y}}_0$, which corresponds to the normalization procedure used in the previous sections. Second, we apply $C = \braket{\hat{\sigma}_{\text y}}_2 / \left| \braket{\hat{\sigma}_{\text y}}_1 \right|$, corresponding to normalization using $\braket{\hat{\sigma}_{\text y}}$ of a reference circuit. Here the absolute sign accounts for the fact that the theoretical OTOC values of Clifford circuits are $\pm 1$. The resulting $C$ values are both plotted alongside exact simulation results in the left panel of Fig.~\ref{fig:s2_7}C. It is easily seen that the second normalization procedure with reference Clifford circuits yields experimental values that are in much better agreement with simulation results. Indeed, the experimental errors $\epsilon$ (Fig.~\ref{fig:s2_7}D) have an RMS value of 0.250 when $C = \braket{\hat{\sigma}_{\text y}}_2 / \braket{\hat{\sigma}_{\text y}}_0$ is used and 0.157 when $C = \braket{\hat{\sigma}_{\text y}}_2 / \left| \braket{\hat{\sigma}_{\text y}}_1 \right|$ is used. Given these observations, we adopt normalization via reference Clifford circuits when measuring quantum circuits dominated by Clifford gates.

Lastly, we note that for the data in Fig.~3 and Fig.~4 of the main text, we apply a ensemble of reference circuits $\hat{U}_\text{ref}$ and use the average value of $\left| \braket{\hat{\sigma}_{\text y}} \right|$ obtained from all $\hat{U}_\text{ref}$ to normalize $\braket{\hat{\sigma}_{\text y}}$ of the actual quantum circuit $\hat{U}$. The typical number of $\hat{U}_\text{ref}$ for each $\hat{U}$ is 10 in Fig.~3 and varies between 15 and 70 for Fig.~4.

\section{Large-Scale Simulation of OTOCs of Individual Circuits} \label{sec:largenumerics}

In the last few years, there has been a constant development of new numerical
techniques to simulate large scale quantum circuits. Among the many promising
methods, two major numerical techniques are widely used on HPC clusters for
large scale simulations: tensor contraction \cite{markov2008simulating,
	chen2018classical, villalonga2020establishing, villalonga2019flexible,
	huang2020classical, gray2020hyper} and Clifford gate expansion
\cite{gottesman1998heisenberg, Aaronson_PRA_2004, Bravyi_PRL_2016,
	bravyi2019simulation}. All the aforementioned methods have advantages and
disadvantages, which mainly depend on the underlying layout of the quantum
circuits and the type of used gates. On the one hand, tensor contraction works
best for shallow circuits with a small treewidth \cite{boixo2017simulation,
	gray2020hyper}. On the other hand, Clifford gate expansion is mainly used to
simulate arbitrary circuit layouts with few non-Clifford gates. Indeed, it is
well known that circuits composed of Clifford gates only can be simulated in
polynomial time \cite{gottesman1998heisenberg}, with a numerical cost which
grows exponentially with the number of non-Clifford gates
\cite{Aaronson_PRA_2004}. Both methods can be used to sample exact and
approximate amplitudes, with a computational cost which
decreases with an increasing level of noise. For instance, approximate
amplitudes can be sampled by slicing large tensor network and contracting only
a fraction of the resulting sliced tensors \cite{markov2018quantum,
	markov2019faster}. The final fidelity of the sample amplitudes is therefore
proportional to the fraction of contracted slices
\cite{villalonga2020establishing, villalonga2019flexible}, which can be tuned
to match experimental fidelity. Similarly, it is possible to sample approximate
amplitudes by only selecting the dominant stabilizer states in the Clifford
expansion \cite{Bravyi_PRL_2016, bravyi2019simulation}.\\

In our numerical simulations, we used tensor contraction to compute approximate
OTOC values, which are then validated using results from the Clifford expansion
for circuits with a small number of non-Clifford gates. Both methods are
described in the following sections.

\subsection{Numerical Calculation of the OTOC Value}

As described in the main text and shown in Fig.~1(A), the experimental OTOC
circuits have density-matrix-like structure of the form $\hat C =
\hat{U}^\dagger\, \hat\sigma^{(Q_b)}\, \hat{U}$, with $\hat\sigma^{(Q_b)}$
being the butterfly operator. In all numerical simulations, we used iSWAPs as
entangling two-qubit gates. Before and after $\hat C$, a controlled-$Z$ gate is
applied between the qubit $Q_1$ (in $\hat C$) and an ancilla qubit $Q_a$
(external to $\hat C$): the OTOC value is therefore obtained by computing the
expectation value of $\langle \hat\sigma_y \rangle$ relative to the ancilla
qubit $Q_a$.  To reduce the computational cost, it is always possible to project
the ancilla to either $0$ or $1$ (in the computational basis). Let us call $\hat
C_0 = \hat C$ ($\hat C_1 = \hat\sigma^{(Q_1)}_z\, \hat C\,
\hat\sigma^{(Q_1)}_z$) the circuit with the ancilla qubit projected on $0$ ($1$).
Therefore, the OTOC value $\langle\hat\sigma_y\rangle$
can be obtained as:
\begin{equation}\label{eq:num_otoc}
\langle\hat\sigma_y\rangle = \mathbb{R}\left[
\langle \psi_1 | \psi_0 \rangle\right],
\end{equation}
with $\ket{\psi_0} = \hat C_0 \ket{+}$ and $\ket{\psi_1} = \hat C_1 \ket{+}$
respectively.

\subsection{Branching Method}

To get exact OTOC value for circuit with a small number of non-Clifford
rotations, we used a branching method based on the Clifford expansion. More
precisely, recalling that OTOC circuits have a density-matrix-like structure,
that is $\hat C = \hat{U}^\dagger\, \hat\sigma^{(Q_b)}\, \hat{U} =
\big(\hat{g}_t^\dagger\cdots\hat{g}_1^\dagger\big)\,\hat\sigma^{(Q_b)}\,
\big(\hat{g}_1\cdots\hat{g}_t\big)$, it is possible to apply each pair of gates
$\big\{\hat{g}_t,\,\hat{g}^\dagger_t\big\}$ to $\hat\sigma^{(Q_b)}$ iteratively and
``branch'' only for non-Clifford gates. 

At the beginning of the simulation, a Pauli ``string'' is initialized to all
identities 
except a $\hat\sigma_x^{(Q_b)}$ operator (the butterfly operator) on the butterfly
qubit $Q_b$ (in this examples, the Pauli $X$ is chosen as butterfly operator),
that is $\mathcal{P} = \hat I^{(1)} \hat I^{(2)} \cdots \hat\sigma_x^{(Q_b)}
\cdots$. Whenever a pair of Clifford operators
$\big\{\hat{g}_t,\,\hat{g}^\dagger_t\big\}$ is applied to $\mathcal{P}$, the
Pauli string is ``evolved'' to another Pauli string. For instance, an
$\text{iSWAP}$ operator evolves the Pauli string $\mathcal{P} = \hat
I\,\hat\sigma_x$ to $\mathcal{P}^\prime = \text{iSWAP}^\dagger\,\big(\hat I\,
\hat \sigma_x\big)\, \text{iSWAP} = -\,\hat\sigma_y \hat\sigma_z$. On the
contrary, if a non-Clifford operator is applied to $\mathcal{P}$,
the Pauli string $\mathcal{P}$ will evolve into a superposition of multiple Pauli
strings, that can be eventually explored as independent branches.
As an example, the non-Clifford rotation $\hat g_t =
\sqrt{W} = \sqrt{X + Y}$ will branch $\mathcal{P} = \hat\sigma_x$ three times
into $\mathcal{P}^\prime_1 = \frac{\hat\sigma_x}{\sqrt{2}}$,
$\mathcal{P}^\prime_2 = \frac{\hat\sigma_y}{\sqrt{2}}$ and
$\mathcal{P}^\prime_3 = \frac{\hat\sigma_z}{2}$ respectively.
Because $\big\{\hat{g}_t,\,\hat{g}^\dagger_t\big\}$ are always applied in pairs
(one gate from $U$ and one gate from $U^\dagger$), the computational complexity
depends on the number $N_D$ of non-Clifford gates in $U$ only ($U$ and
$U^\dagger$ may have a different number of non-Cliffords because of the
different lightcones acting on them. See Fig.~\ref{fig:s2_6}A). More precisely,
our branching algorithms scales as the number of branches $n_b$ induced by the
$N_D$ non-Clifford rotations in $U$, that is $\mathcal{O}\left(n_b\right)$.

After the applications of all $\left\{\hat{g}_i\right\}_{i=1,\,\ldots,t}$ gates,
the OTOC circuit $\hat C$ will be then represented as a superposition of
distinct Pauli strings, each with a different amplitude.
The full states $\ket{\psi_0}$ and $\ket{\psi_1}$ can be then obtained by applying
the initial state $\ket{+}$ to the Clifford expansion of $\hat C$ and,
therefore, the OTOC value from Eq.~(\ref{eq:num_otoc}). To reduce the memory
footprint of our branching method, the initial state $\ket{+}$ is always applied
to all Pauli strings once the last gate in $U$ is applied.  Therefore, our
branching algorithm will output both $\ket{\psi_0}$ and $\ket{\psi_1}$ as a
superposition of binary strings.  Because some of binary string composing
$\ket{\psi_0}$ and $\ket{\psi_1}$ may have a zero amplitude, (due to destructive
interference), the number of binary strings $n_p$ composing $\ket{\psi_0}$ and
$\ket{\psi_1}$ is typically smaller than the number of explored branches $n_b$,
that is $n_p \leq n_b$.

\begin{figure}[!t]
	\centering
	\includegraphics[width=0.85\columnwidth]{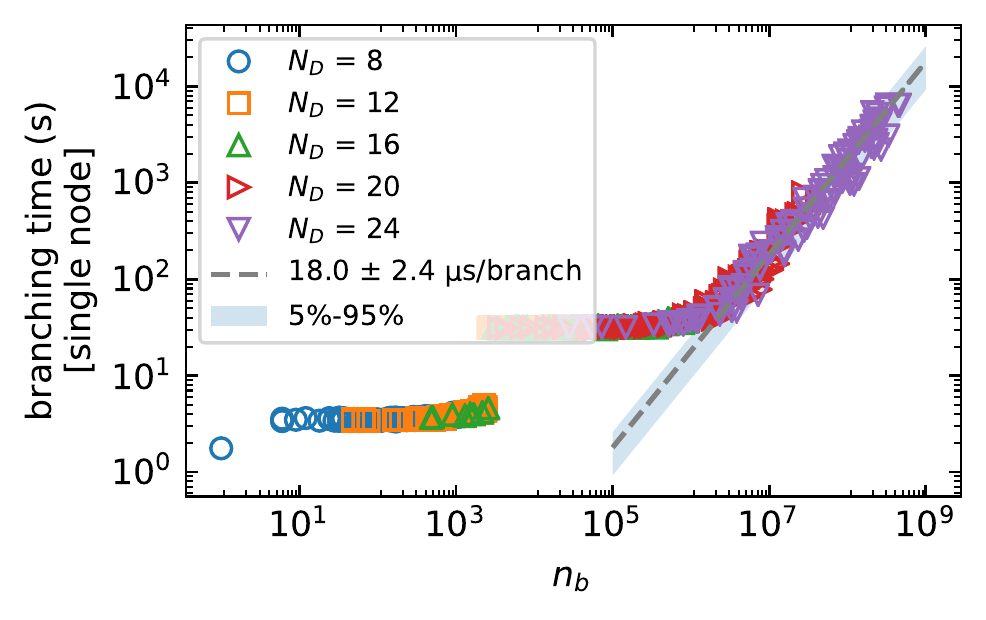}\\
	\includegraphics[width=0.85\columnwidth]{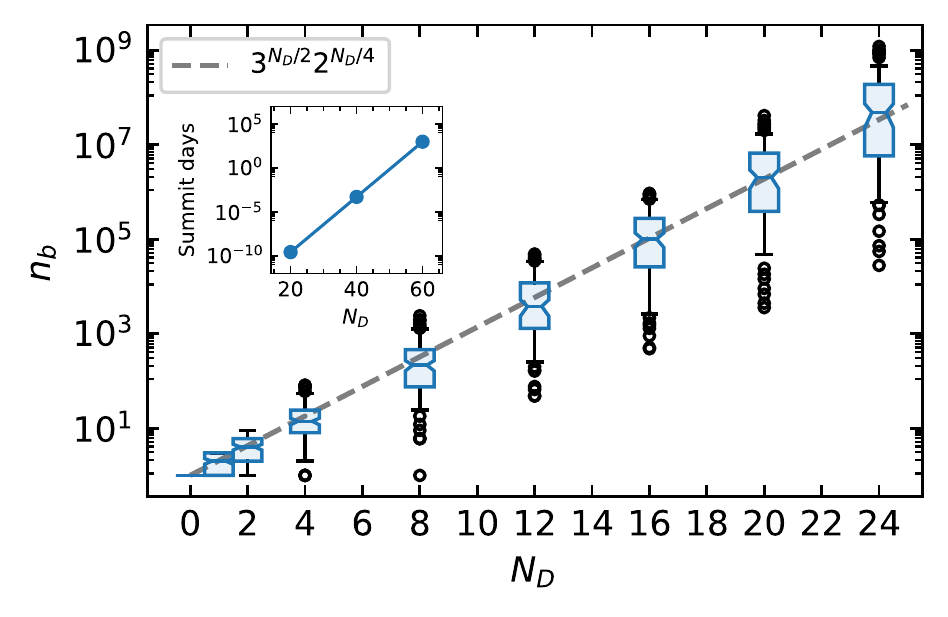}
	\caption{(Top) Runtime (in seconds) to explore a given number
		of branches (results are for nodes with \texttt{2 six-core Intel Xeon X5670@2.93GHz}).
		(Bottom) Number $n_b$ of explored branches by varying the number $N_D$ of non-Clifford
		rotations (boxes extend from the lower to upper quartile values of the data,
		with a line at the median, while whiskers correspond to the $5\%-95\%$
		confidence interval). The inset shows the projected runtime on Summit.}
	\label{fig:nrot_vs_nbranches}
\end{figure}

\begin{figure}[!t]
	\centering
	\includegraphics[width=0.85\columnwidth]{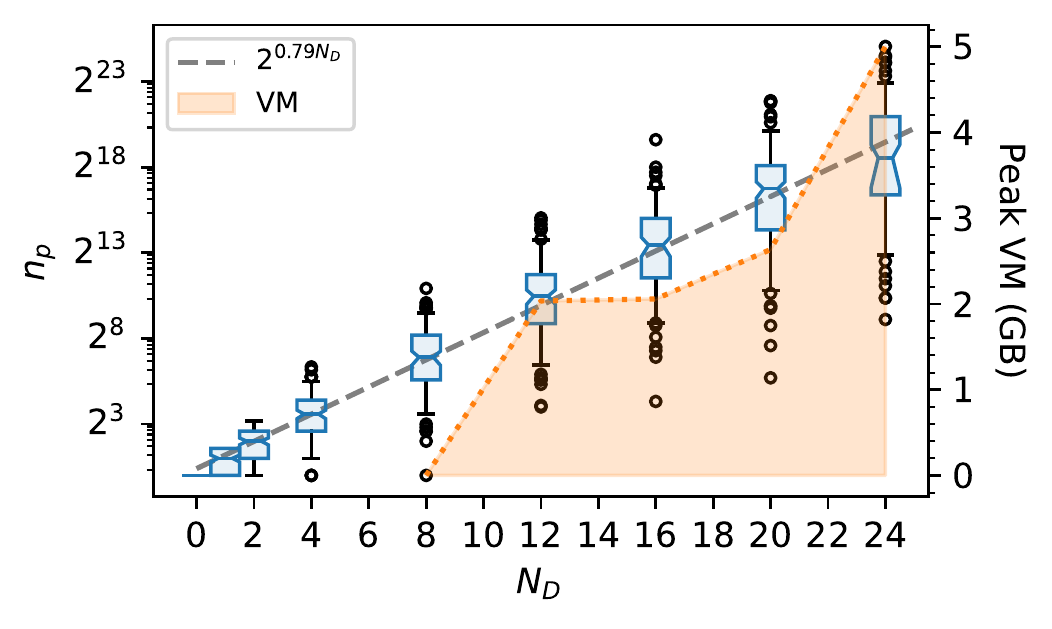}
	\caption{Number $n_p$ of bitstrings in $\ket{\psi_0}$ and $\ket{\psi_1}$ which
		are different from zero after applying the butterfly operator $\hat C = \hat
		U^\dagger \hat\sigma^{(Q_b)} \hat U$ to the initial state $\ket{+}$, by
		varying the number $N_D$ of non-Clifford rotations. The shaded area correspond
		to the peak of the amount of virtual memory (RAM) used to simulate the OTOC
		circuits ($95\%$ among different simulations).}
	\label{fig:nrots_vs_nbitstrings}
\end{figure}

Fig~\ref{fig:nrot_vs_nbranches} (top) shows the number of explored branches
$n_b$ by varying the number $N_D$ of non-Clifford rotations in $\hat U$.
Because the only non-Clifford gates used in the OTOC experimental circuits are
$\sqrt{W^{\pm}}$ and $\sqrt{V^\pm}$, with $W = X + Y$ and $V = X - Y$
respectively, one can compute the expected scaling by assuming that, at each
branching point, there is an homogeneous probability to find any of the four
Pauli operators $\left\{I,\,X,\,Y,\,Z\right\}$. Because non-Clifford rotations
branch only twice on $Z$, thrice on $\left\{X,\,Y\right\}$ and never on $I$, the
expected scaling is $n_b \propto 3^{\frac{n}{2}} 2^{\frac{n}{4}}$, which has
been confirmed numerically in Fig.~\ref{fig:nrot_vs_nbranches} (top).
Fig~\ref{fig:nrot_vs_nbranches} (bottom) shows the runtime (in seconds) to
explore a given number of branches on single nodes of the NASA cluster Merope
\cite{nasa_merope}. While the interface of our branch simulator is completely
written in Python, the core part is just-in-time (JIT) compiled using
\texttt{numba} to achieve $C-$like performance. Our branch simulator also uses
multiple threads (24 threads on the two \texttt{2 six-core Intel Xeon
	X5670@2.93GHz} nodes) to explore multiple branches at the same time and it can
explore a single branch in $\sim\!\! 18\, \mu s$. Because multithreading starts
for $n_b > 10^3$ only, it is possible to see the small jump caused by the
multithreading overhead. The inset of Fig.~\ref{fig:nrot_vs_nbranches} shows the
projected runtime on Summit by rescaling to Summit's
$\text{\texttt{R}}_\text{max}$ \cite{top500}
($\text{\texttt{R}}_\text{max}^\text{Summit} = 200,\!794.9\ \text{TFlops}$,
$\text{\texttt{R}}_\text{max}^\text{Merope (single node)} = 140.64\
\text{GFlops}$), assuming that $\ket{\psi_0}$ and $\ket{\psi_1}$ can be fully
stored in Summit (seed Fig.~\ref{fig:nrots_vs_nbitstrings}). In
Fig.~\ref{fig:nrots_vs_nbitstrings}, we report the number of elements different
from zero in $\ket{\psi_0}$ and $\ket{\psi_1}$. As one can see, the number of
elements different from zeros scales as $2^{0.79\,N_D}$ and can be accommodated
on single nodes (shaded area corresponds to the amount of virtual memory [RAM]
used by the simulator). 
Because branches are explored using a depth-first search strategy, most of the
virtual memory used by our branching algorithm is reserved to store
$\ket{\psi_0}$ and $\ket{\psi_1}$, it would be in principle possible to simulate
between $N_D \approx 50$ and $N_D \approx 60$ non-Clifford rotations on Summit
before running out of virtual memory (Summit has $10\ \text{PB}$ of available
DDR4 RAM among its $4,\!608$ nodes \cite{summit_memory}).

\subsection{Tensor Contraction}

\begin{figure}[!t]
	\centering
	\includegraphics[width=0.85\columnwidth]{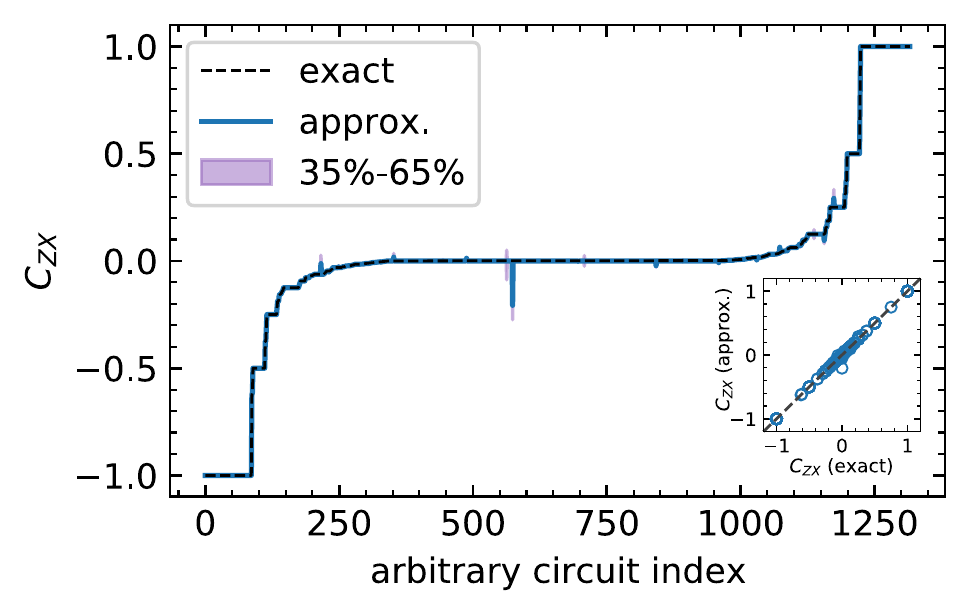}\\
	\includegraphics[width=0.85\columnwidth]{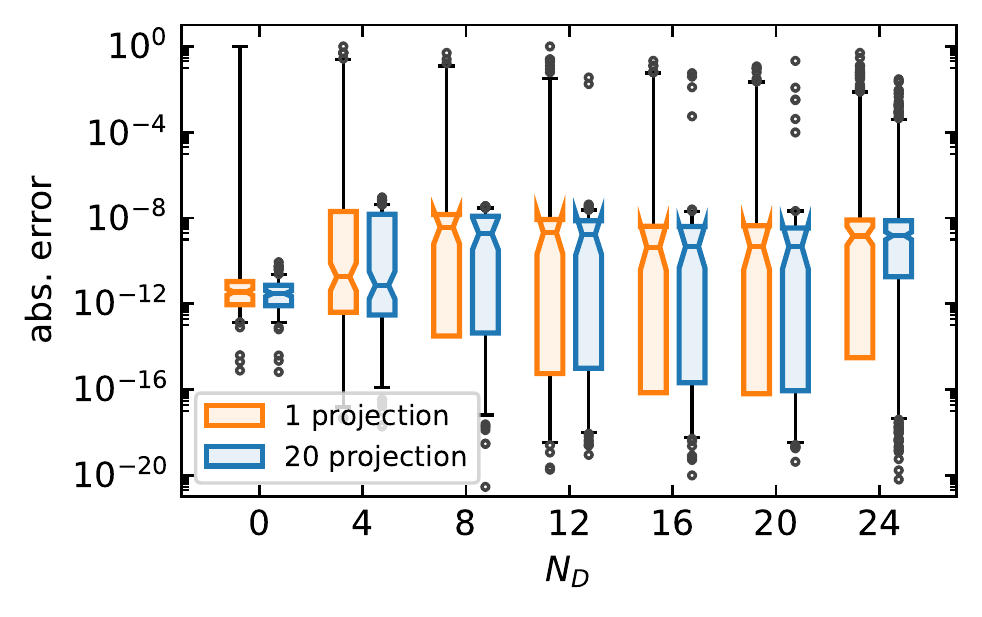}
	\caption{(Top) Comparison between exact and approximate OTOC values (circuits
		are ordered accordingly to the exact OTOC value over different circuits with
		different depths, layouts and numbers of non-Clifford rotations). The Pearson
		coefficient between exact and approximate OTOC values is $R = 0.99987$.
		(Bottom) Absolute error by varying the number $N_D$ of non-Clifford rotations (boxes
		extend from the lower to upper quartile values of the data, with a line at the
		median, while whiskers correspond to the $5\%-95\%$ confidence interval).}
	\label{fig:approx_vs_exact}
\end{figure}

Tensor contraction is a powerful tool to simulate large quantum circuits
\cite{markov2008simulating, chen2018classical, villalonga2020establishing,
	villalonga2019flexible, huang2020classical, gray2020hyper}. In our numerical
simulations, we use tensor contraction to compute approximate OTOC values. It is
well known that approximate amplitudes can be sampled by properly slicing the
tensor network and only contracting a fraction of the sliced tensor networks
\cite{markov2018quantum, markov2019faster, villalonga2020establishing,
	villalonga2019flexible}. However, in our numerical simulations, we used a
different approach to compute approximate OTOC values. More precisely, rather
than computing one (approximate) amplitude at a time using tensor contraction,
we compute (exact) ``batches'' of amplitudes by leaving some of the terminal
qubits in the tensor network ``open''. Let us call $\kappa$ a given projection
of the non-open qubits and us define $\ket{\psi_0^{(\kappa)}}$ and
$\ket{\psi_1^{(\kappa)}}$ the projection of $\ket{\psi_0} = \sum_\kappa
\ket{\psi_0^{(\kappa)}}$ and $\ket{\psi_1} = \sum_\kappa \ket{\psi_1^{(\kappa)}}$
respectively.  Therefore, we can re-define the OTOC value as a ``weighted''
average of partial OTOC values, that is:
\begin{equation}\label{eq:num_otoc_kappa}
\langle\hat\sigma_y\rangle := 
\frac{\sum_\kappa
	\omega_\kappa\langle\hat\sigma_y\rangle_\kappa}{\sum_\kappa
	\omega_\kappa},
\end{equation}
with 
\begin{equation}\label{eq:num_otoc_single}
\langle\hat\sigma_y\rangle_\kappa =
\frac{\mathbb{R}\big[\langle\psi_1^{(\kappa)}|\psi_0^{(\kappa)}\rangle\big]}{\big(\lVert
	\psi_0^{(\kappa)} \rVert^2 + \lVert \psi_1^{(\kappa)} \rVert^2\big)/2}
\end{equation}
and
$\omega_\kappa = \big(\lVert \psi_0^{(\kappa)} \rVert^2 + \lVert
\psi_1^{(\kappa)} \rVert^2\big)/2$. It is interesting to observe that
Eq.~(\ref{eq:num_otoc_single}) corresponds to the correlation coefficient
between two non-normalized states and that the exact OTOC value is equivalent
to the weighted average of single projection OTOC values.  Indeed, when all $\kappa$
are included, Eq.~(\ref{eq:num_otoc_kappa}) reduces to Eq.~(\ref{eq:num_otoc}).
However, because $\langle\sigma_y\rangle$ needs to be $\mathcal{O}(1)$ to be
experimentally measurable, few projection $\kappa$ may actually be sufficient
to get a good estimate of Eq.~(\ref{eq:num_otoc_kappa}).

In our numerical simulations, we left $24$ qubits open and we used $20$ random
$\kappa$ projections to compute an approximate OTOC value using
Eq.~(\ref{eq:num_otoc_kappa}). If not indicated otherwise, medians and
confidence intervals are computed by bootstrapping $1,\!000$ times
Eq.~(\ref{eq:num_otoc_kappa}) using $10$ randomly chosen projections among the
$20$ available.
Fig.~\ref{fig:approx_vs_exact} shows the comparison of approximate and exact
OTOC values, by varying the circuit index (top) and by varying the number of
non-Clifford rotations (bottom). As one can see, there is a great agreement
between approximate and exact results across different circuits with different
depth, layout and number of non-Clifford rotations (top). The bottom part of
Fig.~\ref{fig:approx_vs_exact} shows the absolute error by varying the number of
non-Clifford rotations. In particular, we are comparing results by averaging
over single projections $\kappa$ using Eq.~(\ref{eq:num_otoc_single})
(orange/light gray boxes) or by bootstrapping $1,\!000$ times
Eq.~(\ref{eq:num_otoc_kappa}) using $10$ randomly chosen projections among the
$20$ available (blue/dark gray boxes). As expected, results become less noisy
by increasing the number of used projections.

\begin{figure}[!t]
	\centering
	\includegraphics[width=0.87\columnwidth]{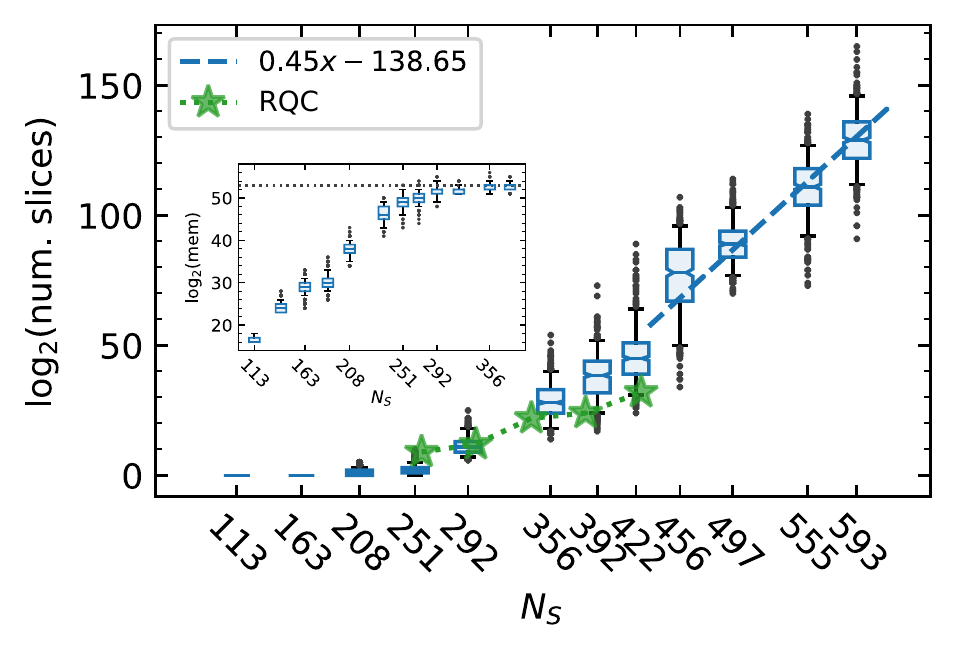}
	\caption{Number of slices required to fit the largest tensor in memory during
		tensor contraction with a threshold of $2^{28}$ elements by varying the number
		$N_S$ of iSWAPs (boxes extend from the lower to upper quartile values of the
		data, with a line at the median, while whiskers correspond to the $5\%-95\%$
		confidence interval). Green stars correspond to expected number of slices to compute single amplitudes of
		random quantum circuits (RQC) with a similar depth \cite{Arute2019, gray2020hyper}.
		(Inset) Number of elements to store in memory for the largest tensor during
		tensor contraction. For sufficiently deep circuits, the number of elements for
		the largest tensor saturates to the size of the Hilbert space $2^n$, with $n =
		53$ the number of qubits in the Sycamore chip.}
	\label{fig:iswaps_vs_n_slices}
\end{figure}

Due to the limited amount of memory in HPC nodes, slicing techniques are
required to fit the tensor contraction on a single node and avoid node-to-node
memory communication overhead \cite{villalonga2019flexible, gray2020hyper}. In
our numerical simulations, we used \texttt{cotengra}~\cite{gray2020hyper} and
\texttt{quimb}~\cite{gray2018quimb} to identify the best contraction (including
slicing) and perform the actual tensor contraction respectively. Because each
different projection $\kappa$ may lead to a slightly different simplification of
the tensor network (in our numerical simulations, we used the rank and column
reduction included in the \texttt{quimb} library), we recompute the best
contraction for each single projection. For each projection (regardless of the
depth/number of iSWAPs) we fixed \texttt{max\_repeats~=~128} in
\texttt{cotengra} and restarted the heuristics $10$ times to identify the
optimal contraction (with a hard limit of $2$ minutes for each run), using $24$
threads on a \texttt{2 six-core Intel Xeon X5670@2.93GHz} node. We found that
the runtime to the best contraction scales as $0.03 N_S - 3.68$
minutes, with $N_S$ the number of iSWAPs in the circuit,
that is less than $20$ minutes for $\sim\!600$ iSWAPs.
Fig.~\ref{fig:iswaps_vs_n_slices} shows the number of slices required to have
the largest tensor in the tensor contraction no larger than $2^{28}$ elements.
Green stars in the figure correspond to the number of slices to compute single amplitudes of random quantum
circuits with a similar number of iSWAPs \cite{Arute2019,
	gray2020hyper}. Because OTOC circuits and the random quantum circuits presented in
\cite{Arute2019} share a similar
randomized structure, the computational complexity mainly depends on the number
of iSWAPs (OTOC slicing is slightly worse because of the open qubits).
We may expect an improvement in the slicing by using novel techniques as the
``subtree reconfiguration'' proposed by Huang \emph{et al.}
\cite{huang2020classical}. It is interesting to observe that, for deep enough
circuits, the number of elements of the largest tensor in the tensor contraction
saturates to the number of qubits (inset).

\begin{figure}[!t]
	\centering
	\hspace*{-35pt}\includegraphics[width=0.9\columnwidth]{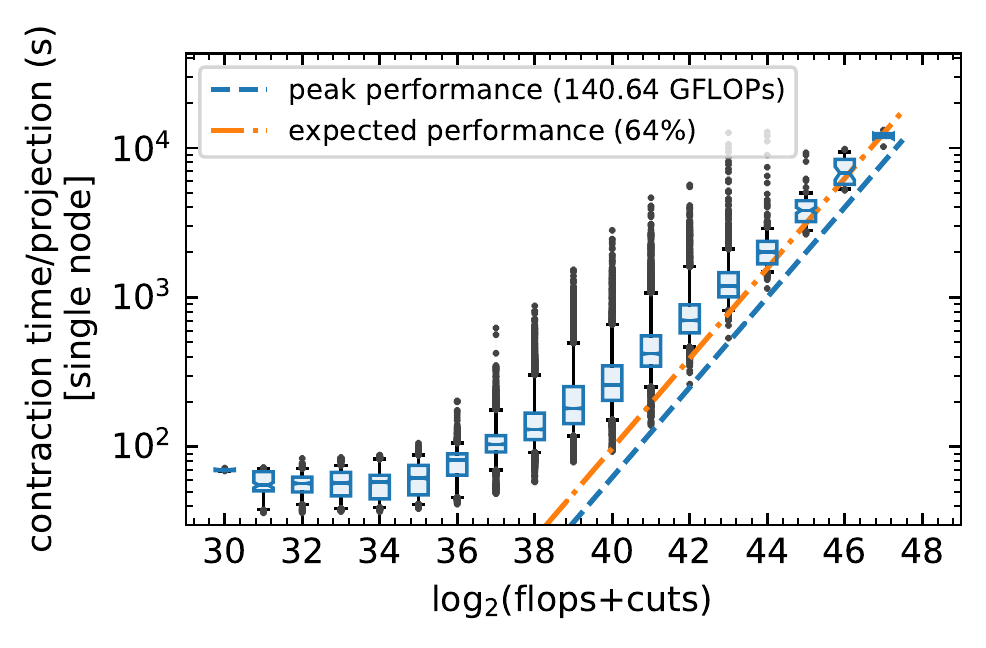}\\
	\hspace*{5pt}\includegraphics[width=\columnwidth]{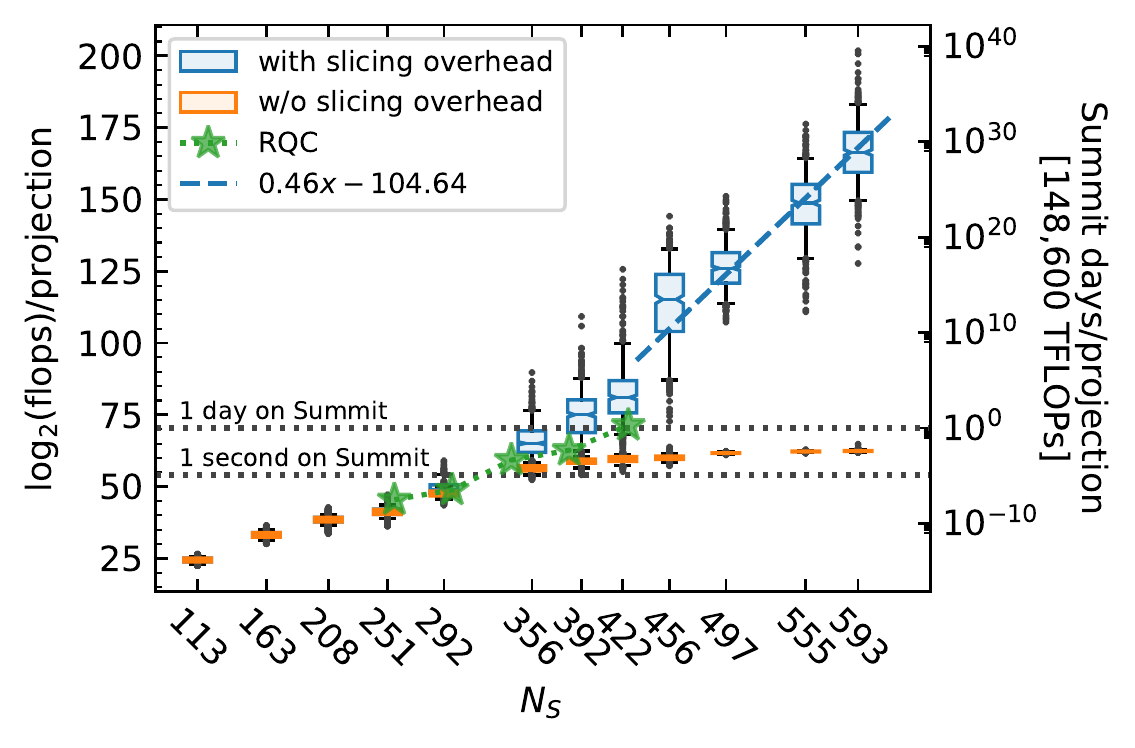}
	\caption{(Top) Runtime (in seconds) to fully contract a single projection by
		varying the required number of FLOPs (results are for single nodes with
		\texttt{2 six-core Intel Xeon X5670@2.93GHz}). Because \texttt{long double}
		are used throughout the simulation, we used the conversion factor $8$ to
		convert FLOPs to actual time. (Bottom) Flops to contract a single projection
		by varying the number $N_S$ of iSWAPs, with and without the slicing overhead.
		Summit's days are obtained by using Summit's $\text{\texttt{R}}_\text{peak}$
		\cite{top500} and the conversion factor $8$ from FLOPs to actual time (because
		\texttt{long double}'s have been used throughout the simulation). Green stars
		correspond to expected number of FLOPs to compute single amplitudes of random quantum circuits (RQC) with
		a similar number of iSWAPs \cite{Arute2019, gray2020hyper}.}
	\label{fig:flops_vs_time}
\end{figure}

Fig.~\ref{fig:flops_vs_time} summarizes the computational cost to compute an
exact single projection $\kappa$. Top panel of Fig.~\ref{fig:flops_vs_time}
reports the actual runtime to contract a tensor network, by varying its expected
total cost (including the slicing overhead) in FLOPs (expected FLOPs are using
by \texttt{cotengra} to identify an optimal contraction). Dashed and dot-dashed
lines correspond to the peak performance and expected performance of a 
\texttt{2 six-core Intel Xeon X5670@2.93GHz} node. The sustained performance of
$64\%$ is consistent with similar analysis on the NASA cluster
\cite{villalonga2019flexible}. Bottom panel reports the number of FLOPs
(with and without the slicing overhead) by varying the number of
iSWAPs. On the right y-axis, it is reported the extrapolated number of days
to simulate OTOC circuits of a given number of iSWAPs, by assuming a
sustained performance of $148,\!600$ TFLOPs. Green stars corresponds to the total
FLOPs (including the slicing overhead) to compute single amplitudes of random quantum circuits
with a similar number of iSWAPs \cite{gray2020hyper}.

\section{Markov population dynamics}

For a broad class of circuit ensembles the average OTOC can be computed efficiently (in polynomial time) on a classical computer. This appears surprising as computing the output of the random circuit is expected to require exponential resource. This contradiction is resolved by demonstrating an exact mapping of the average evolution of OTOC onto a Markov population dynamics process. Such connection was identified for Hamiltonian dynamics~\cite{AleinerIoffe_AP_2016} and subsequently for simplified models of random circuits~\cite{Nahum_PRX_2018, Keyserlingk_PRX_2018} where uniformly random two-qubit gate was assumed.  In practice, we implement a specific gate set consisting of {``}cycles{''} of the form \(\prod_{\langle i j \rangle} \hat{G}_i\hat{G}_j \hat{U}_{ij}(\theta ,\phi )\) applied to non-overlapping pairs of nearest neighbour qubits \(\langle i, j \rangle\). For each pair a cycle consists of two single qubit gates \(\hat{G}_i, \hat{G}_j\) and an entangling two-qubit gate,
\begin{gather}
\hat{U}_{i,j}(\theta ,\phi )= e^{-\frac{i}{2}  \theta  \left(X_i X_j + Y_i Y_j \right)-\frac{i \phi }{2} Z_i Z_j},
\label{eq:2qbit}
\end{gather}
parameterized by fixed angles $\theta, \phi$. The random instances are generated by
drawing single qubit gates from a specific finite set \(\left\{\hat{G}_i\right\}\). We consider two sets corresponding to generators of single qubit Clifford group $\{\sqrt{X^{\pm1}}, \sqrt{Y^{\pm1}}\}$ and the set which in conjunction with any entangling two-qubit gate generates a universal set $\{\sqrt{X^{\pm1}}, \sqrt{Y^{\pm1}}, \sqrt{W^{\pm1}}, \sqrt{V^{\pm1}}\}$ introduced in Sec.~\ref{ext_data}.

Average OTOC can be calculated by first considering the dynamics of the pair of butterfly operators
$\hat{O} (t) = U^\dag \hat{O} U$ evolved under circuit $U$. We introduce the average of the pair of the operators acting
in the direct product of two Hilbert spaces of the two replicas of the same circuit:
\begin{gather}
\hat{\cal O}^{(2)}(t)=\overline{ \hat{O}(t) \otimes \hat{O}(t) } \equiv
\overline{ \hat{O}(t)^{\otimes 2}}
\label{eq:AveragePair}
\end{gather}
where averaging over ensemble of the circuits is denoted by $\overline{(...)}$, and the rightmost equation defines the short-hand notation.

Analogously, one can introduce the higher order averages
\begin{gather}
\hat{\cal O}^{(4)}(t)=\overline{ \hat{O}(t) \otimes \hat{O}(t)\otimes \hat{O}(t)\otimes \hat{O}(t) } \equiv
\overline{ \hat{O}(t)^{\otimes 4}},
\label{eq:AverageFour}
\end{gather}
as a starting point to analyze the circuit to circuit fluctuations of $C(t)$.


Then the value of OTOC is obtained by taking the matrix element
of $C (t) = \bra{\psi}\overline{\hat{M} \hat{O}(t) \hat{M} \hat{O}(t) }\ket{\psi}$
with the
initial state that in our experimental setup is chosen to be $ \ket{\psi} = \bigotimes_{i=1}^n \ket{+}_i$
where $\ket{+}_i$ is the symmetric superposition of the computational basis states. The average and the second moment of $C(t)$ are obtained
as a straightforward convolution of the indices in Eqs.~(\ref{eq:AveragePair}) and (\ref{eq:AverageFour}) respectively. Operator
$\hat{\cal O}^{(4)}$ can be also used to study the effect of the initial state to the OTOC (clearly its average is not sensitive to the initial conditions).

\subsection{Symmetric single qubit gate set}
\label{sec:symmetric}

We first consider average over uniformly random single qubits gate such that for any Pauli matrix,
\begin{gather}
\overline{ \hat{\alpha}_i}\equiv   \overline{\hat{G}^{\dagger } \hat{\alpha}_i \hat{G}} = 0,\label{eq:SymGateSet}
\end{gather}
We will analyze the specific discrete gate sets used in the experiment in Section~\ref{sec:expGateSet}. In this section, we use
the Latin indices to label a qubit, and
the Greek ones to denote the corresponding Pauli matrices $\hat{\alpha}_i = \{ \hat{X}_i, \hat{Y}_i, \hat{Z}_i\}$

For a pair of operators as in Eq.~(\ref{eq:AveragePair}) there exist an analog of scalar product -- a
spherically symmetric combination that does not vanish after averaging,
\begin{gather}
\overline{ \hat{\alpha}_i\otimes\hat{\beta}_j}=
\hat{G}^{\dagger } \hat{\alpha}_i \hat{G} \otimes \hat{G}^{\dagger } \hat{\beta}_j \hat{G} = \delta_{\alpha\beta}\delta_{ij}\mathcal{B}_i, \label{1Qaverage}
\end{gather}
where we introduced {``}bond{''} notation,
\begin{gather}
\mathcal{B}_i \equiv \frac{1}{3}  \hat{\alpha}_i\otimes\hat{\alpha}_i \equiv \frac{1}{3} \left(\hat{X}_i^{\otimes2} + Y_i^{\otimes2} 
+ Z_i^{\otimes2}\right), \label{eq:bond}
\end{gather}
and the summation over the repeated Greek indices is always implied.

\begin{subequations}
	\label{eq:bond4}
	Analogously one can find non-vanishing averages of the Pauli matrices acting in four replicas space of the same
	qubit (we will omit the qubit label). There are in total six bilinear combinations
	\begin{equation}
	\begin{aligned}
	\overline{
		\hat{\alpha}\otimes \hat{\beta}
		\otimes \hat{\openone}\otimes \hat{\openone}
	}
	= \delta_{\alpha\beta}\mathcal{B}^{(12)},\quad & \mathcal{B}^{(12)}= \frac{1}{3}
	\hat{\alpha} \otimes \hat{\alpha}\otimes \hat{ \openone}\otimes \hat{\openone} ,
	\\
	\overline{
		\hat{\alpha}\otimes \hat{\openone}\otimes \hat{\beta}
		\otimes \hat{\openone}
	}
	= \delta_{\alpha\beta}\mathcal{B}^{(13)},\quad & \mathcal{B}^{(12)}= \frac{1}{3}
	\hat{\alpha} \otimes \hat{ \openone}\otimes \hat{\alpha}\otimes \hat{\openone} ,
	\\
	&\vdots
	\\ \overline{
		\hat{\openone}\otimes \hat{\openone}\otimes \hat{\alpha} \otimes \hat{\beta}
	}
	=\delta_{\alpha\beta}\mathcal{B}^{(34)},\quad &
	\mathcal{B}^{(34)} = \frac{1}{3} \hat{\openone}\otimes \hat{\openone}\otimes
	\hat{\alpha} \otimes \hat{\alpha}.
	\end{aligned}
	\label{eq:bond4a}    
	\end{equation}
	
	Four cubic invariants are possible because there is no inversion operation for the spin:
	\begin{equation}
	\begin{aligned}
	\overline{     \hat{ \openone}\otimes
		\hat{\alpha} \otimes \hat{\beta}\otimes \hat{\gamma}
	}
	= \epsilon^{\alpha\beta\gamma}\mathcal{C}^{(1)},
	& \quad\mathcal{C}^{(1)}\equiv \frac{\epsilon^{\alpha\beta\gamma}}{6}
	\hat{ \openone}\otimes \hat{\alpha}\otimes \hat{\beta}\otimes \hat{\gamma},
	\\
	\overline{    
		\hat{\alpha} \otimes  \hat{ \openone}\otimes \hat{\beta}\otimes \hat{\gamma}
	}
	= \epsilon^{\alpha\beta\gamma}\mathcal{C}^{(2)},
	& \quad\mathcal{C}^{(2)}\equiv \frac{\epsilon^{\alpha\beta\gamma}}{6}
	\hat{\alpha}\otimes \hat{ \openone}\otimes  \hat{\beta}\otimes \hat{\gamma},\\       
	&\vdots\\
	\overline{    
		\hat{\alpha} \otimes \hat{\beta}\otimes \hat{\gamma} \otimes\hat{ \openone}
	}
	= \epsilon^{\alpha\beta\gamma}\mathcal{C}^{(4)},        
	& \quad\mathcal{C}^{(4)}\equiv \frac{\epsilon^{\alpha\beta\gamma}}{6}
	\hat{\alpha}\otimes  \hat{\beta}\otimes \hat{\gamma}\otimes \hat{ \openone}.
	\end{aligned}
	\label{eq:bond4b}    
	\end{equation}
	where $ \epsilon_{\alpha\beta\gamma}$ is the
	three-dimensional Levi-Civita symbol. Because they would change sign under the inversion operation,
	the cubic invariants can appear only in pairs on neighboring qubits.
	
	Finally, the three quartic invariants are
	\begin{equation}
	\begin{aligned}
	&\overline{   
		\hat{\alpha} \otimes \hat{\beta}\otimes \hat{\gamma} \otimes\hat{\delta}
	}= \mathcal{D}^{(2)}\Upsilon_{\alpha\beta\gamma\delta}
	+\mathcal{D}^{(3)}\Upsilon_{\alpha\gamma\beta\delta}
	+\mathcal{D}^{(4)}\Upsilon_{\alpha\gamma\delta\beta},
	\\
	&\Upsilon_{\alpha\beta\gamma\delta} = \delta_{\alpha\beta}\delta_{\gamma\delta}
	-\frac{1}{4}\delta_{\alpha\gamma}\delta_{\beta\delta}
	-\frac{1}{4}\delta_{\alpha\delta}\delta_{\beta\gamma},
	\\
	& \mathcal{D}^{(2)}=\frac{2}{15}\hat{\alpha} \otimes \hat{\alpha}\otimes \hat{\beta} \otimes\hat{\beta},
	\\
	& \mathcal{D}^{(3)}=\frac{2}{15}\hat{\alpha}\otimes \hat{\beta}   \otimes\hat{\alpha}\otimes\hat{\beta},
	\\
	& \mathcal{D}^{(4)}=\frac{2}{15}\hat{\alpha} \otimes \hat{\beta} \otimes\hat{\beta}\otimes \hat{\alpha}.
	\end{aligned}
	\label{eq:bond4c}    
	\end{equation}

	This simple form implies the spherical symmetry of the single qubit averaging. For a lower symmetry (e.g.  all the rotations
	of a cube)
	other quartic and cubic invariants
	are possible (like $X^{\otimes 4}+Y^{\otimes 4}+Z^{\otimes 4}$ or $|\epsilon_{\alpha\beta\gamma}|
	\hat{ \openone}\otimes \alpha \otimes \beta\otimes \gamma$) but we will ignore
	them for the sake of simplicity.
\end{subequations}

\subsection{Efficient Population Dynamics for the Averaged OTOC}

We expand average evolution of the pair of butterfly operators (\ref{eq:AveragePair}) as,
\begin{gather}
\hat{\cal O}^{(2)}(t)
= \sum_{\{v_i\}} P_{\{v_i\}} \bigotimes_{i=1}^n  \left( v_i  \hat{\mathcal{B}}_i 
+ u_i \hat{\openone}_i^{\otimes 2}\right). \label{eq:OvsB}
\end{gather}
where normalization condition reads,
\begin{equation}
u_i+v_i=1,
\label{normalization1}
\end{equation}
the variable $v_i = \{0, 1\}, i =1, 2, ..., n$ indicates whether or
not Pauli matrices occupy qubit $i$, and $P_{\{v_i\}}$ are formfactors.
In other words, each $i$th is characterized either by $\rho^{(0)}_i \equiv \openone_i^{\otimes2}$, "vacuum",
or the {``}bond{''} $\mathcal{B}_i$. All other terms in the right hand side of Eq.~(\ref{eq:OvsB}) vanish upon averaging over single qubit gates
see Sec.~\ref{sec:symmetric}.

Application of a two-qubit Sycamore gate to a pair of qubits $\{i, j\}$ is then described by $2^2\times 2^2$ matrix in the space  $\{v_i, v_{j}\}$,
\begin{gather}
\Omega  =\left(
\begin{array}{cccc}
1 & 0 & 0 & 0 \\
0 & 1-a-b & a & b \\
0 & a & 1-a-b & b \\
0 & \frac{b}{3} & \frac{b}{3} & \left(1-\frac{2}{3}b\right) \\
\end{array}
\right),
\label{eq:Omega} \\
a=\frac{1}{3}\left(2\sin ^2\theta +\sin ^4\theta \right), \nonumber \\
b=\frac{1}{3}\left(\frac{1}{2}\sin^22\theta +2\left(\sin ^2\theta  +\cos ^2\theta \right)\right). \nonumber
\end{gather}
where the matrix $\Omega$ acts from the right on four dimensional row vector with the basis $(00),(01),(10),(11)$
The standard \(\sqrt{\text{iSWAP}}\) gate corresponds
to \(\theta =\frac{\pi }{4}\), { }\(a=\frac{1}{12}, b=\frac{1}{2}\), and iSWAP \(\theta=\frac{\pi }{2}\) with \(a=\frac{1}{3}, b=\frac{2}{3}\).

Each time when the two-qubit gate is applied the formfactor $P$ is updated according
to the rules
\begin{gather}
P_{v_1\text{...}v_iv_j\text{...}v_n}(t+1) 
=\sum _{v_i^{\prime }v_j^{\prime }} P_{v_1\text{...}v_i^{\prime }v_j^{\prime }\text{...}v_n} (t)
\Omega_{v_i^{\prime }v_j^{\prime },v_iv_j}. \label{eq:MarkovChain}
\end{gather}
Equations (\ref{eq:Omega})--(\ref{eq:MarkovChain}) are obtained by an application of a two-qubit gate (\ref{eq:2qbit}) with $\phi=0$ to a pair $(i,j)$ of factors in Eq.~(\ref{eq:OvsB})
and averaging the  result using Eqs.~(\ref{eq:SymGateSet}) --  (\ref{1Qaverage}).

Some additional constraints can be extracted from the exact condition
\begin{equation}
\left[\hat{O}(t)\right]^2= \bigotimes_{i=1}^n\hat{\openone}_n.
\label{eq:squared}
\end{equation}
Convoluting operators in Eq.~(\ref{eq:OvsB}), using $\hat{\openone}_i^{\otimes 2}\to \hat{\openone}_i$,
$\hat{\mathcal{B}}_i\to \hat{\openone}_i$ and the normalization condition (\ref{normalization1}), we obtain
the requirement
\begin{equation}
\sum_{\{v_i\}} P_{\{v_i\}}(t)=1.
\label{normalization2}
\end{equation}
Preserving this condition in the update rule (\ref{eq:MarkovChain}) requires
the elements in each row of the matrix $\Omega$ from Eq.~(\ref{eq:Omega}) to add up to one.
Moreover, all the elements of $\Omega$ are non-negative  and therefore not only is $P_{\{v_i\}}(t)$ normalized but it is also non-negative.

Therefore, rules (\ref{eq:MarkovChain}) and (\ref{normalization2}) defines the
Markov process and variables $v_i=0,1$ correspond to the classical population of each qubit.
As the non-populated states $v_i=v_j=0$ do not evolve and the  reproducing  $(01)\to (11)$ and destruction $(11)\to (01)$ processes
are allowed the problem (\ref{eq:MarkovChain}) is nothing but the classical population dynamics.
The formfactors $P\left(v_1\text{...}v_i^{\prime }v_j^{\prime }\text{...}v_n,t\right)$ are interpreted as a distribution
function over an \(n\) bit register $\left\{v_1\text{...}v_n\right\}$, subject to the Markov process defined by the update, Eq.~(\ref{eq:Omega}).

Direct solution of Eq.~(\ref{eq:MarkovChain}) would require $2^n$ real numbers. However, unlike the original
unitary evolution, the classical population dynamics involves only {\em positive} numbers constrained by normalization (\ref{normalization2}).
Such dynamics is very efficiently stimulated using a {\em classical} Monte Carlo type algorithm. 

The butterfly operator is the starting point of the Markov process  Eq.~(\ref{eq:MarkovChain}). At long times the distribution $P_{\{v_i\}}$ converges to the stationary state $\bigotimes_{i=1}^n \rho_i^{(erg)}$, 
where $
\rho_i^{(erg)}
=\frac{1}{4}\left(\openone_i^{\otimes2} + 3\mathcal{B}_i\right)$,
which corresponds to "vacuum" occurring on each site $i$ with probability $p\left(\openone_i^{\otimes 2}\right)=\frac{1}{4}$ and "bond" occurring with probability $p\left(\mathcal{B}_i\right) = 3/4$. OTOC in this limit takes the value corresponding to the random matrix statistics~\cite{KitaevYoshida}. Intermediate dynamics of OTOC between these two limits is fully described by the Markov process Eq.~(\ref{eq:MarkovChain}). It has a form of shock wave spreading from the initial butterfly operator to cover the whole system. Note that the choice of the two qubit gate parameter $\theta$ has a dramatic effect on the intermediate OTOC dynamics. As discussed in the main text $\theta = \pi/2$ corresponds to the probability of butterfly operator to spread equal one, and therefore spreading with maximum velocity equal to the light cone velocity and saturating the Lieb-Robinson bound. At any other  
value of $\theta$ probability to spread is less then one which results in diffusive broadening of the front, and the center of the front propagates with a butterfly velocity that is smaller than the light cone velocity.

\subsection{Sign Problem in the Population Dynamics for OTOC Fluctuations}

As we already mentioned, the analytic calculation of the OTOC for an individual circuit is impossible. One can expect,
however, that it is possible to express the variance of the OTOC in terms of some products of classical propagators similarly to
the analysis of the mesoscopic fluctuations in the disordered metals.
The purpose of this subsection is to show that even such a modest task can not be efficiently undertaken.

The starting point is the expansion in terms of the single qubit rotations invariants (\ref{eq:bond4}). Similarly to 
Eq.~(\ref{eq:OvsB}), we write for $\hat{\cal O}^{(4)}$ of Eq.~(\ref{eq:AverageFour})
\begin{equation}
\begin{split}
\hat{\cal O}^{(4)}(t)
= \sum_{\{{\mathbf V}_i\}} P_{\{{\mathbf V}_i\}} \bigotimes_{i=1}^n   {\mathbf V}_i \cdot \hat{\bf\cal{Q}}_i,
\end{split}
\label{eq:O4}
\end{equation}
where $\hat{\bf\cal{Q}}_i$ is the vector with $14=1+6+4+3$ operator components given by invariants of
Eq.~(\ref{eq:AverageFour}):
\begin{equation}
\begin{split}
&\hat{\bf\cal{Q}}_i
=
\left[\hat{\openone}^{\otimes 4}_i, \mathbf{b},
\mathbf{c},
\mathbf{d}
\right]. \\
&\mathbf{b} = \left(\hat{ \mathcal{B}}^{(12)}_i, ..., \hat{\mathcal{B}}^{(34)}_i\right), \\
&\mathbf{c} = \left(\hat{ \mathcal{C}}^{(1)}_i, ..., \hat{\mathcal{C}}^{(4)}_i\right), \\
&\mathbf{d} = \left(\hat{ \mathcal{D}}^{(2)}_i, \hat{ \mathcal{D}}^{(3)}_i, \hat{\mathcal{D}}^{(4)}_i\right),
\end{split}
\label{eq:Q}
\end{equation}
and ${\mathbf V}_i$ are the $14$ basis unit vectors so that $13$ components equal to zero and the remaining component is  $1$.
In other words, each site can be in $14$ possible states.

A straightforward generalization of the evolution equation (\ref{eq:MarkovChain}) reads
\begin{gather}
P_{{\mathbf V}_1\cdot{\mathbf V}_i{\mathbf V}_j\cdot{\mathbf V}_n}(t+1) 
=\sum _{{\mathbf V}_i^{\prime }{\mathbf V}_j^{\prime }} P_{{\mathbf V}_1\cdot{\mathbf V}_i^{\prime }{\mathbf V}_j^{\prime }\cdot{\mathbf V}_n} (t)
{\mathbf \Omega}_{{\mathbf V}_i^{\prime }{\mathbf V}_j^{\prime },{\mathbf V}_i{\mathbf V}_j}, \label{eq:MarkovChainVariance}
\end{gather}
where ${\mathbf \Omega}$ is now $14^2\times 14^2$ whose explicit form is known but not quite important for the further consideration.

Equation  (\ref{eq:MarkovChainVariance}) involves $14^n$ real numbers. The only feasible path to the solution would
be an efficient Monte Carlo sampling. Naively, one can hope to map the problem to the multicolored population  dynamics.
However, it is not possible as we explain below.

Reliable Monte Carlo sampling requires (1) normalizable and non-negative weights, and (2) absence of the correlations in contribution
of different configurations. Let us demonstrate that both conditions do not hold for the evolution of formfactors in Eq.~(\ref{eq:O4}).
Once again, we use exact Eq.~(\ref{eq:squared}). Convoluting third and fourth replica in $\hat{\cal O}^{(4)}(t)$
we obtain
\begin{equation}
\hat{\cal O}^{(4)}(t)\to  \hat{\cal O}^{(2)}(t) \otimes \hat{\mathbf \openone},\quad
\hat{\mathbf \openone} \equiv \left[ \bigotimes_{i=1}^n \hat{\openone}_i\right].
\label{eq:conv1}
\end{equation}

Convolutions of the four operators (\ref{eq:bond4}) yield
\begin{equation}
\begin{aligned}
\mathcal{B}^{(12)}& \to \mathcal{B}^{(3)}\equiv   \frac{1}{3}
\hat{\alpha} \otimes \hat{\alpha}\otimes \hat{ \openone},\\
\hat{ \openone}^{\otimes 4}, \mathcal{B}^{(34)} & \to  \hat{ \openone}^{\otimes 3},\\
\mathcal{B}^{(13)},\ \mathcal{B}^{(14)}& \to \mathcal{B}^{(2)}\equiv   \frac{1}{3}
\hat{\alpha} \otimes  \hat{ \openone} \otimes\hat{\alpha},\\
\mathcal{B}^{(23)},\ \mathcal{B}^{(24)}& \to \mathcal{B}^{(1)}\equiv   \frac{1}{3}
\hat{ \openone} \otimes   \hat{\alpha} \otimes\hat{\alpha},\\
\mathcal{C}^{(3)},\mathcal{C}^{(4)}& \to \mathcal{C}\equiv \frac{\epsilon^{\alpha\beta\gamma}}{6}
\hat{\alpha}\otimes  \hat{\beta}\otimes \hat{\gamma},
\\
\mathcal{C}^{(1)}&  \to i {B}^{(1)}, \qquad \mathcal{C}^{(2)} \to i {B}^{(2)},
\\
\mathcal{D}^{(2)} & \to \frac{6}{5}\mathcal{B}^{(3)},\\
\mathcal{D}^{(3)} &\to  \frac{2}{5}\mathcal{B}^{(3)}+ \frac{4i}{5}\mathcal{C},\\
\mathcal{D}^{(4)} &\to  \frac{2}{5}\mathcal{B}^{(3)}- \frac{4i}{5}\mathcal{C}.
\end{aligned} 
\label{eq:conv2}
\end{equation}

Let us convolute both sides of Eq.~(\ref{eq:O4}) using the rules (\ref{eq:conv1}) -- (\ref{eq:conv2}).
We find 
\begin{equation*}
\begin{split}
\hat{\cal O}^{(2)}&(t)  \otimes \hat{\mathbf \openone} 
= \sum_{\{{\mathbf V}_i\}} P_{\{{\mathbf V}_i\}} \bigotimes_{i=1}^n\left(q_i^{(1)}+q_i^{(2)}+q_i^{(3)}+q_i^{(4)}\right),\\
q_i^{(1)}&\equiv V_i^{(1)}\hat{ \openone}^{\otimes 3}_i+V_i^{(2)}\mathcal{B}^{(3)}_i,
\\
q_i^{(2)}& \equiv V_i^{(7)}\hat{ \openone}^{\otimes 3}_i+\frac{2}{5}\left(3V_i^{(12)}+V_i^{(13)}+V_i^{(14)}\right)\mathcal{B}^{(3)}_i,
\\
q_i^{(3)}& \equiv \left(V_i^{(3)}+V_i^{(4)}+iV_i^{(8)}\right)\mathcal{B}^{(1)}_i
\\ + &
\left(V_i^{(5)}+V_i^{(6)}+iV_i^{(9)}\right)\mathcal{B}^{(2)}_i,
\\
q_i^{(4)}& \equiv \left(V_i^{(10)}+V_i^{(11)}+\frac{4i}{5}V_i^{(13)}
-\frac{4i}{5}V_i^{(14)}\right)\mathcal{C}_i.
\end{split}
\label{eq:conv3}
\end{equation*}
Here, the superscript in vector $V_i^{(\cdot)}$ enumerates components according to Eq.~(\ref{eq:Q}).
The total result is real as the imaginary terms are always generated in pairs.

Configurations with $V_i^{(1)}=u_i$, $V_i^{(2)}=v_i$, $V_i^{(3,..,14)}=0$, exactly reproduce averaged result (\ref{eq:OvsB}).
It means that all the other terms exactly cancel each other. It is possible only if the corresponding formfactors can be of both signs.
Therefore, any finite unconstrained sampling leads to an arbitrary result (sign  problem). Moreover, the constraints are non-local.
Consider, e.g. cancellation of contribution proportional to $\mathcal{B}^{(1)}_i\mathcal{B}^{(1)}_j$. Cancellation occurs
only if the formfactors different by replacement $V_i^{(8)}V_j^{(8)} \to V_i^{(a)}V_j^{(b)}$,  $a, b = \{3, 4\}$ must be kept precisely the same.
The requirement quickly becomes intractable with the increasing of the number of non-trivial matrices involved into cancellation.

To summarize, the general requirements on the evolution of the $\hat{\cal O}^{(4)}(t)$ lead to the sign and locality problems.
Those two facts render a brute force classical Monte Carlo algorithm impossible. At present, we are not aware of any algorithm enabling
to circumvent those
obstacles.

\subsection{Population Dynamics for iSWAP Gate Sets Implemented in the Main Text} \label{sec:expGateSet}

Circuits with $\theta = \pi/2$ were used to realize Clifford as well as a universal ensemble. It is therefore instructive to study dynamics of the specific gate sets used in the experiment in more detail. Consider conjugation of a pair of Pauli operators $\hat{\alpha}_i \hat{\beta}_j$ by the iSWAP gate $\hat{S}$. It maps the pair onto another pair of Pauli operators $\hat{\gamma}_i \hat{\delta}_j$ according to the following rules,
\begin{gather*}
\begin{array}{|c|c|c|c|c|c|c|c|}
\hline
\hat{\alpha}_i \hat{\beta}_j
& \hat{X}_i \hat{\openone}_j &  \hat{Y}_i \hat{\openone}_j & \hat{Z}_i \hat{\openone}_j & 
\hat{Z}_i  \hat{X}_j& \hat{Z}_i  \hat{Y}_j & \hat{\alpha}_i \hat{\alpha}_j & \hat{X}_i \hat{Y}_j \\
\hline
S^{\dagger} \hat{\alpha}_i \hat{\beta}_j S
& - \hat{Z}_i\hat{Y}_j &  \hat{Z}_i\hat{X}_j & \hat{\openone}_i  \hat{Z}_j & 
- \hat{Y}_i \hat{\openone}_j & \hat{X}_i \hat{\openone}_j & \hat{\alpha}_i \hat{\alpha}_j 
&  \hat{Y}_i \hat{X}_j\\
\hline
\end{array}
\end{gather*}

\subsubsection{Clifford Gate Set}

Clifford gate set used to obtain the data in the main text is drawn form the single qubit gate set $\{\hat{G}\} = \{ \sqrt{X}, \sqrt{X^{-1}}, \sqrt{Y}, \sqrt{Y^{-1}} \}$. Averaging over this gates set of a symmetric pair of Pauli operators $\alpha_i \otimes \alpha_i$ reads,
\begin{gather}
\overline{ \hat{G}^{\dagger}\hat{\alpha } \hat{G} \otimes \hat{G}^{\dagger } \hat{\alpha }\hat{G}} = \Xi _{\alpha },
\end{gather}
where we introduce the basis $\Xi_\alpha = \{ \openone, \mathbb{X},\mathbb{Y}, \mathbb{Z}\}$,
\begin{gather*}
\mathbb{X} = \frac{1}{2}(\hat{Z}^{\otimes2} + \hat{Y}^{\otimes2}),\\
\mathbb{Y} = \frac{1}{2}(\hat{X}^{\otimes2} + \hat{Z}^{\otimes2}),\\
\mathbb{Z} = \frac{1}{2}(\hat{X}^{\otimes2} + \hat{Y}^{\otimes2}).
\end{gather*}
Single qubit gate average can be described in terms of the invariants $\Xi_\alpha \to M^{(S)}_{\alpha \beta} \Xi_\beta$, 
\begin{gather}
M^{(S)} =
\left(
\begin{array}{cccc}
1 & 0 & 0 & 0 \\
0 &  0 & \tfrac{1}{2} & \tfrac{1}{2} \\
0 & \tfrac{1}{2} & 0 & \tfrac{1}{2} \\
0 & \tfrac{1}{2} & \tfrac{1}{2} & 0 \\
\end{array}
\right). \label{eq:CliffordSQ}
\end{gather}
In this new basis $\{\Xi _{\gamma }\Xi _{\delta }\}$ instead of a $4\times4$ matrix $\Omega$ in Eq.~(\ref{eq:Omega}) effect of iSWAP gate is described by $16\times16$ matrix constructed straightforwardly from the rules Eqs.~(\ref{eq:CliffordSQ})~and the rules for iSWAP.

\subsubsection{Universal Gate Set}

\begin{figure}
	\centering
	\includegraphics[width = 1 \columnwidth]{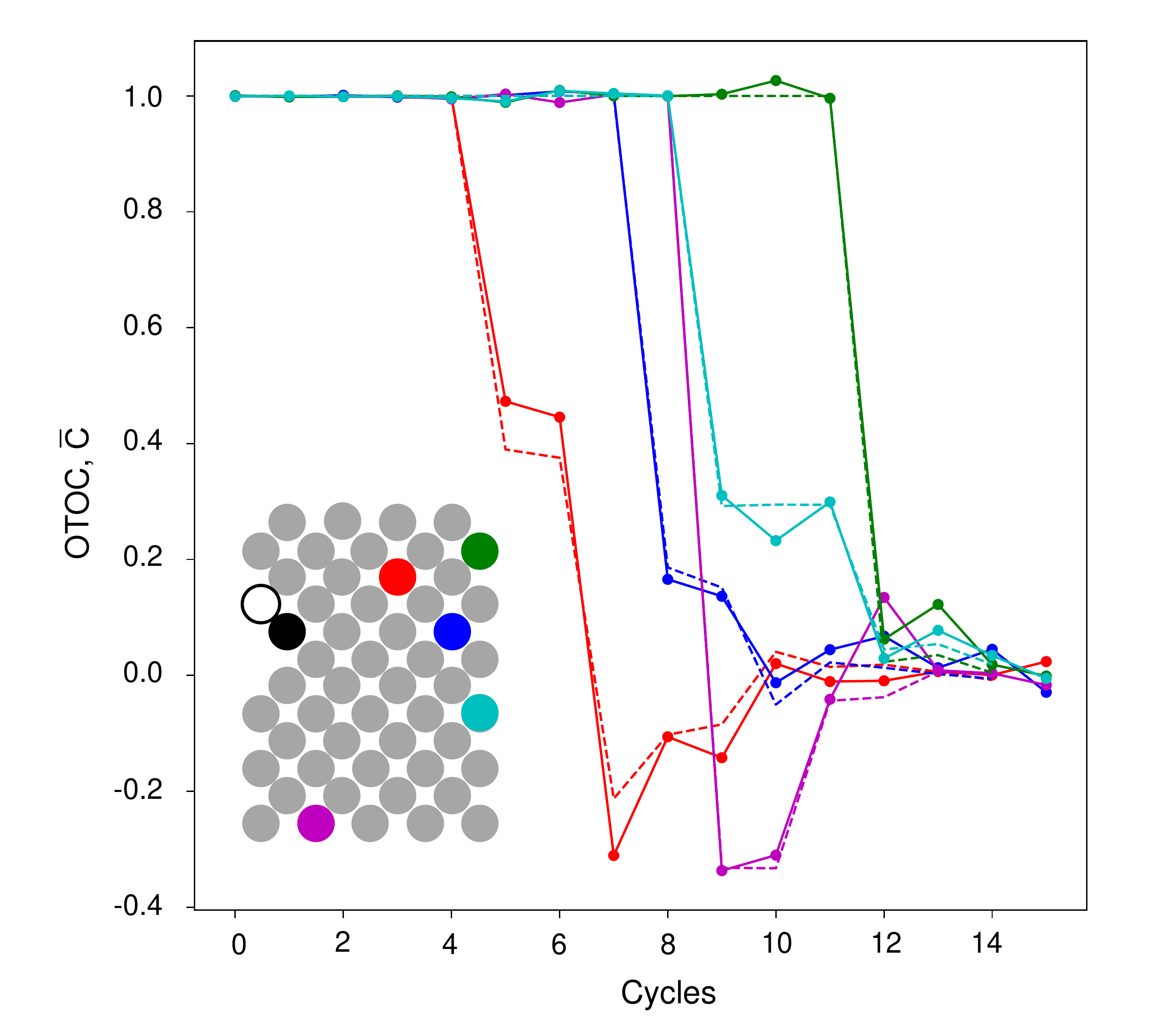}
	\caption{Extended data for Fig.~2 in the main text: Average OTOC for four different locations of butterfly operator, $X$. Solid lines correspond to experimental data and dashed lines to the population dynamics simulation with Eq.~(\ref{eq:UniversalSet}).  }
	\label{fig:iSWAPOTOC}
\end{figure}

The universal gates set used in the main text consists of eight choices of single qubit gates $\{\hat{G}\} = \{ \sqrt{X^{\pm1}}, \sqrt{Y^{\pm1}}, \sqrt{W^{\pm1}}, \sqrt{V^{\pm1}} \}$.
For the butterfly operator we choose $\hat{O}_i = \hat{X}_i$ the dynamics is described in the reduced subspace spanned by the basis $\Lambda = \{ \openone^{\otimes 2}, \hat{X}^{\otimes 2}, \hat{Y}^{\otimes2}, \hat{Z}^{\otimes2} \}$. Average of a pair of Pauli operators over this ensemble reads $\Lambda_\alpha \to M^{(U)}_{\alpha \beta} \Lambda_\beta$,
\begin{gather}
M^{(U)} =
\left(
\begin{array}{cccc}
1 & 0 & 0 & 0 \\
0 &  \tfrac{3}{8} & \tfrac{1}{8} & \tfrac{1}{2} \\
0 & \tfrac{1}{8} & \tfrac{3}{8} & \tfrac{1}{2} \\
0&  \tfrac{1}{2} & \tfrac{1}{2} & 0 \\
\end{array}
\right), \label{eq:UniversalSet}
\end{gather}
Together with the rules for iSWAP gate it is straightforward to generate $16\times 16$ matrix defining the Markov population dynamics process.  Comparison of this noise-free population dynamics prediction for several different circuits with different locations of the butterfly operator $X$ throughout the 53 qubit chip implemented experimentally is shown in Fig.~\ref{fig:iSWAPOTOC}. 

\section{Efficient Population Dynamics for Noisy Circuits} \label{sec:NoisyPopDyn}

\subsection{Inversion Error} \label{InversionError}

The circuit to measure OTOC require inversion of gates $\hat{U}_{i,j}(\theta,\phi )$ which is not perfect. An important source of error is non-invertible phase $\phi$ such that instead of $\hat{U}_{i,j}^\dag(\theta,\phi )$ the gate $\hat{U}_{i,j}(-\theta,\phi )$ is implemented. This inversion error can be included in the Markov population dynamics process in the following way,
\begin{gather*}
\Omega
=
\left(
\begin{array}{cccc}
\cos^2 \phi & 0 & 0 & \sin^2 \phi \\
0 & 1-\tilde{a}-b & \tilde{a} & b \\
0 & \tilde{a} & 1-\tilde{a}-b & b \\
\tfrac{\sin^2 \phi}{9} & \frac{b}{3} & \frac{b}{3} & \tfrac{8+\cos^2\phi}{9}-\frac{2}{3}b \\
\end{array}
\right), \\
\tilde{a} = \frac{1}{3}\left(\cos
^2\phi \sin ^4\theta  +\sin ^2\phi \cos ^4\theta \right).
\end{gather*}

\subsection{Generic Error Model} \label{GeneralError}

\begin{figure}
	\centering
	\includegraphics[width = 1 \columnwidth]{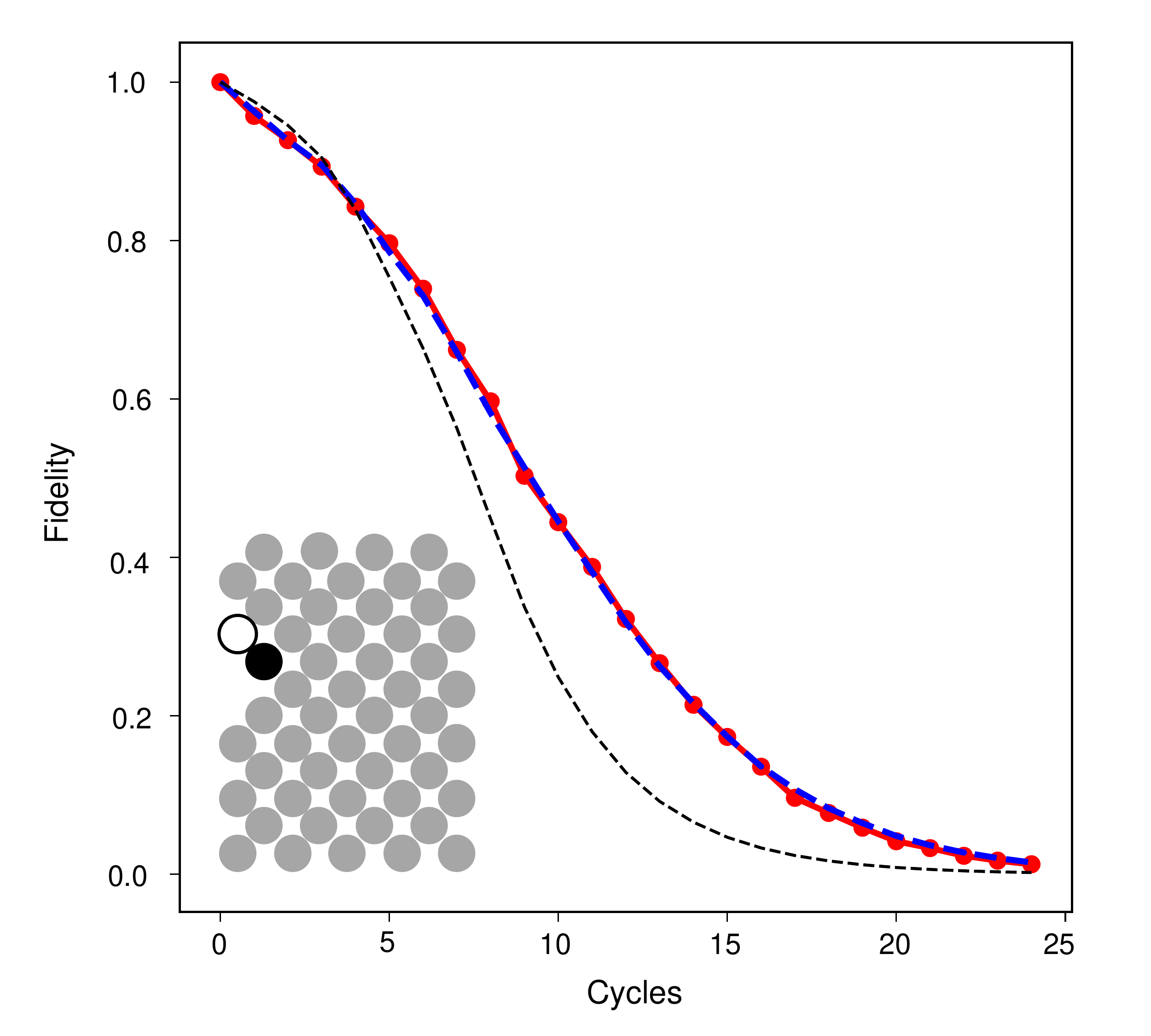}
	\caption{Average fidelity of OTOC, i.e. the circuit with no butterfly applied shown as red line. OTOC fidelity from Monte Carlo simulation of noisy population dynamics Sec.~\ref{sec:NoisyPopDyn}. The Pauli error rate $r_p = 0.013$ is determined by fitting the numerics to the experimental data. This value is within $10\%$ of the gate fidelity determined in a separate experiment benchmarking two-qubit gates, see Sec.~\ref{exp_tech}. This is the single parameter of the model which is used to reproduce the rest of the OTOC data for 51 different locations of butterfly operator on the chip. Black dashed line shows a naive estimate of circuit fidelity via product of fidelities of gates within the light cone. Note that naive fidelity decays much faster with time than OTOC fidelity.  }
	\label{fig:sqrtiSWAPfidelity}
\end{figure}

For an ensemble that satisfies Eq.~(\ref{eq:SymGateSet}) the effect of relaxation and dephasing in the quantum processor can be captured by a one and two qubit depolarizing channel noise model,
\begin{gather}
\rho \to (1-p_1) \rho + \frac{p_1}{3} \sum_{\hat{\alpha} =\hat{X}, \hat{Y}, \hat{Z}} \hat{\alpha} \rho \hat{\alpha},\\
\rho \to (1-p_2) \rho + \frac{p_2}{15} \sum_{\hat{\alpha},\hat{\beta} } \hat{\alpha} \hat{\beta} \rho \hat{\alpha} \hat{\beta}.
\end{gather}
In this case the Markov process Eq.~(\ref{eq:Omega}) can be modified in a straight forward way. For a single qubit depolarizing channel the Markov process Eq.~(\ref{eq:Omega}) is supplemented by the exponential decay rate as follows,
\begin{gather}
\openone_i^{\otimes2} \rightarrow \openone_i^{\otimes2},~\label{eq:B1Qnoise1}\\
\mathcal{B}_i \rightarrow e^{-4p_1/3} \mathcal{B}_i,~\label{eq:B1Qnoise2}
\end{gather}
for each bond per gate cycle. Two-qubit depolarizing channel noise is accounted by supplementing the Markov process by the following,
\begin{gather}
\openone_i^{\otimes2}  \openone_j^{\otimes2} \rightarrow  \openone_i^{\otimes2}  \openone_j^{\otimes2}, ~\label{eq:B2Qnoise1} \\
\mathcal{B}_i \openone_j^{\otimes2} \rightarrow e^{- \frac{16}{15}p_2} \mathcal{B}_i  \openone_j^{\otimes2},~\label{eq:B2Qnoise2} \\
\mathcal{B}_i \mathcal{B}_j \rightarrow e^{- \frac{16}{15}p_2} \mathcal{B}_i  \mathcal{B}_j.~\label{eq:B2Qnoise3}
\end{gather}
Note that the effect of noise on the Markov population dynamics cannot be described by the global depolarizing channel model that is often conjectured for ergodic circuits. Instead the time dependence of OTOC will demonstrate characteristic time dependence that can be used to verify the experimental data.

\begin{figure}
	\centering
	\includegraphics[width = 1 \columnwidth]{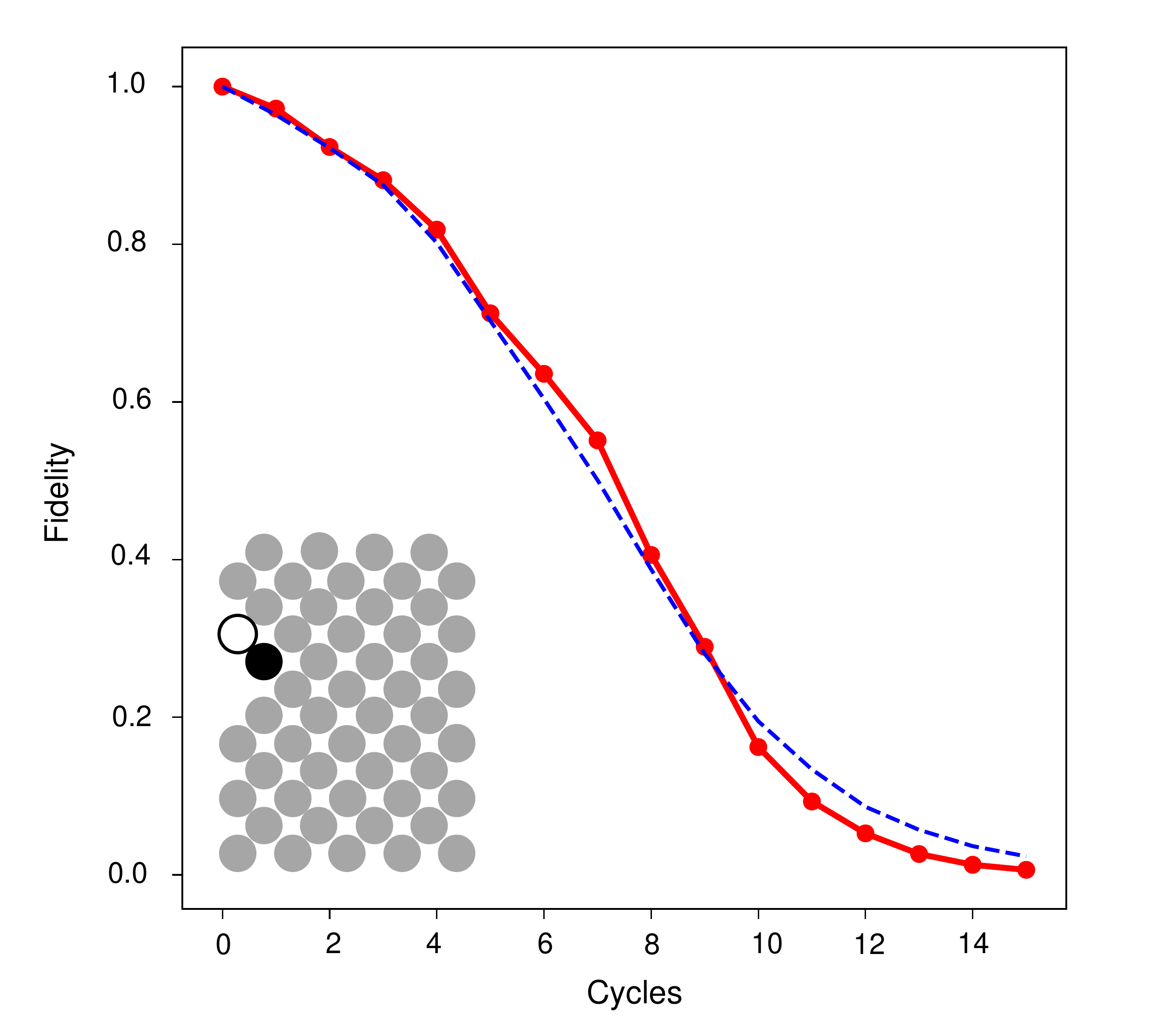}
	\caption{OTOC circuits (with no butterfly applied) for universal iSWAP gate set. Solid red line shows experimental data and dashed blue line is the theoretical fit with a single parameter, Pauli error rate. The extracted Pauli error rate is $r_p = 0.012$ within $20\%$ of the value obtained via two-qubit gate benchmarking.  }
	\label{fig:iSWAPfidelity}
\end{figure}

\begin{figure}[t]
	\centering
	\includegraphics[width = 1 \columnwidth]{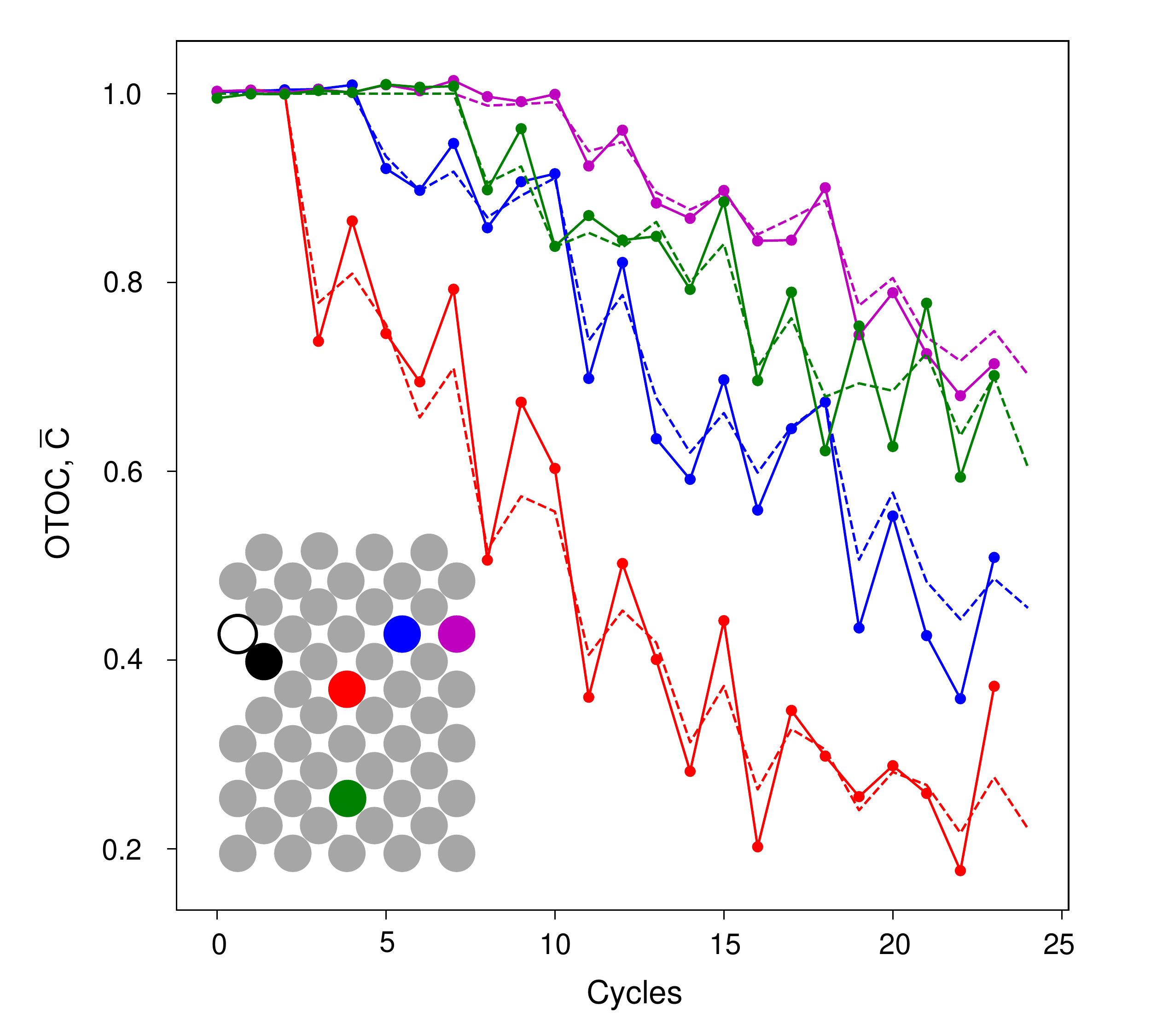}
	\caption{ Extended data for Fig.~2 of the main text: Normalized OTOC for four different locations of butterfly operator $Z$. Solid lines show experimental data, and dashed lines correspond to the noisy population dynamics. }
	\label{fig:sqrtiSWAPOTOC}
\end{figure}

The described procedure allows us to verify the experimental results for average fidelity of OTOC circuits, the circuit with no butterfly applied, by direct comparison to the noisy population dynamics, see Fig.~\ref{fig:sqrtiSWAPfidelity} and Fig.~\ref{fig:iSWAPfidelity}. We use the two-qubit Pauli error as a fitting parameter. Best fit corresponds to errors within $10\%$ of the average error of two qubit cycle measured independently. We use the extracted error to predict values of OTOC for every position of the butterfly with no additional fitting. Comparison of the values of normalized OTOC predicted in this way with experimental data is shown in Fig.~\ref{fig:sqrtiSWAPOTOC}. This procedure introduces substantial noise-dependent bias into the observed OTOC values, which is illustrated in Fig.~\ref{fig:sqrtiSWAPOTOCvsNoiseFree}.

\subsection{Error Mitigation for Average OTOC}

We estimate circuit fidelity from the circuit shown in, Fig.~1~A, B, without the butterfly operator. This fidelity estimate is used for the error mitigation procedure aplied to the data in Fig.~2. In the error-free circuit absence of the butterfly operator means that $U$ and $U^\dagger$ cancel each other exactly resulting in $C_{z0}(t)=1$. In practice, inversion is imperfect as detailed in Sec.~\ref{InversionError}. Both errors in the unitary parameters and the effect of noise are reflected in the time dependence of $C_{z0}(t)<1$ which serves as circuit fidelity, an analog of Loschmidt echo for local operator $\hat{M}$. 

Average over circuits reads,
\begin{gather}
\overline{C}_{0z} = \mathcal{F}_{\openone^{\otimes2}_1} +  \mathcal{F}_{\mathcal{B}_1}, \label{eq:avCz0}\\
\overline{C}_{zz} = \mathcal{F}_{\openone^{\otimes2}_1} - \frac{1}{3} \mathcal{F}_{\mathcal{B}_1}.\label{eq:avCzz}
\end{gather}
where the probabilities of vacuum $\mathcal{F}_{\openone^{\otimes2}_1}$ and bond $\mathcal{F}_{\mathcal{B}_1}$ at the measurement site are described by the population dynamics Eqs.~(\ref{eq:Omega}) with respective decay rates Eqs.~(\ref{eq:B1Qnoise1}-\ref{eq:B2Qnoise3}). Note that the decay rate grows with the extent of spreading of the butterfly operator, resulting in the decay of fidelity that is not a simple exponent. Moreover, in general $\mathcal{F}_{\openone^{\otimes2}_1}, \mathcal{F}_{\mathcal{B}_1}$ is not a simple probability no error occurred, as one could naively expect, see Fig.~\ref{fig:sqrtiSWAPfidelity} for comparison.

\begin{figure}[t]
	\centering
	\includegraphics[width = 1\columnwidth]{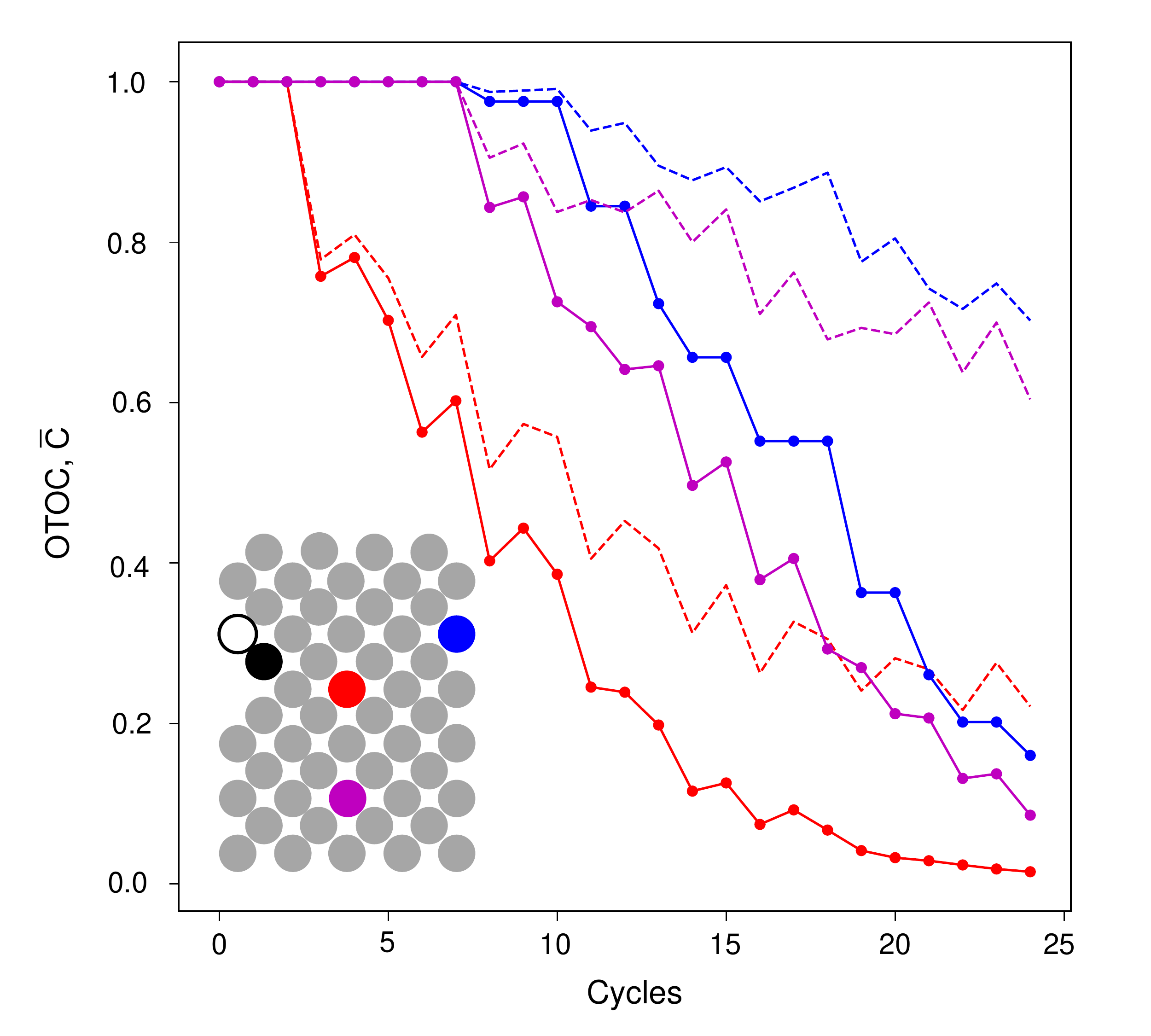}
	\caption{Comparison of noisy OTOC obtained by normalization procedure, dashed lines, to noise-free population dynamics for the same circuit, solid lines. }
	\label{fig:sqrtiSWAPOTOCvsNoiseFree}
\end{figure}

The ratio of $\overline{C}_{zz} /\overline{C}_{0z}$ is compared to normalized data in Fig.~\ref{fig:sqrtiSWAPOTOC}.
Note that this ratio does not correspond to the noise free OTOC, Fig.~\ref{fig:sqrtiSWAPOTOCvsNoiseFree}. This is because the probability of vacuum at measure site $\mathcal{F}_{\openone^{\otimes2}_1}$ decays slower with respect to its noise free value $p(\openone^{\otimes2}_1)$ than the probability of the bond $\mathcal{F}_{\mathcal{B}_1}$. The latter corresponds to the weight of the operators which span the distance between measure and butterfly qubit which are more susceptible to noise. As a result normalized OTOC from noisy circuits overestimates the value of noise-free OTOC. For individual circuits such a simple description is no longer valid, as demonstrated by the dependence of fidelity estimate $C_{z0}(t)$ on the circuit instance, see Sec.~\ref{sec:NoiseNumerics}. In many cases circuit dependent corrections are relatively small and the normalization procedure still works for individual circuits as well. Nonetheless, for our OTOC fluctuations data we develop a more efficient error mitigation procedure described in the following section.

\subsection{Theory of Error Mitigation for Individual Circuits}

We use a more precise error mitigation procedure for individual OTOC measurements presented in Figs.~3, 4 of the main text. These circuits contain iSWAP entangling gate and OTOC value can be expanded in terms of contributions from Clifford circuits. In presence of non-Clifford gates the butterfly operator can be conveniently expanded into Pauli strings $B_i$,
\begin{gather*}
O(t) = \sum w_{\alpha_1 ... \alpha_n} B_{\alpha_1 ... \alpha_n}, \\
B_{\alpha_1 ... \alpha_n} = \hat{\alpha}_1 \otimes ...  \otimes \hat{\alpha}_n
\end{gather*}
For the initial state used in the experimental protocol $\ket{\psi} = \bigotimes \ket{+}_i$, the value of OTOC is expanded as,
\begin{gather}
C = \sum w_{\alpha \alpha_2...\alpha_n} w_{\beta\alpha_2...\alpha_n} \bra{+_1} \sigma^z_1 \sigma^{\alpha}_1 \sigma^z_1 \sigma^{\beta}_1\ket{+}_1. \label{eq:Cexp}
\end{gather}
The real part of the OTOC corresponds to $\alpha = \beta$,
\begin{gather}
\mathrm{Re} C = \sum w_{\alpha_1 ... \alpha_n}^2 \kappa_{\alpha_1},
\end{gather}
where $\kappa_{\alpha} = \{1, -1, -1, 1\}$ for $\alpha = \{0, x, y, z\}$.

For Clifford circuits ideal value of OTOC can be calculated efficiently. It can then be used to calculate circuit fidelity  by comparing data with the expected value. The fidelity of the OTOC in presence of non-Clifford gates is then calculated by sampling a subset of OTOCs for Cliffords that appear in its expansion, see Eq.~(\ref{eq:Cexp}). The fidelity is calculated by averaging over fidelities of individual Clifford contributions. Averaging reduces the circuit specific effect of noise and gives a more accurate estimate of fidelity. 

\section{Numerical Simulation of Error-Limiting Mechanisms for OTOC}~\label{sec:NoiseNumerics}

In this section, we provide numerical simulation results aimed at identifying the potential error sources that limit our experimental accuracy in resolving OTOCs. As demonstrated in Section~\ref{ext_data}, shot noise from finite statistical sampling is unlikely the dominant mechanism. The remaining known error channels are: 1. Incoherent, depolarizing noise in the quantum circuits, which can arise from qubit dephasing or relaxation. 2. Coherent errors in the quantum gates, e.g. the remnant conditional-phase $\phi$ demonstrated in Section~\ref{exp_tech}. We study each of these errors below.

\subsection{Coherent and Incoherent Contributions}

\begin{figure}[!t]
	\centering
	\includegraphics[width=1\columnwidth]{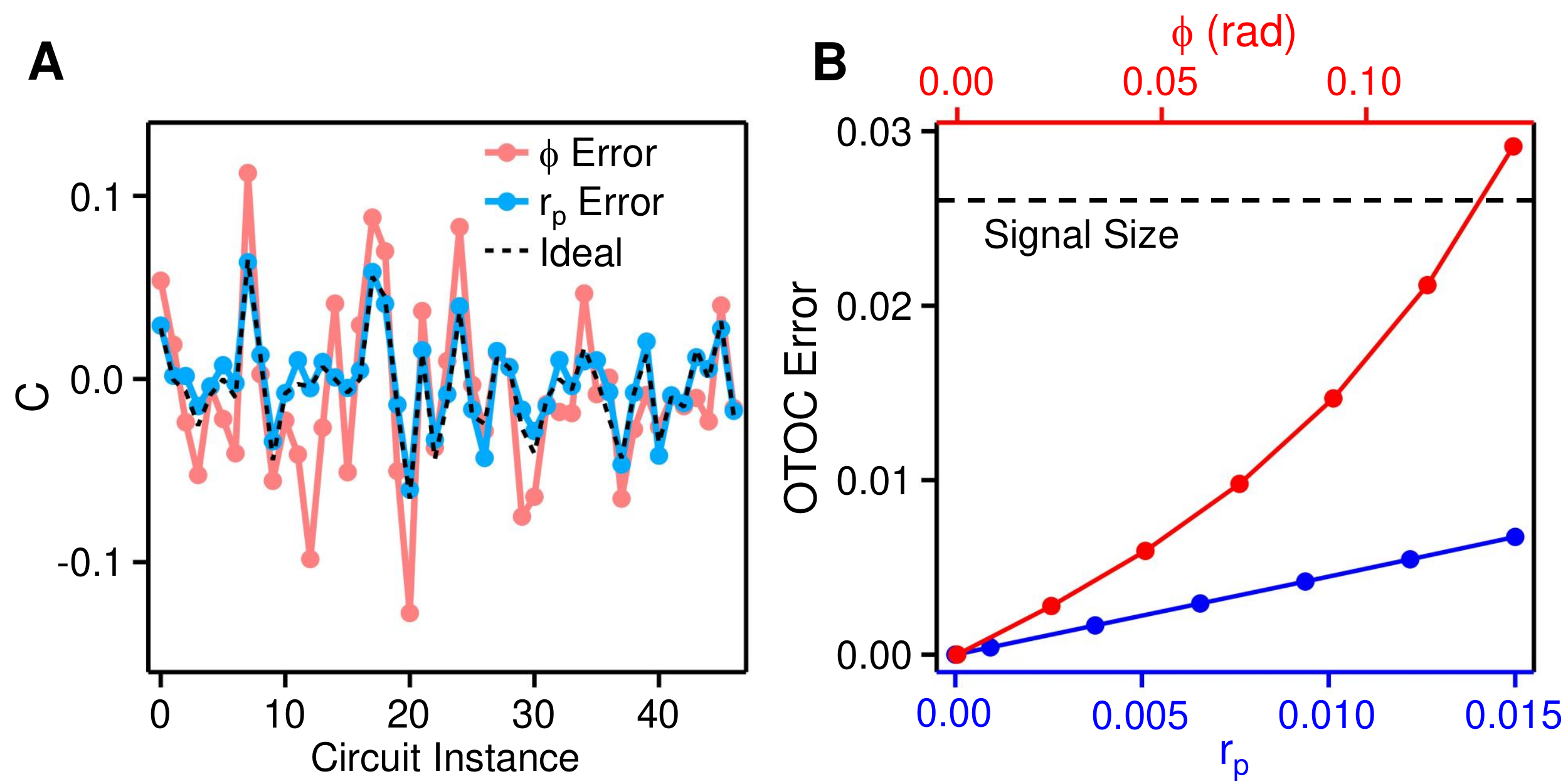}
	\caption{Simulation of incoherent and coherent OTOC errors. (A) OTOCs, $C$, of 47 quantum circuit instances simulated in three different ways: with $\phi$ error, where a conditional-phase of $0.136$ rad is added to every two-qubit gate; with $r_\text{p}$ error, where a Pauli error of 0.015 is added to each two-qubit gate; ideal, where no error is introduced. The qubit configuration used here is a 1D chain of 10 qubits wherein $Q_1$ and $Q_\text{b}$ reside at opposite ends of the chain. The two-qubit gate is iSWAP and $N_\text{D} = 12$ non-Clifford gates are used for each instance. The number of circuit cycles is 34 ($N_\text{S} = 250$). (B) The scaling of OTOC error (i.e. RMS deviation from the ideal OTOC values) against $\phi$ (red) and $r_\text{p}$ (blue). Dashed line shows the size of the OTOC signal (i.e. RMS value of the ideal OTOCs).}
	\label{fig:distortion-curve}
\end{figure}

We first describe how OTOC error from depolarizing noise may be simulated. We consider a depolarizing channel model parameterized by an error probability $p$,
\begin{equation}
\label{eq:depol}
\mathcal{E}(\rho) = (1-p)\rho + p\frac{\mathbf{I}}{d},
\end{equation}
where $d=2^n$ is the dimension, and $\mathbf{I}$ the identity operator. The Kraus operators of this map are all Pauli strings of length $n$, where each non-trivial string has weight $\frac{p}{d^2}$. We consider a model where after each two qubit gate (these typically dominate the loss in fidelity compared to single-qubit gates), a two-qubit depolarizing channel is applied.

In Fig.~\ref{fig:distortion-curve}A, we show a number of instance-dependent OTOC values $C$ simulated using full density matrix calculations and an experimentally measured Pauli error rate $r_\text{p}$ of 0.015 ($r_{\text{p}}=15p/16$ in Eq.~\eqref{eq:depol}, due to the trivial Pauli string in the Kraus map from the identity matrix). Here we have used a smaller number of qubits due to the high cost of density-matrix simulation. To match with experiment, we have adjusted the number of circuit cycles to yield a total number of iSWAP gates close to those shown in Fig.~4 of the main text. The same normalization protocol as the experiment is also adopted, such that $\braket{\hat{\sigma}_\text{y}}$ is simulated with and without the butterfly operator and their ratio is recorded as $C$. The results, compared to the ideal OTOC values also plotted in Fig.~\ref{fig:distortion-curve}B, show little deviation.

Next, we simulate OTOCs of the same circuits with $r_\text{p} = 0$ but an experimentally measured conditional-phase of $\phi = 0.136$ rad on each iSWAP gate and its inverse. The results are also plotted in Fig.~\ref{fig:distortion-curve}A and seen to deviate much more from the ideal values. To quantify these observations, we have plotted the OTOC error as a function of both $r_\text{p}$ and $\phi$ in Fig.~\ref{fig:distortion-curve}B. Here we see that at the experimental limit ($r_\text{p} = 0.015$ and $\phi = 0.136$ rad), the OTOC error is dominated by the contribution from $\phi$ (where it is $\sim$0.03) rather than $r_\text{p}$ (where it is $\sim$0.006). Interestingly, the SNR is about 0.9 for $\phi = 0.136$ rad, which is close to the value measured in Fig.~4 of the main text for $N_\text{S} \approx 250$.

Figure~\ref{fig:distortion-curve}B also provides preliminary indication of how the OTOC accuracy for our experiments may be improved as we decrease both $\phi$ and $r_\text{p}$. We see that the OTOC error is linearly proportional to $r_\text{p}$ whereas its scaling is steeper against $\phi$. In particular, reducing $\phi$ by a factor of 2 leads to an OTOC error that is 3 times lower. Although these results may depend on the number of qubits and the structure of circuits, their similarity to experimental data nevertheless provides hints that reducing the conditional-phases in our iSWAP gates can potentially lead to large improvements in the OTOC accuracy. This can, for example, be achieved through concatenated pulses demonstrated in Ref.~\cite{Foxen_PRL_2020}.

\subsection{Perturbative Expansion of OTOC Error}

\begin{figure}[!t]
	\centering
	\includegraphics[width=1\columnwidth]{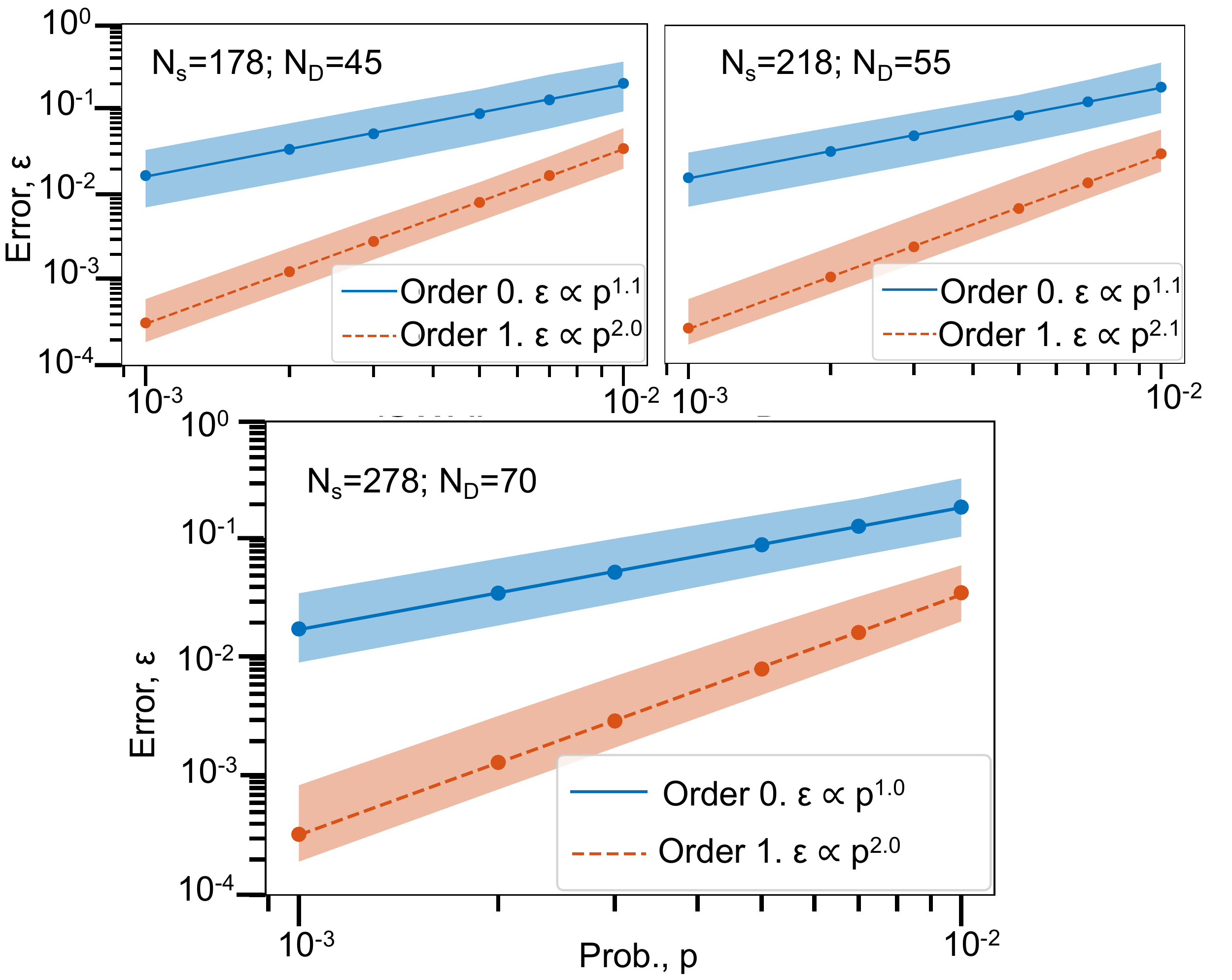}
	\caption{Relative error $\epsilon$ of the 0'th and 1'st order error approximation (from Eq.~\eqref{eq:single-error-approx}) calculated by a pure state simulation, comparing to the exact value from a density matrix calculation, as a function of depolarizing probability $p$ (Eq.\eqref{eq:depol}). Here we compute $C_\text{zx}$, varying the number of iSWAP ($N_s$) and non-Clifford gates ($N_D$), with 11 qubits on a chain. 
		The dots are the median over 48 circuits with lines of best fit given by the legend. The shaded region is the middle 50\% of the data. The scaling of the error is close to $\epsilon \propto p^2$ as expected for the one-error approximation in all three cases. }
	\label{fig:one-err}
\end{figure}

The exact density-matrix calculation used in the preceding section is costly to implement and becomes quickly intractable as the number of qubits increase. One can more systematically estimate the instance specific noise contribution from a perturbation theory expansion of the quantum map Eq.~\eqref{eq:depol}, with an entirely pure state calculation. Let us adopt the notation $\sigma_{m_1,m_2}^{(i,j)} = \sigma_{m_1}^{(i)}\sigma_{m_2}^{(j)}$, where $\sigma_m^{(i)}$ is the $m$'th Pauli operator ($m \in \{0,x,y,z\}$) applied on qubit $i$. We will also call $C[\sigma_{m_1,m_2}^{(i,j)}(d)]$ the OTOC value with the additional `error gate' $\sigma_{m_1,m_2}^{(i,j)}$ inserted at layer $d$ in the circuit. Then, to accuracy $O(p^2)$ one has
\begin{equation}
\begin{split}
\label{eq:single-error-approx}
& C_p = (1-p)^{n_2} C_{\mathrm{ideal}}  \\ 
& + \frac{p}{15}(1-p)^{n_2 - 1} \sum_{\substack{(m_1,m_2)\neq (0,0), \\ \langle i, j\rangle, d}} C[\sigma_{m_1,m_2}^{(i,j)}(d)],
\end{split}
\end{equation}
where $C_{\mathrm{ideal}}$ is the OTOC value in the limit of no noise, $C_p$ the first order approximation of $C_{\mathrm{ideal}}$ in parameter $p$, and $n_2$ the total number of two-qubit gates in the circuit, occurring over pairs defined by $\langle i, j\rangle$ (i.e. Eq.~\eqref{eq:single-error-approx} is a sum over all error terms in the circuit). For convenience, we have also redefined $p \rightarrow 15p/16$ from Eq.~\eqref{eq:depol} due to the trivial contribution from the all zero Pauli string. 

Eq.~\eqref{eq:single-error-approx} can be used to separate out the ideal OTOC value of an individual circuit, from the noise contribution. This noise contribution can be computed using $15\times n_2$ circuit simulations (inserting each Pauli pair at all two-qubit gate locations in the circuit), though in some cases symmetries can be used to reduce this. For example, for the normalization curve $C_\text{z0}$, only around half of this is required, since for each error in the reverse circuit $U^\dag$, there is an equivalent one in $U$. 
Of course one can continue this expansion to arbitrary order, however the number of terms quickly becomes infeasible to compute (the $k$'th order contains $15 \times {n_2 \choose k}$ terms).


In Fig.~\ref{fig:one-err} we show the error scaling (comparing to an exact density matrix computation) for the zero'th (i.e. only the first term in Eq.~\eqref{eq:single-error-approx}) and first order approximation for $C_{\text{zx}}$, giving scaling in the error as expected; the error $\epsilon$ is computed as $|C_p(\text{exact}) - C_p(\text{approx})|/|C_p(\text{exact})|$, where $C_p(\text{approx})$ is from Eq.~\eqref{eq:single-error-approx}, and $C_p(\text{exact})$ the value from the density matrix calculation with noise rate $p$. Moreover, the accuracy of the approximation remains fairly consistent, for a fixed noise level ($p$) for the three different circuit sizes (using a similar number of iSWAP gates as in the main text).

Here we have outlined a protocol of simulating the instance specific error contribution to the OTOC. This can be used to more accurately separate the instance specific OTOC value, systematic errors, and contributions from gate noise. Of course, the overhead in the simulation is a factor of $15 n_2$, which can itself become challenging in deep enough circuits, although statistical sampling may be feasible in some cases.

\twocolumngrid


\begin{thebibliography}{68}
\bibitem [{foo()}]{footnote}%
  \BibitemOpen
  \href@noop {} {\bibinfo  {journal} {$C$ in some cases can have a
  non-negligible imaginary part which is measured through the $\langle
  \hat{\sigma}_\text{x} \rangle$ of $Q_\text{a}$. For this work, we only
  measure the real part of $C$ which is equal to $\langle \hat{\sigma}_\text{y}
  \rangle$ of $Q_\text{a}$}\ }\BibitemShut {NoStop}%
\end{thebibliography}
\end{document}